\documentclass[a4paper,12pt]{article}
\pdfoutput=1

\usepackage[height=23truecm,width=16truecm,top=3.05truecm]{geometry}

\usepackage[utf8]{inputenc}
\usepackage{graphicx}
\usepackage{amsmath}
\usepackage{amssymb}
\usepackage{hyperref}
\hypersetup{colorlinks,citecolor=TokiwaIro,linkcolor=RuriIro,urlcolor=RuriIro,bookmarksopen,bookmarksnumbered,pdfstartview=FitH,linktocpage}
\usepackage{cite}
\usepackage{braket}
\usepackage{bm}
\usepackage{bbm}
\usepackage{color}
\usepackage{float}
\usepackage{hhline}
\usepackage{multirow}
\usepackage{indentfirst}
\usepackage{latexsym}
\usepackage{mathtools}
\usepackage{cancel}
\usepackage{slashed}
\usepackage{colortbl}
\usepackage[mathscr]{euscript}
\usepackage{enumitem}
\usepackage{subcaption}
\usepackage{comment}
\usepackage{amsthm}

\usepackage{arydshln}
\usepackage{ulem}

\numberwithin{equation}{section}

\allowdisplaybreaks

\definecolor{RuriIro}{rgb}{0.,0.28,0.60}
\definecolor{TokiwaIro}{rgb}{0.,0.39,0.16}

\graphicspath{{./fig/}}

\newcommand{\nn}{\nonumber}

\newcommand{\mc}{\mathcal}
\newcommand{\mr}{\mathrm}
\newcommand{\ms}{\mathsf}

\newcommand{\mbb}{\mathbb}

\newcommand{\del}{\partial}
\newcommand{\ol}{\overline}

\newcommand{\rv}{\mathrm{v}}

\newcommand{\conv}{\mathrm{conv}}
\newcommand{\body}{\text{-body}}

\newcommand{\cl}{\mathrm{cl}}

\begin{document}

\begin{titlepage}

\begin{flushright}
KOBE-COSMO-22-12\\
KEK-TH-2448\\
KUNS-2940
\end{flushright}

\vspace{1cm}

\begin{center}

{\Large \bfseries
Quantum current dissipation in superconducting strings and vortons
}

\vspace{1cm}

\renewcommand{\thefootnote}{\fnsymbol{footnote}}
{%
\hypersetup{linkcolor=black}
Yoshihiko Abe$^{1}$\footnote[1]{yabe@diamond.kobe-u.ac.jp},
\ 
Yu Hamada$^{2}$\footnote[2]{yuhamada@post.kek.jp},
\ 
Kota Saji$^{3}$\footnote[3]{ksaji@gauge.scphys.kyoto-u.ac.jp}
and
Koichi Yoshioka$^{3}$\footnote[4]{yoshioka@gauge.scphys.kyoto-u.ac.jp}
}%
\vspace{5mm}

{\itshape%
$^{1}${Department of Physics, Kobe University, Kobe 657-8501, Japan}\\
$^{2}${KEK Theory Center, IPNS, Tsukuba, Ibaraki 305-0801, Japan}\\
$^{3}${Department of Physics, Kyoto University, Kyoto 606-8502, Japan}
}%

\vspace{8mm}

\abstract{% -----
\noindent
In this work, the current stability is discussed for cosmic strings with the bosonic superconductivity. 
A non-vanishing curvature of string generally induce the quantum instability of the current-carrying particle. Its decay rates are explored for various types of model parameters, curved string shapes, and decay processes. 
As a cosmological application, the stability is examined for superconducting strings in the string network and also for cosmic vortons by evaluating their cosmological evolution. 
The zero mode and hence the vorton cannot be stable in various cases, e.g., with a hierarchy between the current-carrying particle mass off the string and the string tension or with sizable couplings of the current-carrying particle to light species such as the Standard Model particles.
}% -----

\end{center}
\end{titlepage}

\renewcommand{\thefootnote}{\arabic{footnote}}
\setcounter{footnote}{0}
\setcounter{page}{1}

\tableofcontents

\bigskip

\section{Introduction}

Topological defects appear in many particle physics models whose 
vacua have non-trivial topology. 
The dynamics of the defects play important roles in the history of 
our Universe. A well-known example is cosmic string (or vortex 
string)~\cite{Vilenkin:2000jqa}, which appears when a model has 
non-simply connected vacuum. 
The cosmic string has been studied for long decades as the probes 
of high-energy physics beyond the Standard Model (SM) of particle 
physics. Furthermore, it can also produce stochastic gravitational 
waves and is expected to be observed in the future gravitational 
wave detectors (e.g., 
Refs.~\cite{Caldwell:2022qsj,LISACosmologyWorkingGroup:2022jok} and 
references therein).

Recently, current-carrying cosmic strings have gathered renewed 
attention. Such strings can be regarded as superconducting 
one-dimensional objects (if the current is charged under the 
electromagnetism), and hence are called superconducting 
strings~\cite{Jackiw:1981ee,Witten:1984eb,Ostriker:1986xc}. 
Since its existence was pointed out in \cite{Witten:1984eb}, it
has been realized that superconducting strings are ubiquitous in 
various extensions of the SM, e.g., the gauged $U(1)_{B-L}$ 
model~\cite{Witten:1984eb,Jeannerot:1996yi} and the axion 
models~\cite{Lazarides:1984zq,Ganoulis:1989hz,Iwazaki:1997bk,Fukuda:2020kym,Abe:2020ure,Agrawal:2020euj}. 
The current-carrying particle can be fermionic or bosonic 
depending on what the string couples with.
The superconductivity often provides non-trivial phenomena 
compared to the case of conventional cosmic strings. 
For instance, the currents that strings carry affect their motions 
and make the string network dynamics different from the 
conventional one via the reconnection 
process~\cite{Bettencourt:1994kc,Bettencourt:1996qe,Copeland:2006eh,Copeland:2006if,Salmi:2007ah,Bevis:2008hg,Bevis:2009az,Hiramatsu:2013yxa,Hiramatsu:2013tga,Martins:2020jbq,Martins:2021cid}.
The current can also affect gravitational wave spectrum from cosmic strings~\cite{Babichev:2003au}, and see also Refs.~\cite{Auclair:2022ylu,Rybak:2022sbo} for the recent works.
In addition, a loop of superconducting string can be a stable object, 
called the vorton~\cite{Davis:1988ij,Davis:1988jq}, which is a 
dark matter candidate in our 
Universe~\cite{Brandenberger:1996zp,Martins:1998th,Martins:1998gb,Cordero-Cid:2002hmv,Peter:2013jj,Auclair:2020wse}. 
The classical stability of the vorton is understood as the balance 
between the centrifugal repulsive force and the attractive 
force associated with the string tension.

Recently, the authors of Ref.~\cite{Ibe:2021ctf} discussed the 
instability of localized zero modes (the current-carrying particles) 
in superconducting strings and vortons, taking into account 
the effects of string bending.
They studied the axion superconducting string in the KSVZ axion 
model~\cite{Kim:1979if,Shifman:1979if}, where the zero mode comes 
from the KSVZ heavy quarks which are assumed to couple to the 
SM particles. Their conclusion is that the decay of the fermionic 
zero mode into light SM  particles is not suppressed and that 
the vorton cannot be long-lived.

On the other hand, it is still unclear whether the bosonic zero 
mode is (un)stable in a similar fashion, while the bosonic 
superconductivity is realized in various models as well. 
In fact, some of the present authors pointed out in 
Ref.~\cite{Abe:2020ure} that the axion string in the DFSZ axion 
model~\cite{Zhitnitsky:1980tq,Dine:1981rt} can be superconducting 
due to the zero modes consisting of the charged Higgs 
and $W$ bosons in addition to fermionic carriers.\footnote{%
Similar cosmic strings appear in the two Higgs doublet model 
with a global $U(1)$ symmetry~\cite{Eto:2018hhg,Eto:2018tnk,Eto:2021dca}.}
Since the decay phenomena of bosonic zero mode may be different 
from that of the fermionic one, it is necessary to clarify the 
charge/current stability in order to discuss the fate 
of superconductivity. 

In this paper, we examine the (in)stability of bosonic 
superconductivity in the $U(1)\times \tilde U(1)$ model. 
Similarly to the result of \cite{Ibe:2021ctf}, a curved
string induces the decay of zero mode to the outside of the 
string. We calculate its decay rate for various cases of 
model parameters, types of curved shapes, and decay processes. 
As a cosmological application, we consider the string network and 
the cosmic vorton. It is found that the zero modes on 
strings and vortons are unstable in many cases. 
For instance, when there is a hierarchy between the mass 
scales of the string tension and the current-carrier particle off the string, 
the zero mode obtains a sufficient energy to escape from the 
trapping potential and easily decays. In addition, when a field 
producing the zero mode has sizable couplings to light particles, 
the corresponding decay mode is not suppressed and the superconducting 
currents in strings and vortons do not survive to date.

The rest of the paper is organized as follows. 
In Section~\ref{sec:U(1)xtU(1)-model}, the $U(1)\times\tilde{U}(1)$ 
model and its superconducting string solution are briefly reviewed. 
The classical vorton stability is also discussed.  
In Section~\ref{sec:decay}, we derive the decay width of zero modes 
from curved strings and give its approximate formulae. 
In Section~\ref{sec:application}, we consider the zero mode decay, 
that is, the superconducting current dissipation in the string 
network and cosmic vortons, and examine whether they are stable 
in various parameter space. 
Section~\ref{sec:conclusions} is devoted to our conclusion. 
In Appendix~\ref{app:details}, we present the scalar part 
Lagrangian and the mass matrix for quantum fluctuations around 
the superconducting string background. 
In Appendix~\ref{app:propagator}, the formal scalar propagators 
on the background string solution are presented. 
The cosmological temperature dependence of the decay width is 
shown in Appendix~\ref{app:Gamma-temperature}.

\bigskip

\section{Bosonic superconducting string}
\label{sec:U(1)xtU(1)-model}

In this section, we introduce the model of bosonic superconducting 
string with $U(1) \times \tilde{U}(1)$ symmetry, where the 
former $U(1)$ and latter $\tilde U(1)$ are gauged and global, 
respectively. 
The global $\tilde U(1)$ symmetry is spontaneously broken in the vacuum 
leading to the global string, which can be superconducting under 
the gauged $U(1)$ symmetry. (The gauged $\tilde U(1)$ case was first 
discussed in Ref.~\cite{Witten:1984eb}.) 
We also give a brief review on the classical vorton stability.

\subsection{String and zero mode}

\begin{table}
  \centering
  \begin{tabular}{|c||c|c|} \hline
    & $U(1)$ (gauge) & $\tilde{U}(1)$ (global)\\
  \hhline{|=#=|=|}
  $\Phi$ & $0$ & $+1$ \\
  \hline
  $\Sigma$ & $+1$ & $0$ \\
  \hline
  $r \to 0$ & broken & unbroken \\
  \hline
  $ r \to \infty$ & unbroken & broken \\
  \hline
  \end{tabular}
  \caption{The quantum charges of scalar fields and the 
  symmetry configuration. The global $\tilde{U}(1)$ symmetry is 
  restored at $r \to 0$ and the gauge $U(1)$ is broken 
  around $r \sim 0$ due to the existence of string.}
  \label{tab:charge}
  \bigskip
\end{table}

The model contains two complex scalars $\Phi$ and $\Sigma$ whose 
quantum charges are $(0,+1)$ and $(+1,0)$ under the gauged 
$U(1)$ and global $\tilde{U}(1)$ symmetries, see 
Table~\ref{tab:charge}. The Lagrangian is given by
\begin{align}
  \mc{L} \,=\, |\del_\mu \Phi|^2 + |D_\mu \Sigma|^2 - V(\Phi, \Sigma),
  \label{eq:Lagrangian-U1xtU1}
\end{align}
where the covariant derivative $D_\mu=\partial_\mu - i g A_\mu $
with $A_\mu$ the $U(1)$ gauge field. The scalar potential is
\begin{align}
  V(\Phi, \Sigma) \,=\, 
  \frac{\lambda_\phi}{4} \bigl( |\Phi|^2 - v_\phi^2 \bigr)^2 +
  \frac{\lambda_\sigma}{4} \bigl( |\Sigma|^2 - v_\sigma^2 \bigr)^2 
  + \kappa |\Phi|^2 |\Sigma|^2.
  \label{eq:potential}
\end{align}
The classical equations of motion (EOMs) for the scalar fields 
are given by
\begin{align}
  & \square\,\Phi 
  - \frac{\lambda_\phi}{2} \bigl( v_\phi^2 - |\Phi|^2 \bigr) \Phi 
  + \kappa |\Sigma|^2 \Phi \,=\, 0,
  \\
  & \square\,\Sigma 
  - \frac{\lambda_\sigma}{2} \bigl(v_\sigma^2 
  - |\Sigma|^2 \bigr) \Sigma 
  + \kappa |\Phi|^2 \Sigma \,=\, 0 .
  \label{eq:EOM-sigmahat}
\end{align}

We assume that the potential has the minimum $|\Phi|=v_\phi$, 
$|\Sigma|=0$ that is lower than an extremum $|\Phi|=0$, $|\Sigma|=v_\sigma$.
That implies an inequality among the couplings, 
$\lambda_\phi v_\phi^4>\lambda_\sigma v_\sigma^4$. We also assume 
in the minimum the $U(1)$ gauge symmetry is unbroken, 
which means the following relation
\begin{align}
  m_\sigma^2 \,=\, 
  \kappa v_\phi^2 - \frac{\lambda_\sigma}{2} v_\sigma^2 \,>\, 0, 
  \label{eq:m_sigma}
\end{align}
where $m_\sigma^2$ is the mass-squared parameter for $\Sigma$ in 
the vacuum. Around the vacuum expectation value $v_\phi$, 
the radial component of $\Phi$ has the mass 
$m_\phi=\lambda_\phi^{1/2} v_\phi$ and the angular one is a massless 
Nambu-Goldstone boson. 
See Appendix~\ref{app:details} for the full scalar 
Lagrangian and the explicit form of mass matrix.

\begin{figure}[t]
  \centering
  \includegraphics[width=0.48\textwidth]{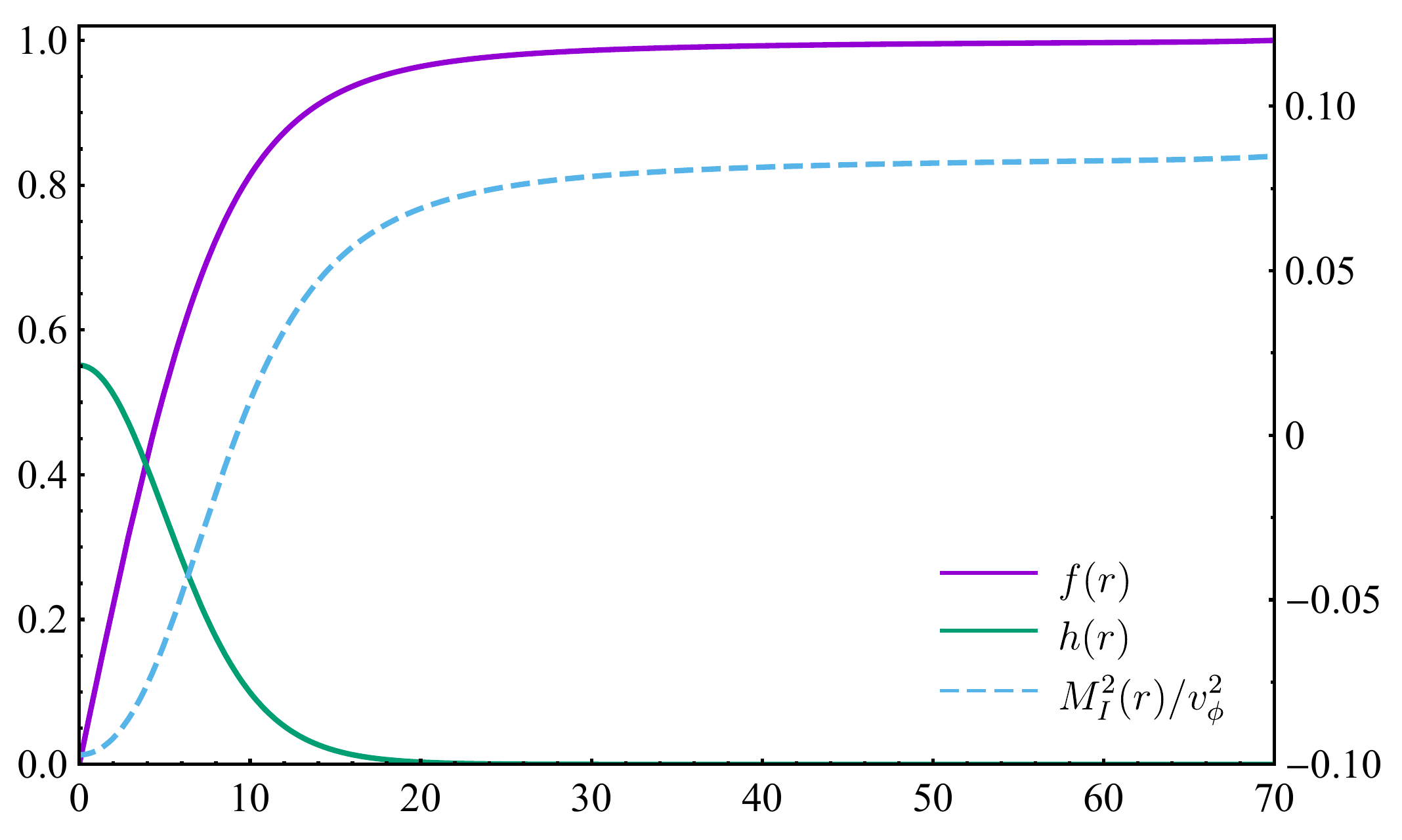}
  \quad
  \includegraphics[width=0.48\textwidth]{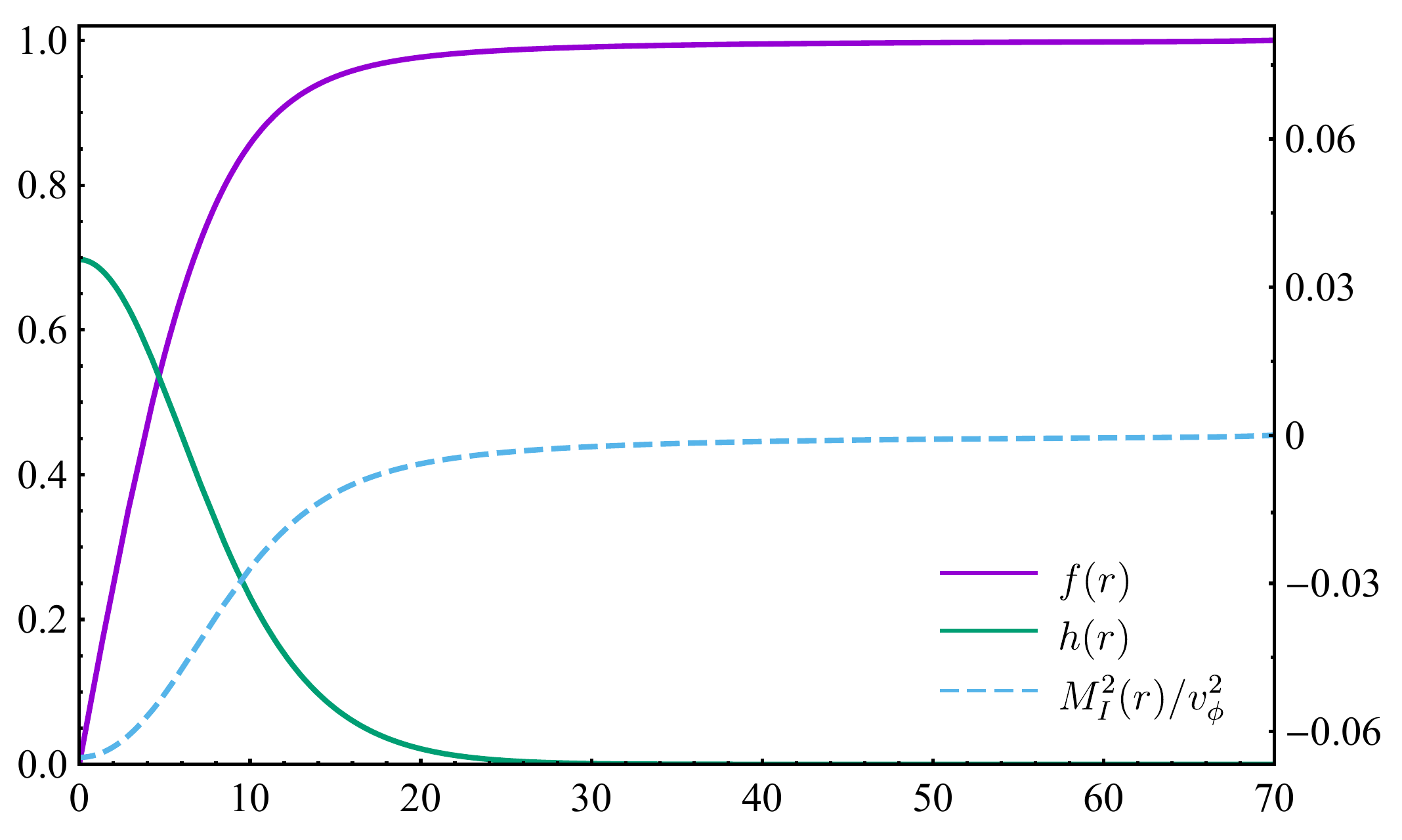}
  \caption{%
  Typical forms of the string profile functions in 
  the $U(1) \times \tilde{U}(1)$ model. The mass parameters are 
  chosen as $m_\phi = m_\sigma$ (left) and 
  $m_\phi \gg m_\sigma$ (right). The purple and green solid 
  lines show the profiles of $\Phi$ and $\Sigma$, 
  respectively, and their values are indicated by the left vertical axis. 
  The blue dashed line means the 
  position-dependent mass $M^2_I(r)$ of the imaginary part 
  of $\Sigma$, whose values are indicated by the right 
  vertical axis (see Section~\ref{sec:decay} for the details).
  }
  \label{fig:profile}
  \bigskip
\end{figure}

We consider a string located on the $z$-axes whose ansatz is
\begin{align}
  \Phi &\,=\, \hat{\phi}(r,\theta) \,=\, v_\phi f(r) e^{i \theta},
  \label{eq:ansatz-phihat}
  \\
  \Sigma &\,=\, \hat{\sigma}(r,\theta) \,=\, v_\sigma h(r),
  \label{eq:ansatz-sigmahat}
\end{align}
where $r$ and $\theta$ are the polar coordinates transverse 
to the string, and $f(r)$, $h(r)$ are the profile functions 
satisfying the following boundary conditions
\begin{align}
  f(0) = 0, 
  \qquad
  \del_r h(r) \bigr|_{r=0} = 0,
  \qquad
  f(\infty) = 1,
  \qquad
  h(\infty) = 0.
  \label{eq:bc}
\end{align}
Substituting the ansatz \eqref{eq:ansatz-phihat} and 
\eqref{eq:ansatz-sigmahat} into the EOMs, we have
\begin{align}
  & \frac{d^2 f}{d r^2} + \frac{1}{r} \frac{d f}{d r} 
  - \frac{1}{r^2}f + \frac{\lambda_\phi v_\phi^2}{2} (1 - f^2) f 
  - \kappa v_\sigma^2 f h^2 \,=\, 0,
  \\
  &\frac{d^2 h}{d r^2} + \frac{1}{r} \frac{d h}{d r} + 
  \frac{\lambda_\sigma v_\sigma^2}{2} (1 - h^2) h 
  - \kappa v_\phi^2 f^2 h \,=\, 0.
\end{align}
Typical numerical solutions of these equations are shown in 
Fig.~\ref{fig:profile}. At a large distance from the string 
core, $\hat{\sigma}$ goes to zero, that is, the $U(1)$ gauge 
symmetry is not broken. 
Inside the string core, $\hat{\phi}$ must vanish due to the 
requirement of regularity of the solution. 
This leads to a negative mass-squared term 
for $\Sigma$ (see \eqref{eq:potential}), which triggers the 
condensation of $\Sigma$ inside the string. 
As a consequence, the $U(1)$ gauge symmetry is spontaneously 
broken only inside the string and it exhibits 
superconductivity~\cite{Witten:1984eb}.

When the magnitude of the negative mass-squared term is small, 
$\Sigma$ does not condensate to avoid a large kinetic energy.
The coupling conditions for the condensation around the string 
is summarized \cite{Vilenkin:2000jqa} as 
\begin{align}
  1 \,>\, \frac{\lambda_\sigma v_\sigma^4}{ \lambda_\phi v_\phi^4}
  \,\gtrsim\, \frac{4\kappa}{\lambda_\sigma} 
  \,>\, \frac{2 v_\sigma^2}{ v_\phi^2} .
  \label{eq:parameter-condition}
\end{align}
We have already discussed the first and third inequalities 
(Eq.~\eqref{eq:m_sigma} and above it), the condition that at 
the minimum of the scalar potential, the $U(1)$ gauge symmetry 
is unbroken. The second inequality in \eqref{eq:parameter-condition} 
comes from the requirement for developing the condensation that 
the effective mass squared of $\Sigma$ becomes negative around the 
string, whose size is roughly determined by $m_\phi$. It is 
noted that the last requirement (the second inequality) is not 
necessarily needed for the condensation when the couplings 
satisfy a relation $m_\phi\gg m_\sigma$. In view of 
\eqref{eq:m_sigma}, the mass of $\Sigma$ has two contributions 
to cancel and is suppressed. Such a situation can often be realized, 
for example, in the DFSZ axion model with a large hierarchy 
between the symmetry-breaking scales. On the other hand, it is 
easily found from \eqref{eq:parameter-condition} 
that the region $m_\phi\ll m_\sigma$ is not suitable for the 
condensation to occur.

It is well known that the above string solution has 
a $U(1)$ modulus, i.e., there is a zero mode fluctuation around 
the string solution. The zero mode is parametrized by the 
following scalar and gauge field configurations:
\begin{align}
  \Sigma &\,=\, \hat{\sigma}\, e^{ i s(r,\theta) \eta(z,t)} ,
  \label{eq:Sigma-zero mode}  \\
  \Phi &\,=\, \hat \phi ,
  \label{eq:Phi-zero mode}  \\
  A_\mu &\,=\, \tfrac{1}{g} \eta(z,t)\partial_\mu s(r,\theta) .
  \label{eq:A-zero mode}
\end{align}
A vanishing $s\eta$ corresponds to the string solution itself. 
Substituting this configuration into the EOMs, the localized 
mode $\eta(z,t)$ is found to obey the two-dimensional wave equation
\begin{align}
  (\del_t^2 - \del_z^2) \eta(z,t) &\,=\, 0,  \\
  \Bigl[\frac{1}{r}\del_r(r\del_r) 
  + \frac{1}{r^2}\del_\theta^2 \Bigr] s(r,\theta)
  &\,=\, 2g^2\hat{\sigma}^2 s(r,\theta) ,
\end{align} 
which describes the massless mode traveling along the straight 
string with the speed of light. This propagating mode is nothing but 
the charge/current carrier on the superconducting string since 
the $U(1)$ Noether current is written as
\begin{equation}
  J_\mu^{U(1)} \,=\, i g \Sigma^\dagger D_\mu \Sigma + \mr{h.c.} \,=\, 
  -2 \hat{\sigma}^2 s(r,\theta) \del_\mu \eta(z,t).
\label{current}
\end{equation}

\subsection{Classical vorton stability}

Let us consider a loop of superconducting string with the radius $R$.
The scalar field characterizing the (chiral) zero mode is 
parametrized as 
\begin{equation}
  \hat{\sigma}(r,\theta) \exp [- i (t-z) Q / R],
  \label{eq:chiral-zeromode}
\end{equation} 
where $z$ is the coordinate along the string (with a 
periodicity $z \sim z+2 \pi R$) and $Q$ denotes the 
total $U(1)$ charge on the loop. For simplicity, we here 
assume the zero mode on the loop is chiral, i.e., the zero mode 
travels only in one direction. At this stage, we do not consider 
any leakage of the charges and current from the string, and 
thus they are trapped on the loop.

Then the total energy of this loop is evaluated as 
\begin{align}
  E_{\rv} \,\approx\, 2 \pi R \mu + \frac{2 \pi \mc{S} Q^2}{R} ,
  \label{eq:Evorton}
\end{align}
where $\mu$ is the string tension of the underlying vortex and 
$\mc{S} = \int d^2 x \, \hat{\sigma}^2$, which should be 
$\mathcal{O}(1)$. The first term is the tension of string loop 
and the second one comes from the contribution of the current 
traveling along the loop. For the gauged $U(1)$ symmetry case,
this is the Coulomb potential induced by the superconducting 
current. If $U(1)$ is global, this potential is regarded as a 
centrifugal-force potential induced by the zero mode.
The energy \eqref{eq:Evorton} means that, due to the 
existence of current, this string loop feels a repulsion 
force to prevent it from shrinking. 
The radius of the loop is stabilized at a certain value $R_0$
satisfying the stationary condition $d E_{\mr{v}} / d R = 0$,
which reads
\begin{align}
  R_0 \,=\, Q \sqrt{\frac{\mc{S}}{\mu}},
  \qquad
  E_{\rv}|_{R=R_0} \,=\, 4 \pi Q \sqrt{\mc{S} \mu}. 
  \label{eq:vorton_radius}
\end{align}
This (classically) stabilized loop of superconducting string is 
called the vorton \cite{Davis:1988ij}, which behaves as 
a particle-like soliton with finite size $R_0$.
In the $(1+1)$-dimensional sense, the zero-mode energy on the stable 
vorton is $Q/R = \sqrt{ \mu / \mc{S}}$, which is typically 
the symmetry-breaking scale associated with the string generation.

The detailed analyses of classical vorton stability in the bosonic 
model are found in \cite{Lemperiere:2003yt,Battye:2008mm,Radu:2008pp,Garaud:2013iba,Battye:2021sji,Battye:2021kbd}. 
Here the vorton is considered to be classically stable as long as 
$R_0$ is much larger than the string width. 
On the other hand, the quantum stability has not yet established, 
while Ref.~\cite{Davis:1988ij} studies a simplified calculation of 
the tunneling process of the zero mode.
In the following, we discuss another type of quantum decay 
of the zero mode from string curves, which results in 
the current and charge dissipation from superconducting strings. 
This may correspond to a bosonic version of \cite{Ibe:2021ctf}.
That enables us to investigate alternative quantum stability of the 
vorton and its cosmological application. It is the main focus 
of this paper.

\bigskip

\section{Zero mode decay}
\label{sec:decay}

We have reviewed on the superconducting string solution and 
the zero mode as current carrier. 
This is the classical solution of the EOMs containing the 
propagating mode on the string. In the quantum description, 
it is regarded as a coherent state,
\begin{align}
  \ket{\text{string + zero mode}} 
  \,=\, \exp\left[\,i\! \int\! d^4 x \left( \Sigma \Pi_\Sigma + 
  \Phi \Pi_\Phi + A_i \Pi^i_A + \mr{h.c.}\right) \right]
  \ket {\text{vac}}, 
  \label{eq:coherent-state}
\end{align}
where $\Sigma$, $\Phi$ and $A_i$ are the classical solutions 
describing the string and zero mode 
given by \eqref{eq:Sigma-zero mode}--\eqref{eq:A-zero mode}, 
while $\Pi_\Sigma$, $\Pi_\Phi$ and $\Pi_A^i$ are the quantum 
operators that are canonical conjugate to $\Sigma$, $\Phi$ and $A_i$, 
respectively. Is such a state stable? 
In other words, can the zero mode decay into other particles?
As long as a string is straight and the system on it is invariant 
under the straight $z$-direction translation, the zero mode which 
behaves as a massless particle in the $(t,z)$ space, does not 
decay to any light particles because such a process is 
forbidden kinematically. However, in realistic situations 
such as the early Universe, strings can randomly move and hence are 
not generally straight. Thus the momentum along the string
direction is not conserved and the decay can occur in principle. 
We study this type of quantum decay processes of the bosonic 
zero mode, i.e., the current dissipation of superconducting string 
in the $U(1)\times \tilde{U}(1)$ model.\footnote{% -----
One may wonder bosonic results should be equivalent to
fermionic one~\cite{Ibe:2021ctf} in view of the $(1+1)$-dimensional duality.
This is however not the case since the decay processes considered in this paper are essentially the $(3+1)$-dimensional system.
% -----
}

\subsection{Lagrangian in the string background}

We consider the situation that a string is modified by the 
classical deformation $\delta \phi$ around the straight 
configuration $\hat{\phi}$,
\begin{align}
  \phi_\mathrm{cl} \,=\, \hat{\phi} + \delta \phi .
\end{align}
Here and hereafter we will use barred symbols for denoting the 
modified string configuration. With this modification, 
$\hat{\sigma}$ and $\hat{A}_\mu$ should also become
\begin{align}
    \sigma_\cl \, = \, \hat{\sigma} + \delta \sigma,
    \qquad 
    A^\cl_\mu \, = \, \hat{A}_\mu + \delta A_\mu.
\end{align}
The induced $\delta \sigma$ and $\delta A_\mu$ are 
determined such that they satisfy the EOMs with $\delta \phi$. 
In this work, we suppose the weak gauge coupling 
limit $g\to0$ (the so-called neutral limit). 
That is expected to be quantitatively justified as long as the 
real value of $g$ is small~\cite{Peter:1992ta}. Thus we hereafter 
drop the gauge field $A_\mu$, which results in that the zero mode 
is just the Nambu-Goldstone mode of $\sigma$ instead of 
the collective excitation \eqref{eq:Sigma-zero mode} and 
\eqref{eq:A-zero mode}.

\begin{figure}[t]
  \centering
  \includegraphics[width=0.3\textwidth]{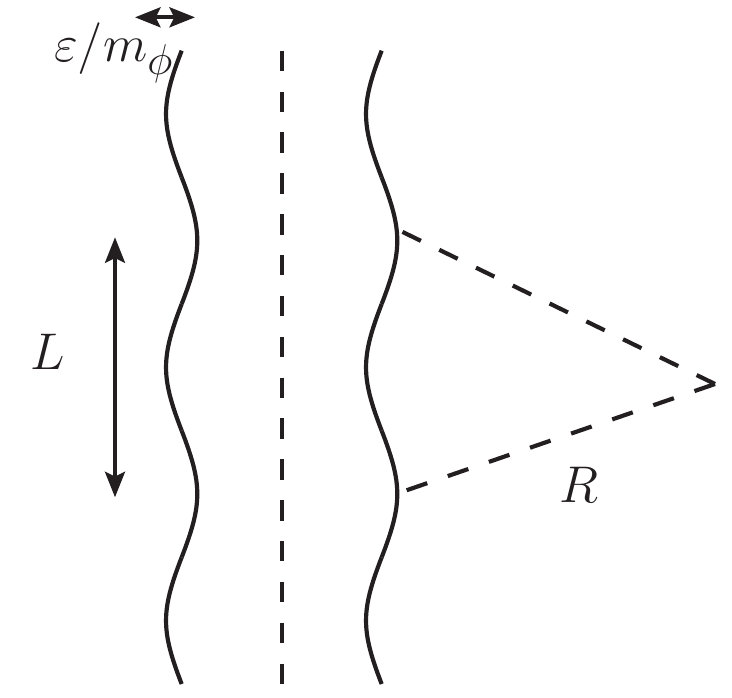}
  \hspace*{20mm}
  \includegraphics[width=0.3\textwidth]{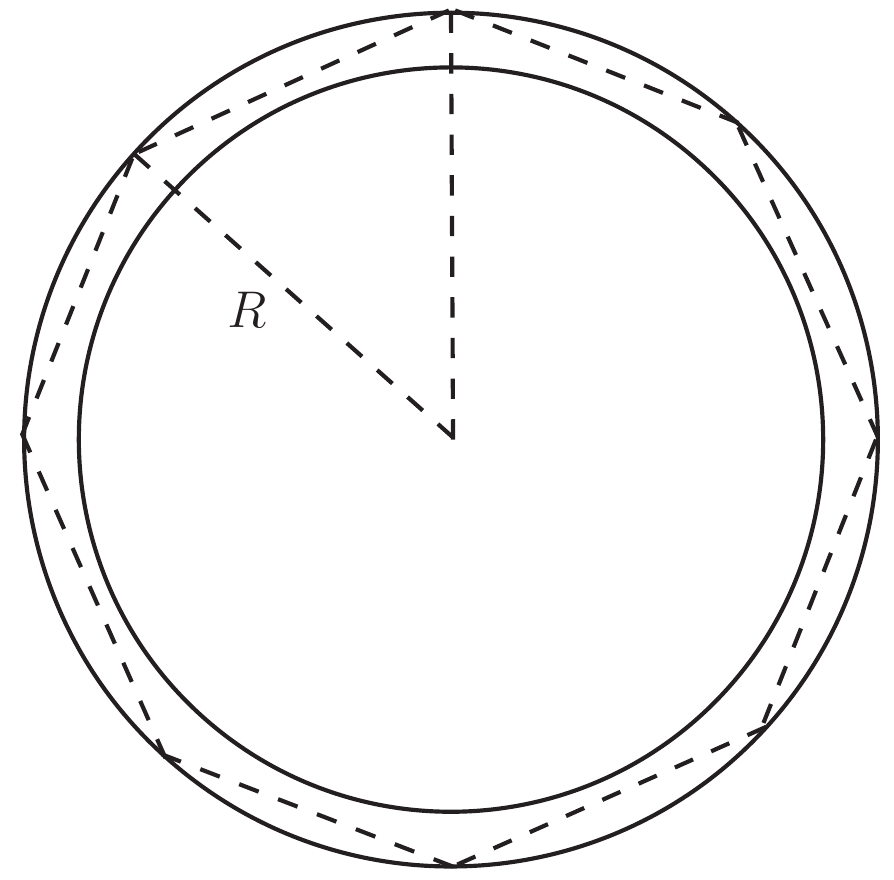}
  \caption{The schematic pictures of curved string  
  and its loop-like object. The typical curvature radius 
  and the period of curves are denoted by $R$ and $L$.}
  \label{fig:string-fluctuation}
  \bigskip
\end{figure}

The classical string deformation is parametrized as a shift into 
the transverse direction such that the center of straight string 
(on the $z$ axis) is shifted position-dependently as 
$(x,y,z)=(0, \zeta(z), z)$ (Fig.~\ref{fig:string-fluctuation}).
Then $\delta \phi$ and $\delta \sigma$,
are given as
\begin{align}
  \delta \phi \,=\, \frac{d \hat{\phi}}{d r} \sin\theta\, \zeta(z),
  \qquad
  \delta \sigma \,=\, \frac{d \hat{\sigma}}{d r} \sin\theta\, \zeta(z),
  \label{eq:fluctuation-phi-sigma}
\end{align}
where $\delta \sigma $ is found from the EOMs in  
the first order of the string deformation.
The function $\zeta(z)$ is assumed to be Fourier-expanded 
with the period $L$
\begin{align}
    \zeta(z) \,=\, \sum_n \zeta_n e^{ - i k_n z},  \qquad
    k_n = \frac{2 \pi n}{L}  \quad  (n \in \mbb{Z}),
    \label{eq:zeta-Fourier}
\end{align}
with $\zeta_{-n} = \zeta_n^*$.

We study the decay process of zero modes in the background of this curved string.
The quantum fluctuations of scalar fields around classical $\phi_\cl$ and $\sigma_\cl$ contain the zero mode and nonzero mode particles and are expressed without bars as
\begin{align}
    \Phi \,= \, \phi_\cl + \phi,
    \qquad 
    \Sigma \,=\, \sigma_\cl + \sigma
    \, = \, \sigma_\cl + \frac{1}{\sqrt{2}} (\sigma_R + i \sigma_I).
  \label{eq:fluctuation}
\end{align}
In these expressions, $\phi_\cl$ and $\phi$ are complex, and $\sigma_\cl$, $\sigma_{R,I}$ are real.
Note that we here describe the zero mode in the linear 
representation for $\Sigma$ instead of the nonlinear one given 
in \eqref{eq:Sigma-zero mode} since the latter cannot be applied 
at infinity where $\Sigma=0$, resulting in potential difficulties
for the calculation.
Both $\eta$ in \eqref{eq:Sigma-zero mode} and $\sigma_I$ in \eqref{eq:fluctuation} parametrize the Nambu-Goldstone mode on the string obeying the two-dimensional massless Klein-Gordon equation, which is regarded as the charge carrier.

We  regard $\delta \phi$ 
and $\delta \sigma$ as classical but perturbative.
Ignoring the interaction terms with 
$\delta\sigma, \delta \phi$ and other scalars, 
the Lagrangian for the quantum fluctuation $\sigma_I$ is given by
\begin{align}
  \mc{L}_I \,=\, 
  \frac12 (\del_\mu \sigma_I)^2 
  - \frac{\lambda_\sigma}{4} ( \hat{\sigma}^2 - v_\sigma^2) \sigma_I^2
  - \frac{\kappa}{2} |\hat{\phi}|^2 \sigma_I^2 .
  \label{eq:quadratic-Lagrangian}
\end{align}
This is regarded as the free part of the Lagrangian for $\sigma_I$.
Performing the mode expansion in the string background, we have
\begin{align}
  \sigma_I(x) \,=\, \sum_l \frac{1}{2\pi} 
  \int\! \frac{d^2 \ms{p}}{(2\pi)^2} \int \frac{d \omega^2}{2\pi} 
  \Bigl[
    a_{l, \omega^2}(\ms{p}) \chi_l(r,\omega^2) 
    e^{i l \theta} e^{- i \ms{p} \cdot \ms{x}} + \mr{h.c.}
  \Bigr]. 
  \label{eq:sigma-I-mode-expandsion}
\end{align}
Here $l$ denotes the angular momentum around the string and takes an 
integer value, and $\ms{p}$ is the two-dimensional momentum 
in $(t,z)$ space ($\ms{p}^2 = \omega^2$ if on shell).
The function $\chi_l(r,\omega^2)$ is the eigenfunction of the EOM 
in the transverse direction
\begin{align}
  \Bigl[ - \frac{1}{r} \del_r r \del_r + \frac{l^2}{r^2} 
  + M_I^2(r) \Bigr] \chi_l(r, \omega^2)
  \,=\, \omega^2 \chi_l(r, \omega^2). 
  \label{eq:EOM-mode-function}
\end{align}
The position-dependent mass parameter is defined as
\begin{align}
    M_I^2 (r) \,=\, \frac{\lambda_\sigma}{2} 
    \bigl( \hat{\sigma}^2 - v_\sigma^2 \bigr) + \kappa |\hat{\phi}|^2, 
    \label{eq:M_I^2}
\end{align}
and is determined by the background string solution. The typical 
behavior of $M_I^2(r)$ are shown in Fig.~\ref{fig:profile}. 
While the eigenvalues $\omega^2$ are dependent on the angular 
momentum, we will drop the explicit index $l$ for notational 
simplicity. The normalization of the eigenfunctions is
\begin{align}
  \int \frac{d r\,r}{2 \pi} \chi_l(r, \omega^2) \chi_{l}(r, \omega'^2 ) 
  \,=\, \delta (\omega^2 - \omega'^2).
\end{align}
For the discrete part of the spectrum, which we label by $n$, 
a similar normalization can be obtained by the replacement 
$\omega^2\to n/L^2$
and $\delta(\omega^2 - \omega'^2) \to \delta_{nn'}L^2$.
From the mode expansion \eqref{eq:sigma-I-mode-expandsion}, the 
canonical commutation relation 
$[\sigma_I(t,\bm{x}), \dot{\sigma}_I(t,\bm{y})] = 
i \delta^3(\bm{x}- \bm{y})$ leads to the commutation relation 
for the creation/annihilation operator
\begin{align}
  [a_{l,\omega^2} (p_z) , a^\dagger_{l',\omega'^2}(q_z)]
  \,=\, (2 \pi)^3 (2 p^0) \delta_{l,l'} \delta(\omega^2 -\omega'^2)
  \delta(p_z - q_z),
\end{align}
where $p^0$ is the zeroth momentum component of $\sigma_I$ and given 
by $p^0 = \sqrt{(p^z)^2 + \omega^2}$ from the on-shell condition.
The total scalar Lagrangian in the modified string background is 
given in Appendix~\ref{app:details} including the other scalars.

The lowest mode with $l=\omega^2=0$, which is denoted 
by $\sigma_I^0$, travels along the string with the speed of 
light. This is nothing but the zero mode given 
in Sec.~\ref{sec:U(1)xtU(1)-model}. That is easily seen by 
recalling that $\hat \sigma$ satisfies the EOM \eqref{eq:EOM-sigmahat}, 
leading to $\chi_0(r, \omega^2=0)\propto \hat \sigma$ and
\begin{equation}
  \hat \sigma + i \sigma_I^0 \,=\, 
  \hat \sigma \left( 1 + \frac{i}{2\pi} \int\! 
  \frac{d^2 \ms{p}}{(2\pi)^2}  \left[a_{0, 0}(\ms{p}) 
  e^{- i \ms{p} \cdot \ms{x}} + \mr{h.c.}\right] _{\ms{p}^2=0}\right), 
\end{equation}
which is the zero mode \eqref{eq:Sigma-zero mode} rewritten in 
the linear representation. The other modes with $\omega^2\neq 0$ 
describe massive modes in the sense that they have the 
two-dimensional momentum $\ms{p}^2=\omega^2 \neq 0$. 
The zero mode is normalizable and localized around the string 
with an asymptotic tail $\exp(- m_\sigma r)$. On the other hand, 
the massive mode with the eigenvalue $\omega^2$ can be either 
normalizable or non-normalizable depending on 
whether $\omega$ is smaller or larger than $m_\sigma$.

The current dissipation of superconducting string is caused by 
the zero mode decay into non-normalizable (i.e., not localized) 
massive modes. As stated above, a straight string does not allow 
this type of decay kinematically. Therefore the curves of 
strings play an important role to examine the stability of superconductivity.
For the perturbative analysis, the interaction terms relevant to the zero mode and first-order deformations $\mathcal{O}(\delta \phi,\delta\sigma)$ can be read off by substituting \eqref{eq:fluctuation} into the Lagrangian \eqref{eq:Lagrangian-U1xtU1}
\begin{align}
  \mc{L}_\text{int} &\,=\, - \frac{\lambda_\sigma}{2} 
  (\hat{\sigma} \delta \sigma) \sigma_I^2 
  - \frac{\kappa}{2} ( \hat{\phi}^* \delta \phi 
  + \hat{\phi} \delta \phi^*) \sigma_I^2  \nonumber \\
  &\qquad\qquad 
  -\frac{\lambda_\sigma}{2\sqrt{2}}(\delta\sigma) \sigma_R \sigma_I^2
  -\frac{\kappa}{2}(\delta\phi^*)\phi\sigma_I^2  
  -\frac{\kappa}{2}(\delta\phi)\phi^*\sigma_I^2 .
  \label{eq:vertex}
\end{align}
The first line contains the conversion vertex between the zero mode 
and massive modes of $\sigma_I$. The second line is the 3-point 
vertices among the zero mode and other scalars. Since the zero 
mode is a NG boson and has odd parity associated with the 
CP symmetry, other vertices such as the first power of the zero 
mode are absent. It is noted that the vertices in the 
second line induce little effect on the zero mode decay. This is 
due to several reasons that the scalars involved are heavy far 
from the string, the couplings are suppressed with the string 
profile, and the phase space limitation. Then we are 
lead to studying the zero mode decay with the conversion vertices 
(the first line in \eqref{eq:vertex}), which are described by the 
string classical deformations. 
Using \eqref{eq:fluctuation-phi-sigma} and \eqref{eq:M_I^2}, we find 
the conversion vertex written as
\begin{equation}
  \mc{L}_\text{int} \,\supset\, 
  \mc{L}_\conv \,=\, \frac{1}{2}c(r, \theta, z)\sigma_I^2  \label{eq:conversion-vertex}
\end{equation}
with
\begin{align}
  c(r, \theta, z) \,\equiv\, 
  - \sin \theta\, \zeta(z) \frac{d M_I^2(r)}{d r} .
\end{align}

Within the present $U(1)\times\tilde{U}(1)$ bosonic 
superconductivity model, there is no effective two-body decay 
channel of the zero mode going out from the string due to 
kinematical reasons, heavy mass scales, and suppressed couplings, 
as stated above. On the other hand, in some extended models with 
the bosonic superconductivity, the current couples 
to light particles which gives a new decay process of the zero mode. 
For example, in the DFSZ axion model, the zero mode consists of 
the charged Higgs and $W^\pm$ bosons and couples to the Standard 
Model fermion pairs due to the Yukawa couplings.
To capture this aspect, let us consider the situation that 
a pair of bulk fermions interact to the scalar field via\footnote{%
Such interaction may come from the Yukawa coupling of $\Sigma$ and 
fermions with different $U(1)$ charges. The gauge anomaly is 
supposed to be cancelled by additional fermions with 
opposite charges, but its detail is qualitatively not essential 
to the following analysis of zero mode decay.}
\begin{align}
  \mc{L}_\lambda \,=\, -\lambda \sigma_I \ol{\psi} \psi' ,
\end{align}
which generates the zero mode decay 
$\sigma_I^0 \to \ol{\psi}\psi'$. Here the finial state fermions 
are assumed to be massless for simplicity.

\subsection{Decay width}

\begin{figure}[t]
  \centering
  \raisebox{11mm}{\includegraphics[width=0.3\textwidth]{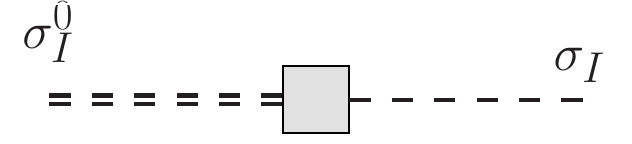}}
  \qquad\quad
  \includegraphics[width=0.37\textwidth]{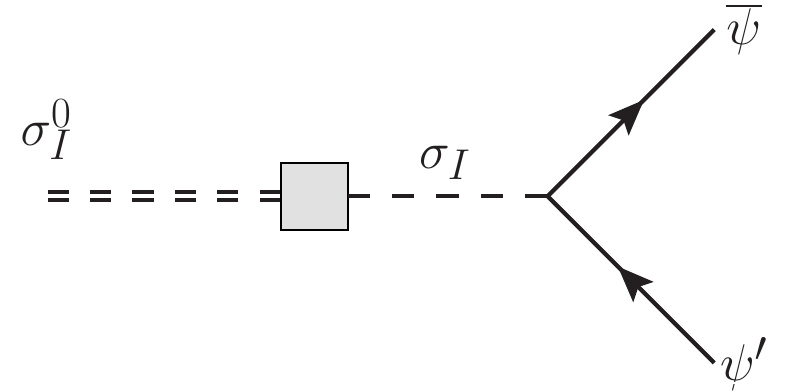}
  \caption{The zero mode decay processes. The gray box denotes 
  the vertex from the string curve $c(r,\theta,z)$. 
  (Left) The conversion to the bulk massive mode. 
  (Right) The two-body decay to light modes. This process is 
  not present in the $U(1)\times \tilde{U}(1)$ model, but naturally 
  occurs, e.g.\ for the DFSZ superconducting string.}
  \label{fig:diagram}
  \bigskip
\end{figure}

We evaluate the decay width of zero mode using the 
vertices \eqref{eq:conversion-vertex}. Let us consider two 
processes shown in Fig.~\ref{fig:diagram}: One is 
that the zero mode is converted to non-localized massive modes 
(called bulk modes) leaving from the string. The other is that 
the zero mode decays into a pair of light particles mediated 
by the bulk modes. The detail of each process will be discussed 
below. We assume that the total decay width is given by
\begin{align}
  \Gamma \,=\, \Gamma_{\conv} + \Gamma_{2\body},
\end{align}
where $\Gamma_{\conv}$ and $\Gamma_{2\body}$ denote the 
partial decay widths of the conversion and two-body processes, 
respectively. Other decay processes, 
if any, give additive contributions to $\Gamma$.

\subsubsection{Conversion}

For the conversion process, consider the zero mode with the 
two-momentum $\ms{p} = (p^0, p^z) = (E, E)$ converting to 
the bulk mode with the four-momentum $q$. As stated before, 
this process is described by the transition from the initial 
state that is a coherent state including the zero mode and 
curved string like \eqref{eq:coherent-state},
into the final state consisting of bulk modes and the string 
deformed due to a back reaction. That requires to introduce 
a collective coordinate describing the string position 
and to quantize the collective coordinate and the zero 
mode simultaneously.
It is however difficult to perform such a calculation 
particularly for the case of curved string. Therefore we 
take an approximation to ignore the string sector of the Hilbert 
space, that is, the string background is fixed 
throughout this process and the initial and final states are 
created by $a_{0,0}^\dagger(\ms{p})$ and 
$a_{l,\omega^2\neq 0}^\dagger (\ms{q})$, respectively, which are 
the Fock space elements in the straight string background.

Thus the amplitude of the conversion process is given by 
\begin{align}
  i \mc{T}_{\conv} 
  &\,=\, \Bigl\langle 0\Big|a_{l,\omega^2\neq 0} (\ms{q}) 
  \exp\!\Big[i \mbox{\large $\int$} d^4x\, \tfrac{1}{2}c(r,\theta,z) 
  \sigma_I^2(x)\Big] a_{0,0}^\dagger(\ms{p})\Big|0\Bigr\rangle  \\
  &\,=\, i\! \int d^4x \; \chi_{l} (r, \omega^2) 
  e^{i \ms{q} \cdot \ms{x}} e^{-i l \theta} c(r, \theta, z) 
  \chi_0(r, 0) e^{- i \ms{p} \cdot \ms{x}}
  \\
  &\,=\, \sum_n \zeta_n (2 \pi)^2 \delta(p^0 - q^0) 
  \delta(p^z - q^z - k_n)
  \int \!d r \, r\, \chi_{0}(r,0) \chi_l (r, \omega^2) 
  \frac{d M_I^2}{d r} \pi (\delta_{l,1}- \delta_{l,-1}) ,
  \label{eq:T-conv}
\end{align}
where $\ket{0}$ satisfies $a_{l,\omega^2}\ket{0}=0 $.
The conversion width is given by the phase space integration 
of the squared amplitude as 
\begin{align}
  \Gamma_{\conv} &\,=\, \frac{1}{2 p^0} \int\! 
  \frac{d q^z d \omega^2}{(2\pi)^3\, 2 q^0} \sum_l
  \frac{|\mc{T}_{\conv}|^2}{\ms{T} \ms{L}_z}  
  \,=\, \frac{\pi}{2 E} \sum_n |\zeta_n|^2 
  \biggl| \int d r \, r\, \chi_0(r,0) \frac{d M_I^2}{d r}
  \chi_1(r,\ms{p}_n^2) \biggr|^2,
  \label{eq:conversion-width-1}
\end{align}
where $\ms{p}_n = (p^0, p^z - k_n) = (E, E - k_n)$ and 
$\ms{p}_n^2 = 2 E k_n - k_n^2$. The system size in the time and 
string direction are introduced 
by $2\pi\delta(p^0 = 0) = \ms{T}$ and 
$2 \pi \delta (p^z = 0) = \ms{L}_z$. 
Note that the $z$-direction momentum is not conserved because the 
translational invariance is violated due to the $z$-dependent 
string curves. In other words, the in-state particle obtains 
the momentum $-k_n$ from curves. For the zero mode to 
convert into bulk modes, it satisfies the kinetic 
condition $\ms{p}_n^2\geq m_\sigma^2$ that implies the 
energy threshold
\begin{align}
  E \,\geq\, \frac{k_n^2+m_\sigma^2}{2k_n} .
  \label{eq:threshold}
\end{align}
The sum in \eqref{eq:conversion-width-1} is limited to the 
eigenstates which satisfies this bound for a fixed 
initial energy $E$.

This process seems to violate the  conservation of the $U(1)$ charge defined in \eqref{current}.
The  initial state has the charge $\int d^3x  J_0^{U(1)} = \int dxdy\,\hat\sigma^2 E \ms{L}_z (\equiv Q_{init})$
while the final bulk state has the unit charge. The mismatch does not 
necessarily mean any inconsistency but comes from an 
approximation ignoring the string sector as mentioned. 
If one carries out the calculation within the full analysis, 
a back-reacted string in the final state should have explicit time dependence and carry the 
compensating charge $Q_{init}-1$, and hence this process corresponds to the 
unit-charge leakage from the string.

\subsubsection{Two-body decay}

We also have the two-body decay of zero mode 
$\sigma_I^0 \to \ol{\psi}\psi'$ caused by the vertices 
$\mc{L}_\conv$ and $\mc{L}_\lambda$. For this process, consider 
the zero mode with the two-momentum $\ms{p} = (p^0, p^z) = (E, E)$ 
decaying to the fermion pair with the four-momentum $q_{1,2}$. 
Again we make the approximation in which the string background is 
fixed throughout this process and the initial and final states 
are given by $a_{0,0}^\dagger(\ms{p})$ and the fermion creation 
operators.

The amplitude for this two-body decay is written by 
\begin{align}
  i \mc{T}_{2\body} \,=\, i\lambda \int d^4 x d^4 x' \, 
  \chi_{0}(r, 0) e^{- i \ms{p} \cdot \ms{x}} c(r, \theta, z) 
  G(x, x') e^{i q_1 x'} \ol{u}(q_1) e^{i q_2 x'} v(q_2) ,
\end{align}
where $G(x,x')$ is the propagator for $\sigma_I$ in the string
background (see Appendix \ref{app:propagator} for details). Its 
Fourier transformation with respect to the coordinates 
$t$, $z$, $\theta$ is 
\begin{align}
  G(x, x') \,=\, \int\!\frac{d^2 \ms{k}}{(2\pi)^3}\,
  \sum_l G^l_{\ms{k}}(r, r') e^{- i \ms{k} 
  \cdot (\ms{x} - \ms{x}')} e^{i l (\theta- \theta')} .
\end{align}
The two-body decay width is obtained by the phase space integral 
of this squared amplitude, exactly written as 
\begin{align}
  \Gamma_{2\body} &\,= \int\! \frac{d^3 q_1}{(2\pi)^3 2 q_1^0} 
  \frac{d^3 q_2}{(2\pi)^3 2 q_2^0} 
  \frac{|\mc{T}_{2\body}|^2}{2p^0\ms{T} \ms{L}_z}   \\ 
  &\,=\, \frac{|\lambda|^2}{32 \pi E} \sum_n |\zeta_n|^2 
  \int d q^2 (\ms{p}_n^2 - q^2)  \biggl| \int d r d r' \, r r' 
  \chi_0(r,0) \frac{d M_I^2(r)}{d r} 
  G_{\ms{p}_n}^1(r,r') J_1(qr') \biggr|^2.
\end{align}
Here $J_1$ is the Bessel function of first kind of order one, 
$2 \pi J_1(r) = \int_0^{2\pi} d\theta \, e^{i (\theta-r \sin \theta)}$,
and $q$ means the total momentum of two fermions transverse 
to the string direction. We expect that the Yukawa interaction 
occurs essentially outside the string, and hence we take the 
position $r'$ to be far from the string core. With the 
spectral representation of the propagator
\begin{equation}
  G_{\ms{k}}^l(r,r') \,=\, \int 
  \frac{d \omega^2}{2\pi} \frac{1}{\omega^2 - \ms{k}^2} 
  \chi_l(r,\omega^2) \chi_l(r',\omega^2) ,
\end{equation}
and the asymptotic form $\chi_1 (r',\omega^2) \sim \pi^{1/2}
J_1(\sqrt{\omega^2-m_\sigma^2}r')$, we obtain the decay width as
\begin{align}
  \Gamma_{2\body} \,\approx\, \frac{|\lambda|^2}{32 \pi E}
  \sum_n |\zeta_n|^2 \int\! d \omega^2\, 
  \frac{\ms{p}_n^2 - \omega^2 + m_\sigma^2}{(\omega^2 - \ms{p}_n^2)^2}
  \biggl|
    \int d r\, r \chi_0(r,0) \frac{d M_I^2(r)}{d r} \chi_1(r, \omega^2)
  \biggr|^2,
  \label{eq:two-body-width}
\end{align}
Once the couplings in the model are fixed, the eigenfunctions and 
mass parameters are determined and then the decay widths are 
evaluated using \eqref{eq:conversion-width-1} and 
\eqref{eq:two-body-width}.

In the case that the energy threshold condition 
\eqref{eq:threshold} is met, for example, by a large initial 
energy of the zero mode, the intermediate state can be on shell 
at specific values of $\omega$ (the poles of 
the propagator $G_{\ms{p}_n}^1$). The on-shell decay width 
may be evaluated by a replacement
\begin{align}
  \frac{1}{(\omega^2 - \ms{p}_n^2)^2} \,\to\,
  \frac{1}{(\omega^2 - \ms{p}_n^2)^2 + \omega^2 \Gamma_\omega^2}.
\end{align}
Here we assume that the width is given by 
$\Gamma_\omega=|\lambda|^2 m_\sigma/8\pi$ as the bulk mode mass 
in four-dimensional viewpoint is approximated 
by $m_\sigma$ outside the string.

\subsection{Square well potential}

\begin{figure}[t]
  \centering
  \includegraphics[width=0.4\textwidth]{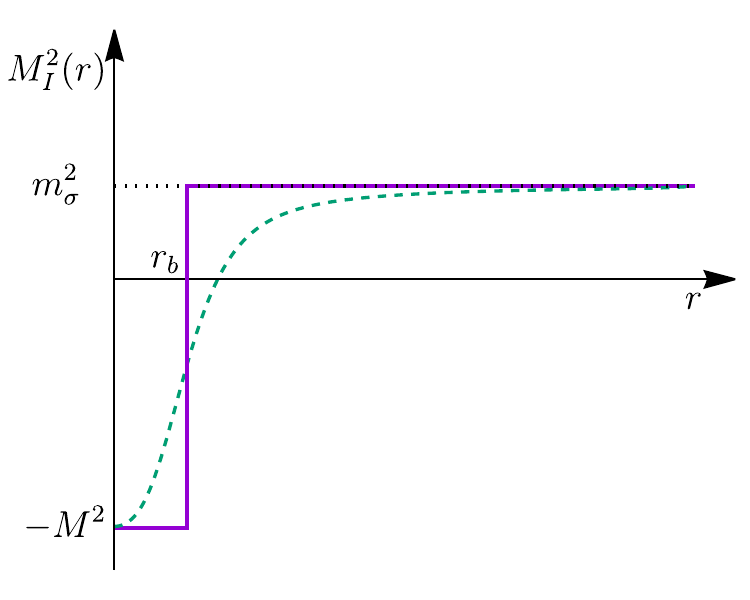}
  \caption{The schematic picture of square well potential 
  as an approximation of the position-dependent mass 
  function $M_I^2(r)$.}
  \label{fig:well-potential-for-M_I}
  \bigskip
\end{figure}

Some analytical expressions for the partial decay widths can be 
obtained by the approximation that the position-dependent mass 
function $M_I^2(r)$ is regarded as a squared well potential 
shown in Fig.~\ref{fig:well-potential-for-M_I}:
\begin{equation}
  M_I^2(r) \,=\, -M^2\theta(r_b-r) + m_\sigma^2 \theta(r-r_b) ,
 \label{eq:approMI}
\end{equation}
where $\theta(r)$ is the Heaviside step function. 
The parameters $M^2$ and $r_b$ represent the depth and width of 
the well potential, respectively, and we have
\begin{equation}
  M^2 \,\sim\, \lambda_\sigma v_\sigma^2, \qquad 
  r_b \,=\, m_\phi^{-1},
\end{equation}
which are typically expected from the form of string solution 
discussed in the previous section. The asymptotic value 
$m_\sigma^2$ is the mass of $\sigma$ at large distance 
from the string and is defined in \eqref{eq:m_sigma}.

In this approximation, the eigenfunctions are explicitly given by 
the (modified) Bessel functions as 
$\chi_0(r,0) \sim \sqrt{4\pi} m_\sigma K_0 (m_\sigma r)$ and 
$\chi_1(r,\omega^2) \sim \sqrt{\pi} J_1 (\sqrt{\omega^2 - m_\sigma^2}\,r)$ 
for $r \geq r_b$, and the integrals in \eqref{eq:conversion-width-1} and \eqref{eq:two-body-width} can be performed. That leads to the analytic formulae
\begin{align}
  \Gamma_\conv &\,\approx\, \frac{2 \pi^3}{E} (M^2 + m_\sigma^2)^2 
  (m_\sigma r_b)^2 K_0(m_\sigma r_b)^2 
  \sum_ n |\zeta_n|^2 J_1(\sqrt{\ms{p}_n^2 - m_\sigma^2}\,r_b)^2, 
  \label{eq:analytic_Gamma_conv}
  \\
  \Gamma_{2\body} &\,\approx\, \frac{\pi|\lambda|^2}{8E}
   (M^2 + m_\sigma^2)^2 (m_\sigma r_b)^2 K_0(m_\sigma r_b)^2
  \sum_n |\zeta_n|^2    \nonumber \\
  & \qquad\qquad \times
  \int_{m_\sigma^2}^{m_\sigma^2+ \ms{p}_n^2}\! d \omega^2\,
  \frac{\ms{p}_n^2 - \omega^2 + m_\sigma^2}{(\omega^2 - \ms{p}_n^2)^2} 
  J_1 (\sqrt{\omega^2 - m_\sigma^2}\,r_b)^2 .
  \label{eq:analytic_Gamma_2body} 
\end{align}
Here $K_0$ is the modified Bessel function of second kind. 
These expressions are useful for discussing the cosmological 
consequences of superconducting strings and the zero mode behavior 
in the next section.

\bigskip

\section{Application to cosmology}
\label{sec:application}

As an application of the zero mode decay processes in the 
previous section, we examine the stability of superconducting 
strings and vortons against the string curves.

Instead of specifying the couplings in the Lagrangian, we 
replace them with the mass scale parameters defined as
\begin{align}
 m_\sigma^2 \,=\, 
  \kappa v_\phi^2 - \frac{\lambda_\sigma}{2} v_\sigma^2 , 
  \qquad
  m_\phi^2 \,=\, \lambda_\phi v_\phi^2 ,
  \qquad
  M^2 \,=\, \frac{1}{2}\lambda_\sigma v_\sigma^2 .
\end{align}
The physical meanings of these mass parameters are as follows:
$m_\sigma$ determines the tail behavior of zero mode wavefunction 
(the mass of $\sigma$ far from the string), $m_\phi$ represents 
the width of string, and $M$ means the depth of the well 
potential that the zero mode feels. With the coupling conditions for 
superconductivity~\eqref{eq:parameter-condition}, 
we find $m_\sigma^2>0$ and $M^4\gtrsim m_\phi^2(M^2+m_\sigma^2)$, 
which necessarily implies $M\gtrsim m_\phi$. In addition, unless 
no huge hierarchy exists among the dimensionless couplings, 
\eqref{eq:parameter-condition} also implies $M\lesssim m_\phi$. 
These fact leads us to a reasonable assumption $M=m_\phi$, 
which then means $m_\phi\gtrsim m_\sigma$ from 
\eqref{eq:parameter-condition}. In the following analysis, 
we thus assume 
\begin{align}
  m_\sigma \,<\; m_\phi \,=\, M 
  \label{eq:dimful-condition}
\end{align}
for the parameter region preferable for the superconductivity. 
As for the magnitude of $m_\sigma$, we have checked any 
desired (small) value can be realized by appropriately choosing 
the coupling constants under the conditions of superconductivity.

The curved string is generally possible to take 
various forms expressed by the function $\zeta(z)$, whose 
Fourier expansion is given in \eqref{eq:zeta-Fourier}. In the 
following analysis, we implicitly consider the $n=1$ mode only. 
The contributions from other modes with $n>1$ can be evaluated 
by taking the $n$ dependence of $k_n$ into account. In general, 
the energy condition like \eqref{eq:threshold} leads to an upper 
bound on $n$, and the dominant effect usually comes from a 
smaller $n$ unless the string curve is singular. 
We denote the width of this mode as 
\begin{equation}
  |\zeta_1| \,=\, \frac{\varepsilon}{m_\phi},
\end{equation}
where $\varepsilon$ is the relative size 
of curve compared to the string width and should be smaller than
unity for the perturbative analysis to be valid.
The width of curve is also rewritten by using the curvature 
radius $R$ (see Fig.~\ref{fig:string-fluctuation}) in the large 
radius limit, 
\begin{align}
  \frac{\varepsilon}{m_\phi} \,\approx\, \frac{L^2}{4\pi^2 R}.
  \label{eq:modulation_zeta}
\end{align}
If a cosmic string is present in the early Universe at the 
temperature $T$, a typical size of its curvature radius is given 
by the inverse of the Hubble parameter $H(T)$ (assuming the 
scaling regime). From the Friedmann equation in the radiation 
dominated era, we have 
\begin{align}
  H(T) \,=\, \sqrt{\frac{\pi^2 g_*}{90}} \frac{T^2}{M_P},
\end{align}
where $M_P = 2.44 \times 10^{18}~\mathrm{GeV}$ is the reduced 
Planck scale and $g_*$ the effective degrees of freedom for the 
energy density at $T$. We here assume for simplicity that 
$g_*$ is a constant of order $10^2$.

\subsection{Parameter dependence}

Before discussing the cosmological implications, we illustrate 
the parameter dependence of the zero mode decay width. As 
a benchmark case, we fix the curvature radius by the Hubble 
parameters, $R = H(T)^{-1}$. The typical length of $L$ is also 
fixed by the relation \eqref{eq:modulation_zeta}. 
We then have three parameters in addition to the mass scale 
parameters $m_\sigma$ and $m_\phi$: 
the initial energy $E$ of zero mode, the curve size 
$\varepsilon$, and the Yukawa coupling $\lambda$ for the 
two-body decay.

\begin{figure}[t]
  \centering
  \includegraphics[width=0.48\textwidth]{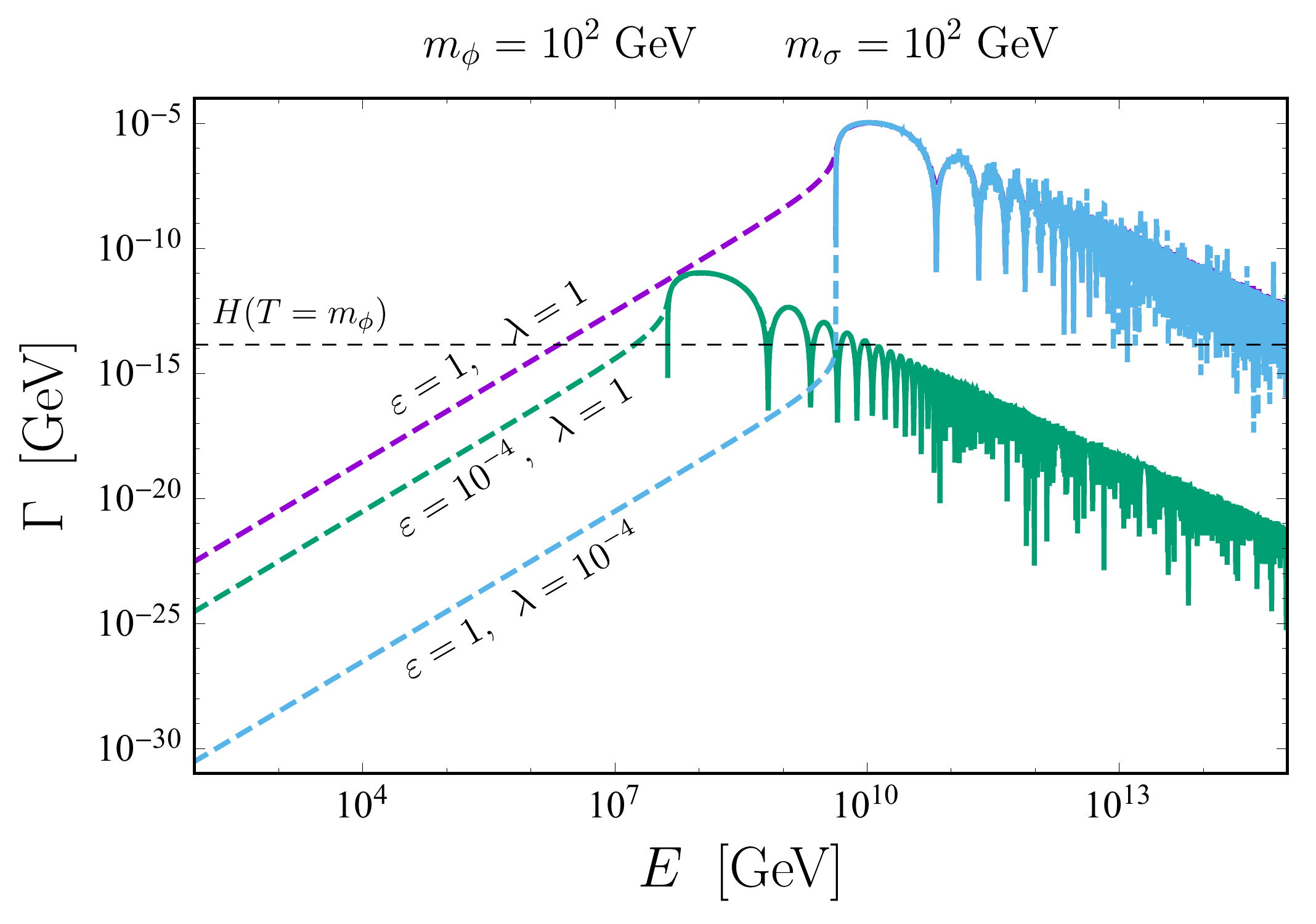}
  \quad
  \includegraphics[width=0.48\textwidth]{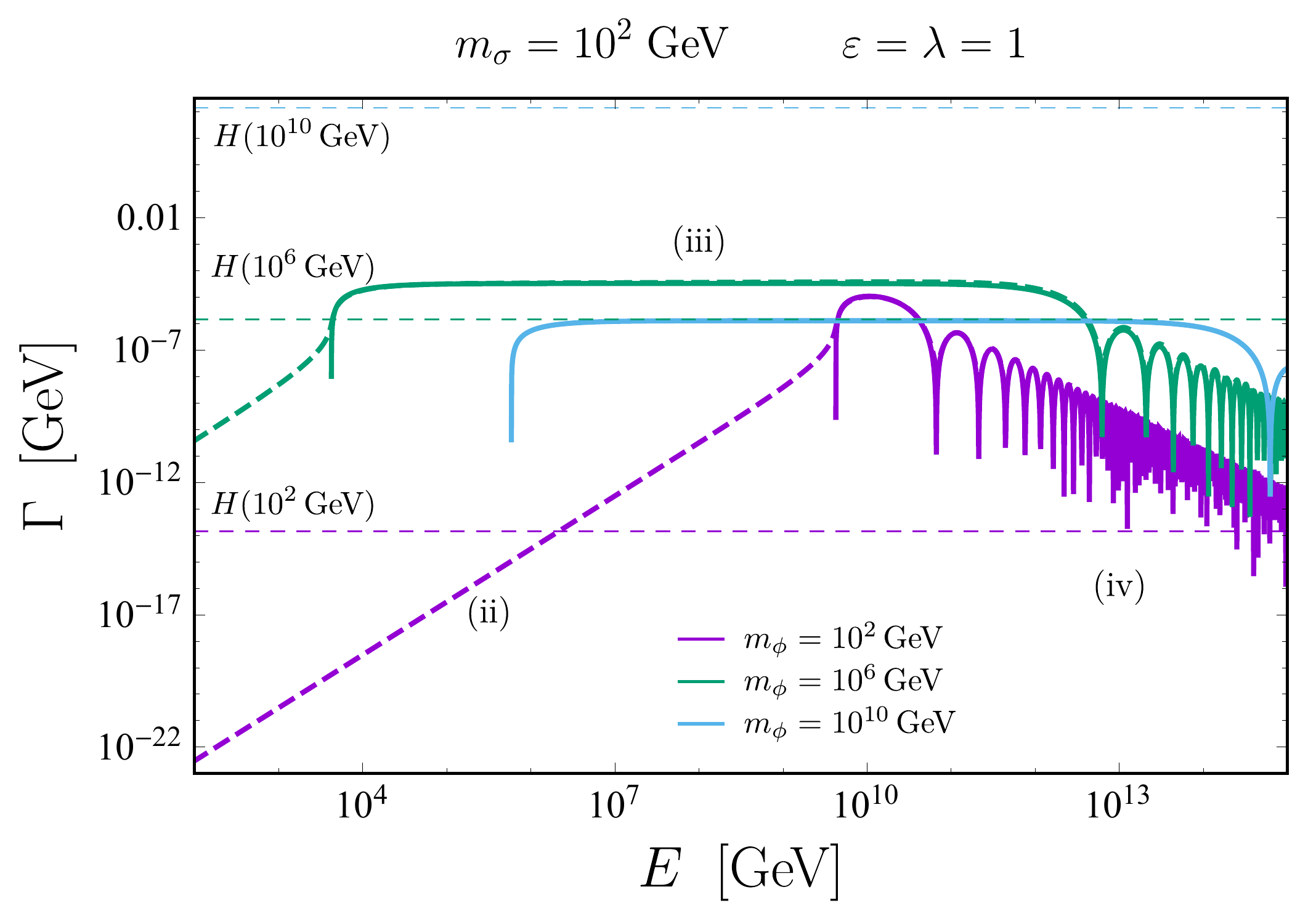}
  \caption{The initial energy dependence of the zero mode decay 
  width at the temperature $T=m_\phi$ with various parameter sets. 
  (Left) The purple, green, blue lines correspond to the parameters 
  $(\varepsilon,\lambda)=(1,1)$, $(10^{-4},1)$, and $(1,10^{-4})$, 
  respectively. The solid and dotted parts mean $\Gamma_{\conv}$ and 
  $\Gamma_{2\body}$ (the purple-solid and blue-solid lines are 
  overlapped). The mass parameters are fixed 
  $m_\phi = m_\sigma = 10^2~\mr{GeV}$. 
  (Right) The purple, green, blue lines correspond to 
  $m_\phi=10^2$, $10^6$ and $10^{10}$~GeV, while 
  $m_\sigma$ is $10^2~\mr{GeV}$. The other couplings are fixed 
  as $\varepsilon = \lambda = 1$. In both figures, the horizontal 
  dashed lines show the Hubble parameter at $T=m_\phi$. 
  The regions (ii), (iii), (iv) imply the classification of the energy dependence of the width. See the text for details.}
  \label{fig:gammaE}
  \bigskip
\end{figure}

\paragraph{$E$ dependence :} 
First we examine the energy $E$ dependence. In  
Fig.~\ref{fig:gammaE}, the partial decay widths $\Gamma_{\conv}$ 
and $\Gamma_{2\body}$ are shown by the solid and dotted 
lines, respectively. We here show two typical cases that 
the mass scales inside and outside the string are of the same 
order $m_\phi = m_\sigma$ (left panel) and 
hierarchical $m_\phi\gg m_\sigma$ (right panel).
The energy dependence is found to be classified into four regions: 
(i) $E\lesssim k_n$, 
(ii) $k_n\lesssim E \lesssim m_\sigma^2/k_n$, 
(iii) $m_\sigma^2/k_n\lesssim E \lesssim m_\phi^2/k_n$, and 
(iv) $m_\phi^2/k_n\lesssim E\,$ 
in the case $m_\sigma>k_n$, which is possible if the mass 
scales satisfy $m_\phi M^2\lesssim \varepsilon\,m_\sigma^2M_P$. 
For $m_\sigma<k_n$, a similar classification 
appears by exchanging $m_\sigma^2/k_n$ and $k_n$, and then 
the region (ii) almost vanishes.

\begin{itemize}[leftmargin=7mm]
\item Region (i) : Any decay is forbidden kinematically.\\
In the words of the 2-dimensional transverse 
momentum $\ms{p}_n=(E,E-k_n)$, 
this region means $\ms{p}_n^2<0$. Due to a relatively large size 
of curve, the final (total) momentum becomes tachyonic. 

\item Region (ii) : Only the two-body decay is possible.\\
In this region, $\ms{p}_n^2<m_\sigma^2$ and the energy threshold 
\eqref{eq:threshold} is not met. Therefore only a decay to 
lighter modes (than $m_\sigma$) can occur. With the series 
expansion for the Bessel functions, the width of zero mode 
is found to have the parameter dependence
\begin{align}
  \Gamma \,\propto\, \varepsilon^{1/2}\lambda^2 E^2 
  M^7/(m_\phi^{9/2}m_\sigma^2),
\end{align}
and increases as the energy $E$ (the dotted lines in the figures). 
It weakly depends on the curve size $\varepsilon$. 
Later, we will consider the case $E=m_\phi$ for superconducting 
current, which belongs to this region (ii) 
if $m_\phi^3M^2\lesssim\varepsilon\,m_\sigma^4M_P$.

\item Region (iii) : Both the conversion and two-body decay are 
possible.\\
The energy threshold condition is satisfied and 
$\ms{p}_n^2>m_\sigma^2$. The two-body decay is dominated by the 
poles of the propagator (in the string background) as long as 
the width is perturbatively narrow. The zero mode decay width is 
roughly given by 
\begin{align} 
  \Gamma \,\propto\, \varepsilon^{3/2}\,M^5 
  m_\sigma^2/m_\phi^{11/2}.
\end{align}
This is free from $\lambda$ as it should be. Further $\Gamma$ is 
a constant against the initial energy $E$ (the plateau seen 
in the right panel). 
The plateau generally appears when the region (iii) is 
effective, i.e., $m_\sigma<m_\phi$. For the hierarchical mass 
case $m_\sigma\ll m_\phi$, the height of plateau is proportional 
to $m_\sigma^2/m_\phi^{1/2}$ and the decay width becomes small. 
The case $E=m_\phi$ for superconducting current belongs to this 
region (iii) if $m_\phi^3M^2\gtrsim\varepsilon\,m_\sigma^4 M_P$.

\item Region (iv) : All the decay widths are suppressed.\\
In this region, the transverse momentum of final states is 
$\ms{p}_n^2>m_\phi^2$. For such higher energy in the radial 
direction, their Compton wavelengths become shorter 
than the string width, and hence the overlapping integral of 
wavefunctions in $\Gamma$ is oscillationally 
suppressed.\footnote{% -----
In the framework with the momentum conservation, an alternative 
view of this suppression is obtained from the dynamics of 
Nambu-Goldstone boson~\cite{Bando:1999di,Kugo:1999mf} 
appearing as a result of the translation symmetry breaking due 
to the presence of string.
% -----
}
The decay width decreases with energy as 
\begin{align}
  \Gamma \,\propto\, \varepsilon^{9/4}
  M^{7/2}m_\sigma^2/(E^{3/2}m_\phi^{13/4}).
\end{align}
\end{itemize}

The energy threshold \eqref{eq:threshold} for the conversion and 
on-shell decay corresponds to the boundary between the solid and 
dotted lines in Fig.~\ref{fig:gammaE}. It has the complex 
dependence on mass  parameters. For a smaller value of $m_\phi$, 
that is, $m_\phi < \varepsilon m_\sigma^2 M_P/M^2$, 
the threshold scale is 
$\varepsilon^{1/2}m_\sigma^2 M_P^{1/2}/Mm_\phi^{1/2}$ (a 
decreasing function of $m_\phi$). For larger $m_\phi$, it is 
$Mm_\phi^{1/2}/\varepsilon^{1/2}M_P^{1/2}$ (increasing 
with $m_\phi$). Therefore the threshold energy has the minimum 
around $m_\phi\sim \varepsilon m_\sigma^2 M_P/M^2$. 
For a larger $m_\phi$ than this minimum, the off-shell 
two-body decay becomes almost ineffective (see the blue 
line in the right panel). 
Above the threshold, the conversion and on-shell decay processes 
open up. That is possible for the zero mode with $E=m_\phi$ only 
when $m_\phi$ is larger than $(\varepsilon m_\sigma^4M_P/M^2)^{1/3}$.

\begin{figure}[t]
  \centering
  \includegraphics[width=0.48\textwidth]{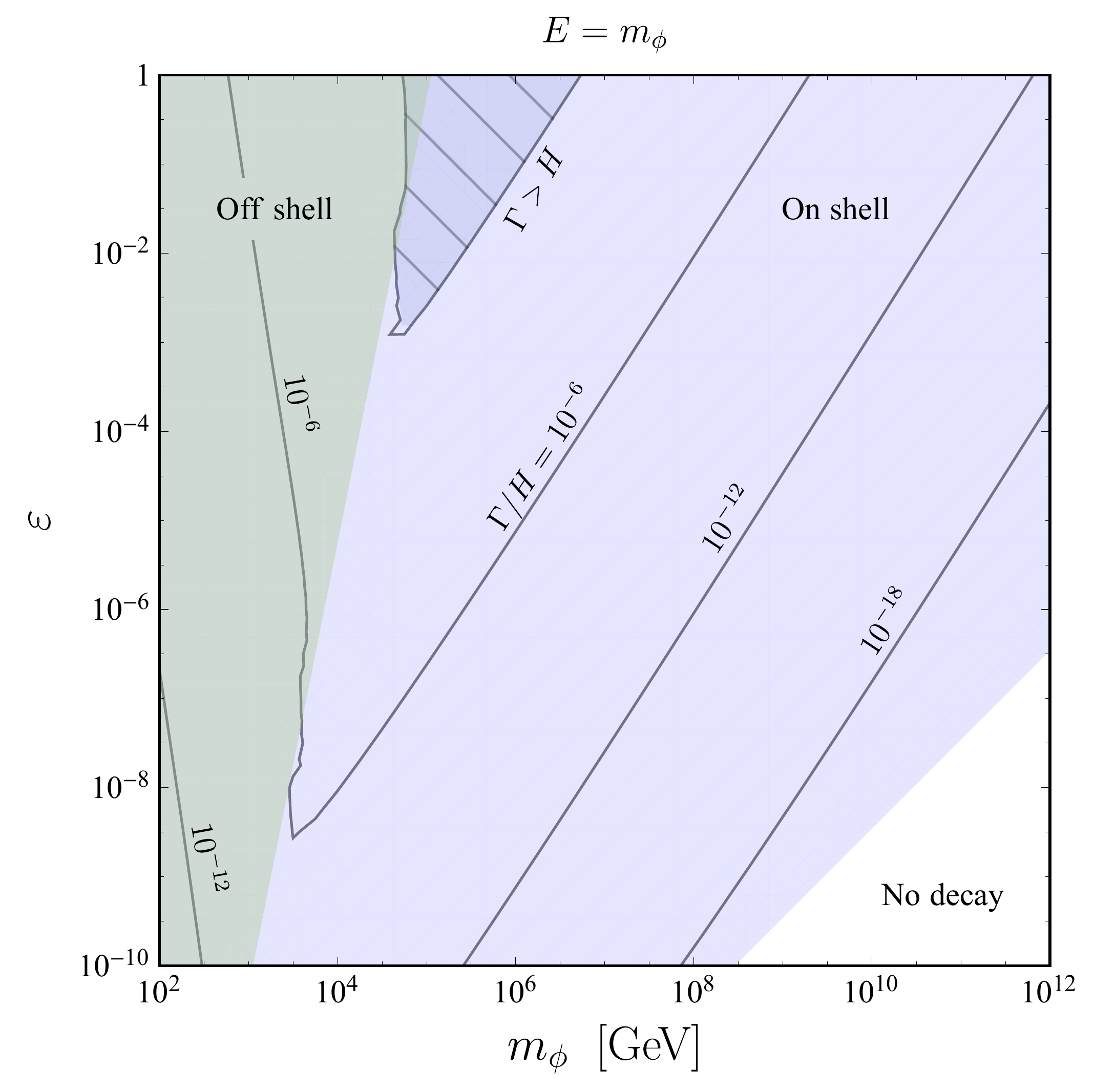}
  \quad
  \includegraphics[width=0.48\textwidth]{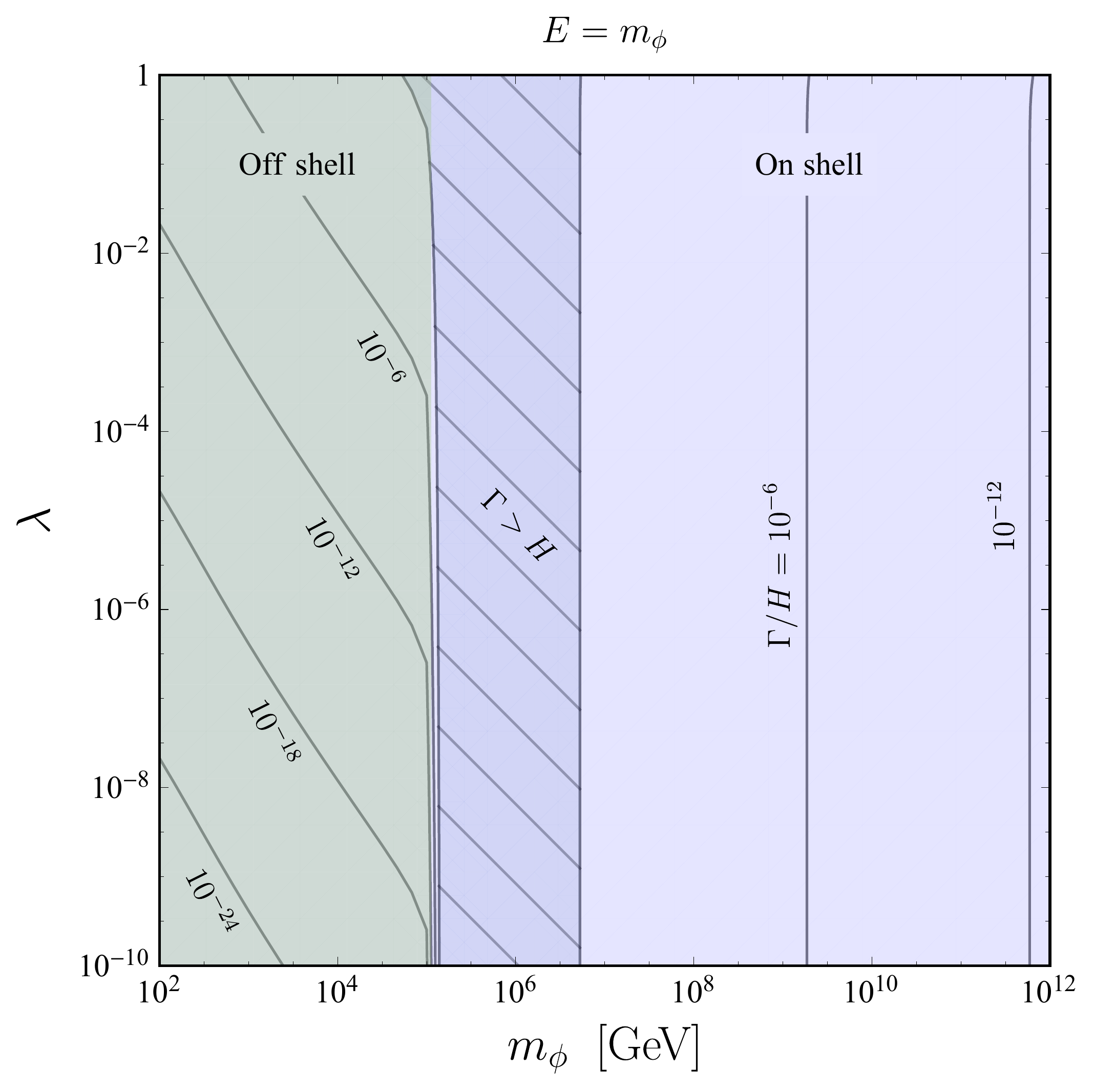}
  \caption{The coupling dependence of the zero mode decay width.
  The blue (``on shell'') region means the initial energy is 
  above the threshold. 
  In the green (``off shell'') region, only the two-body decay 
  is possible. The on-shell decay width does not depend 
  on $\lambda$ as it should be. For the ``no decay'' region, 
  see the text for detail.}
  \label{fig:epVsmf}
  \bigskip
\end{figure}

\paragraph{$\varepsilon$, $\lambda$ dependence :} 
Next we show the coupling $\varepsilon$, $\lambda$ dependence. 
Fig.~\ref{fig:epVsmf} is the contour plot of the decay width $\Gamma$ 
over the Hubble parameter $H$. The left panel is for the curve 
size $\varepsilon$ and the right one for the Yukawa coupling 
$\lambda$ to light fermions. In the region $\Gamma>H$ (the shaded 
region), the zero mode immediately decays in the early Universe 
and the superconducting current is unstable. The initial energy and 
the temperature are fixed at $E = T = m_\phi$ in the figure. 
In the blue region, the threshold condition \eqref{eq:threshold} is 
satisfied and the conversion (on-shell decay) is possible. On the 
other hand, in the green region, only the two-body decay via 
off-shell mediator is allowed. The boundary of these two colored 
regions is given by the equation 
$m_\phi M^2=\varepsilon\,m_\sigma^2M_P$ 
as noted above, which is independent of $\lambda$. 

From the left panel of Fig.~\ref{fig:epVsmf}, one can see that 
the decay width tends to decrease for a smaller size of 
string curve. In the straight string limit, the superconducting 
current becomes stable~\cite{Ibe:2021ctf}. 
Around the right-lower corner where $\varepsilon$ is small and 
$m_\phi$ is large, $k_n\propto (m_\phi/\varepsilon)^{1/2}$ 
becomes large and then $\ms{p}_n^2<0$. Therefore any decay 
process is forbidden there. 

The Yukawa coupling dependence is read from the right panel of 
Fig.~\ref{fig:epVsmf}. For a smaller coupling, the two-body 
decay decreases but the conversion does not. In the tiny coupling 
limit, only the on-shell decay process is valid. 
We will discuss in the following analysis the two distinct cases: 
(i) both of the decay processes are possible and 
(ii) only the conversion is included.
When the two-body decay process is effective, we will 
set $\lambda=1$ as a benchmark value.

\subsection{Stability of zero mode in string network}

In the early Universe, conventional cosmic strings 
without superconductivity are considered to form a network and 
obey the scaling behavior thanks to the reconnection process.
We now discuss whether the zero modes on superconducting strings 
can affect the dynamics of its network. 
In order to affect the string dynamics, the current that the zero 
modes carry should be comparable to the mass scale of the 
string tension, $\sqrt{\mu} \sim m_\phi$, and hence the zero mode 
should have an energy of the order of $m_\phi$ since the amount of 
the current is roughly estimated as $\sim E$. 
Thus we study the stability of the zero mode with the 
energy $E=m_\phi$ in the string network.
In general, the current on the string is not chiral but a 
superposition of chiral zero modes in the string network.
For instance, a superconducting string moving in the 
magnetic field background gets a current without 
charges~\cite{Witten:1984eb,Ostriker:1986xc}, which is 
described by a superposition of the chiral zero modes 
traveling in the opposite directions. While such a case admits 
yet another channel for the current dissipation given by the 
collision of zero modes, we do not consider this type of 
process focusing the decay of single chiral zero mode.

In the analysis below, any generation process of the 
superconducting current is not needed to be specified. Whether 
a sufficient current can be generated depends on the time 
evolution of strings (and other background fields).
Further, we focus on the stability of zero modes 
and assume that the dynamics of string network is 
approximately given by the conventional scaling solution. 
If zero modes are sufficiently stable, one may need to solve 
the coupled equations describing the dynamics both of the string 
network and the zero mode current in terms of, e.g., the 
velocity-dependent one-scale (VOS) model for current-carrying 
strings~\cite{Martins:2020jbq,Martins:2021cid,Auclair:2022ylu}.

In the present case, there are two types of string curves.
First, in the scaling regime, strings have one typical macroscopic 
length scale which is comparable to the Hubble length. 
Thus the curvature radius and the period of curves are 
considered as the Hubble length, $R \sim L \sim H(T)^{-1}$. 
For such a macroscopic curve, the perturbative analysis 
generally breaks down. We therefore fix $R = H(T)^{-1}$ and 
take $L$ such that $\varepsilon=1$. That may give an approximate 
lower bound of the decay rate in the string network. 
Secondly, strings can receive an energy from thermal 
plasma (if they interact with plasma particles) which gives 
microscopic random curves of strings. This thermal 
curve is roughly given by $\delta \phi \sim T \sin (T z)$ 
where the typical momentum scale is $\mc{O}(T)$. That corresponds 
to the function
\begin{align}
  \zeta(z) \,\sim\, \frac{T}{m_\phi v_\phi} \sin (T z).
  \label{eq:thermal_fluc}
\end{align}
In order for the perturbative analysis to be valid, the 
thermal contribution should satisfy
$L \gg 1/m_\phi$ and $\epsilon \sim T/v_\phi \lesssim 1$, 
leading to
\begin{align}
  T \ll m_\phi, 
  \label{eq:thermal_cond}
\end{align}
which is achieved by taking into account the thermal curves 
only when $T < 10^{-3} m_\phi$. In the following, we evaluate the 
decay width for each type of curves separately. That is justified 
at the leading order of perturbations.

\begin{figure}
  \centering
  \includegraphics[width=0.48\textwidth]{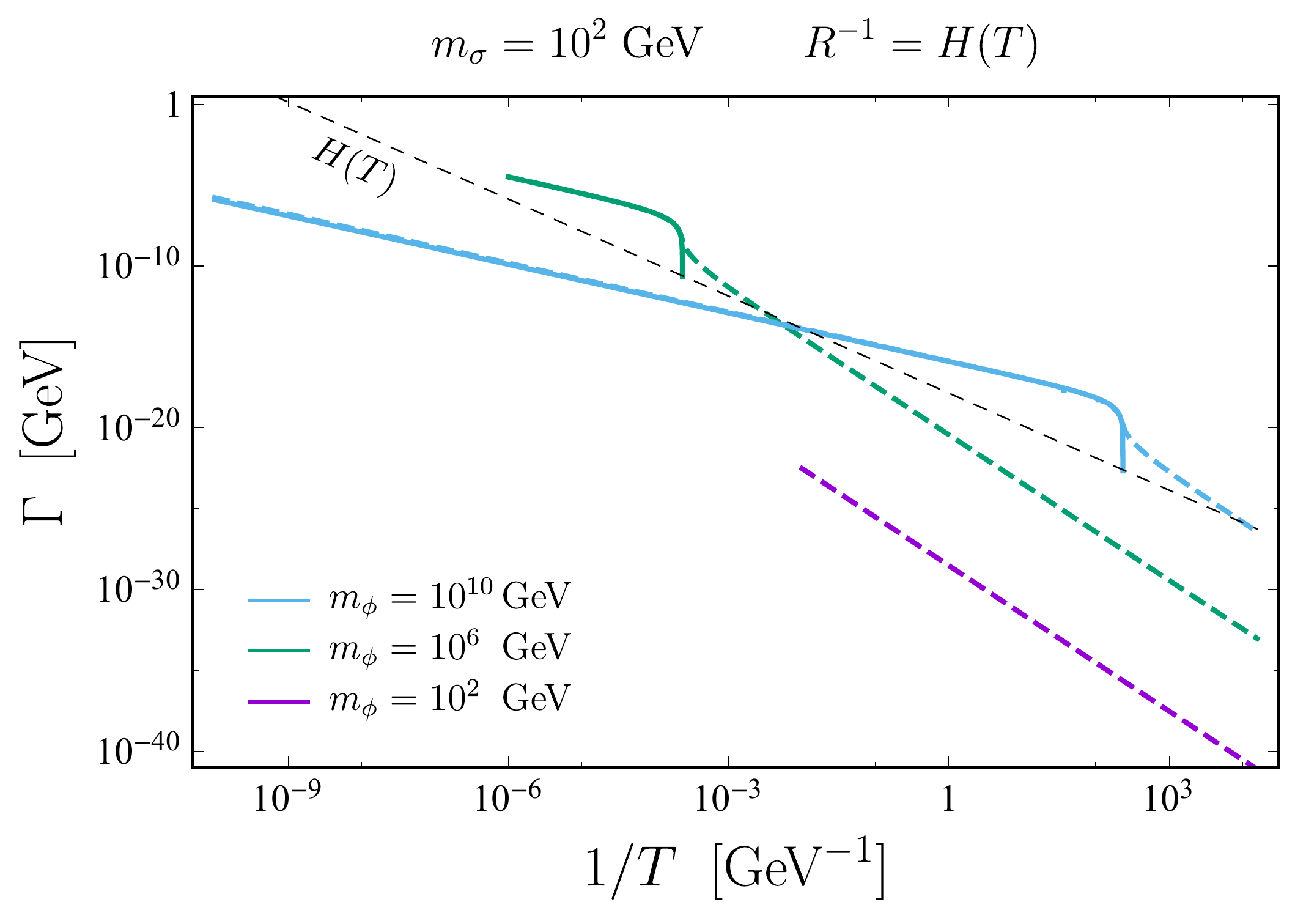}
  \quad
  \includegraphics[width=0.48\textwidth]{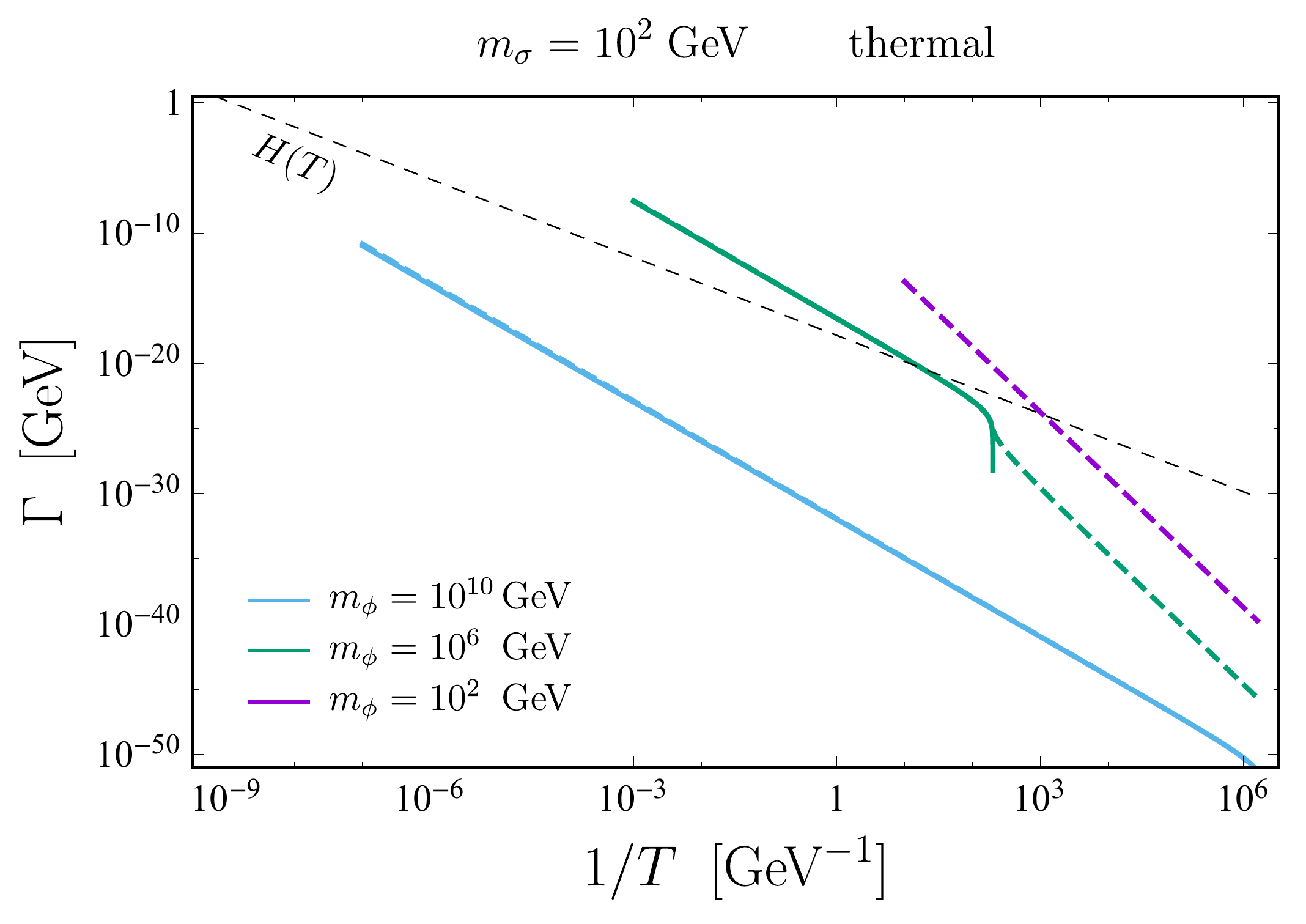}
  \caption{%
  The temperature dependence of the zero mode decay width with 
  $E = m_{\phi}$ and $m_\sigma = 10^2~\mr{GeV}$. 
  The colored solid and dashed lines correspond to the conversion 
  $\Gamma_{\conv}$ and the two-body decay $\Gamma_{2 \body}$, 
  respectively. (Some dashed and solid lines are overlapped.)
  The dashed black lines denote the Hubble parameter $H(T)$.
  (Left) The dependence comes from $R=H^{-1}$ (the scaling regime).
  The Yukawa coupling is set to $\lambda=1$. 
  (Right) The thermal curve with which the analysis is 
  assumed to be valid at $T<10^{-3}m_\phi$.
  }
  \label{fig:Tdepend}
  \bigskip
\end{figure}

These two types of string curves give different temperature 
dependence of the zero mode decay width. We show them 
in Fig.~\ref{fig:Tdepend} for a benchmark point where 
$E = m_\phi$ and $m_\sigma = 10^2~\mr{GeV}$. The left and right 
panels correspond to the curve with $R = H^{-1}$ and the 
thermal one, respectively. In the left panel, one can see that 
the behavior of $\Gamma$ changes at the transition 
temperature $T_\text{trs}^\mr{Hub}\sim 
m_\sigma^2 M_P^{1/2}/m_\phi^{3/2}$.
The asymptotic behaviors at high and low temperatures can be 
seen by substituting $R$ and $L$ into the approximate formulae 
\eqref{eq:analytic_Gamma_conv} and \eqref{eq:analytic_Gamma_2body}.
In the high temperature region, two partial decay 
widths $\Gamma_{\conv}$ and $\Gamma_{2\body}$ agree since the latter 
is dominated by the on-shell production of massive modes, and 
they become
\begin{align}
  \Gamma_{\conv}^\mr{Hub} \,\approx\, 
  \Gamma_{2\body}^\mr{Hub} \,\approx\,  
  T\frac{m_\sigma^2}{m_\phi^{3/2} M_P^{1/2}}
  \Big[\!\log\Big( \frac{m_\sigma}{m_{\phi}} \Big)\Big]^2
  \qquad \text{for } T \gtrsim T_\text{trs}^\mr{Hub} ,
  \label{eq:Gamma-2body-conv-Hub}
\end{align}
where we have used the series expansions for the Bessel 
functions $J_1(x)\sim x$ and $K_0(x)\sim (\log x)^2$, dropping 
$\mathcal{O}(1)$ numerical coefficients.
In the low temperature region, the on-shell production is 
kinematically forbidden and hence only the two-body decay 
$\Gamma_{2\body}$ is nonzero, leading to
\begin{align}
  \qquad \Gamma_{\conv}^\mr{Hub} \,=\,0, \qquad 
  \Gamma_{2\body}^\mr{Hub} \,\approx\, 
  T^3 \frac{\lambda^2 m_\phi ^{3/2} }{m_\sigma^2 M_P^{3/2}}
  \Big[\!\log\Big( \frac{m_\sigma}{m_{\phi}} \Big)\Big]^2 
  \qquad \text{for } T \lesssim T_\text{trs}^\mr{Hub}.
  \label{eq:Gamma-2body-Hub}
\end{align}

For the thermal curves (the right panel in 
Fig.~\ref{fig:Tdepend}), the behavior of $\Gamma$ changes at 
$T_\text{trs}^\mr{th}\sim m_\sigma^2/m_\phi$. The asymptotic 
behaviors are obtained in a similar way as above. The decay width 
via the thermal curve is evaluated by substituting 
\eqref{eq:thermal_fluc} into \eqref{eq:analytic_Gamma_conv} and 
\eqref{eq:analytic_Gamma_2body}. In the high temperature region, 
two partial decay widths $\Gamma_{\conv}$ and $\Gamma_{2\body}$ 
also agree and are given by
\begin{align}
  \Gamma_{\conv}^\text{th} \,\approx\, 
  \Gamma_{2\body}^\text{th} \,\approx\, 
  T^3\frac{m_\sigma^2}{m_\phi^4}
  \Big[\!\log\Big( \frac{m_\sigma}{m_{\phi}} \Big)\Big]^2 
  \qquad \text{for } T \gtrsim T_\text{trs}^\mr{th} ,
  \label{eq:Gamma-2body-conv-thermal}
\end{align}
while in the low temperature region, only the latter is non-vanishing
\begin{align}
  \qquad \Gamma_{\conv}^\text{th} \,=\, 0 , \qquad
  \Gamma_{2\body}^\text{th} \,\approx \, 
  T^5\frac{\lambda^2}{m_\sigma^2 m_\phi^2}
  \Big[\!\log\Big( \frac{m_\sigma}{m_{\phi}} \Big)\Big]^2 
  \qquad \text{for } T \lesssim T_\text{trs}^\mr{th} .
  \label{eq:Gamma-2body-thermal}
\end{align}
(The above expressions are numerically confirmed 
in Appendix~\ref{app:Gamma-temperature}.)
Note that in both temperature regions the decay widths via the 
thermal curves have steeper slopes with respect to $1/T$ than 
that of the Hubble parameter. Therefore in the case of thermal 
curve the stability of zero mode is determined at higher 
temperature, which means the ratio $\Gamma/H$ 
is ``ultraviolet (UV)-dominated''.

It may be worthwhile commenting on how the decay width 
depends on $m_\phi$. At first, the partial 
widths $\Gamma_\conv$ and $\Gamma_{2\body}$ contain 
several parts $(M^2+m_\sigma)^2$, $r_b^2$, 
and $J_1(\sqrt{\ms{p}_n^2 - m_\sigma^2} r_b)^2$ 
(or $J_1(\sqrt{\omega^2 - m_\sigma^2} r_b)^2$), each of 
which comes from the potential depth, the $r$-integration 
measure, and the wavefunction of the bulk mode $\chi_1$, 
giving the dependence $m_\phi^4$, $m_\phi^{-2}$, and 
$\ms{p}_n^2 m_\phi^{-2}$, respectively, 
for $m_\phi \gg m_\sigma$. The reason why the last one is 
inversely proportional to $m_\phi$ is that larger 
$m_\phi$ (smaller $r_b$) reduces the overlap between 
the zero mode and the bulk mode wavefunctions.
Thus the dependence from the potential shape disappears 
and the widths become
$\Gamma_\conv \propto |\zeta_n|^2 \ms{p}_n^2/E 
\approx \epsilon^2 k_1 / m_\phi^2$ and 
$\Gamma_{2\body} \propto |\zeta_n|^2 \ms{p}_n^6/E 
\approx \epsilon^2 E^2 k_1^3 /m_\phi^2$, where we have 
dropped $m_\sigma$ and assumed the off-shell contribution 
from the bulk modes for $\Gamma_{2\body}$, leading to 
the suppression of $\Gamma_{2\body}$ for smaller $E$ and $k_1$.
From these expressions, we can intuitively state that 
the decay widths are reduced for larger $m_\phi$ with 
fixed $E$, $k_1$ and $\epsilon$ because the string becomes 
too heavy to be deformed. If one sets $E=m_\phi$, however, 
a more careful treatment is needed particularly 
for $\Gamma_{2\body}$. Indeed, by taking $E=m_\phi$, 
the explicit $m_\phi$-dependence is canceled and the 
dependence comes only through $\epsilon$ and $k_1$,
which may give positive powers of $m_\phi$ depending on the 
types of string curves. This is why $\Gamma_{2\body}$ 
contains the positive power as $m_\phi^{3/2}$ 
in Eq.~\eqref{eq:Gamma-2body-Hub}.

Now, we present the parameter space in which the zero mode 
affect the string dynamics. We consider the both types of 
string curves. At a temperature $T$, the zero mode 
immediately decays if $\Gamma(T) > H(T)$ holds, resulting in 
the usual string network without any significant current.
We also assume that the current and charge are not 
supplied after the decay.

\begin{figure}[p]
  \centering
  \includegraphics[width=0.48\textwidth]{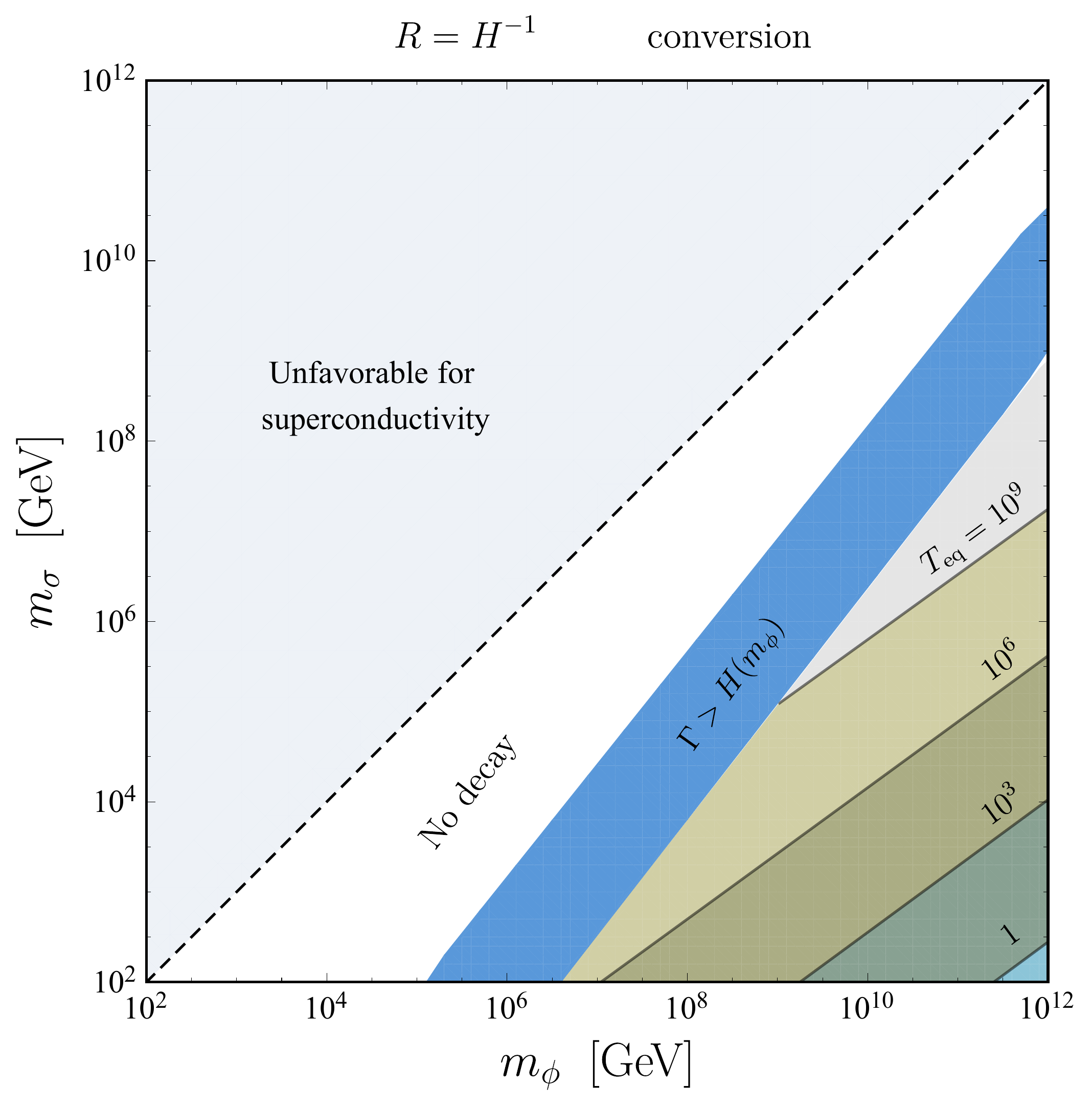}
  \quad
  \includegraphics[width=0.48\textwidth]{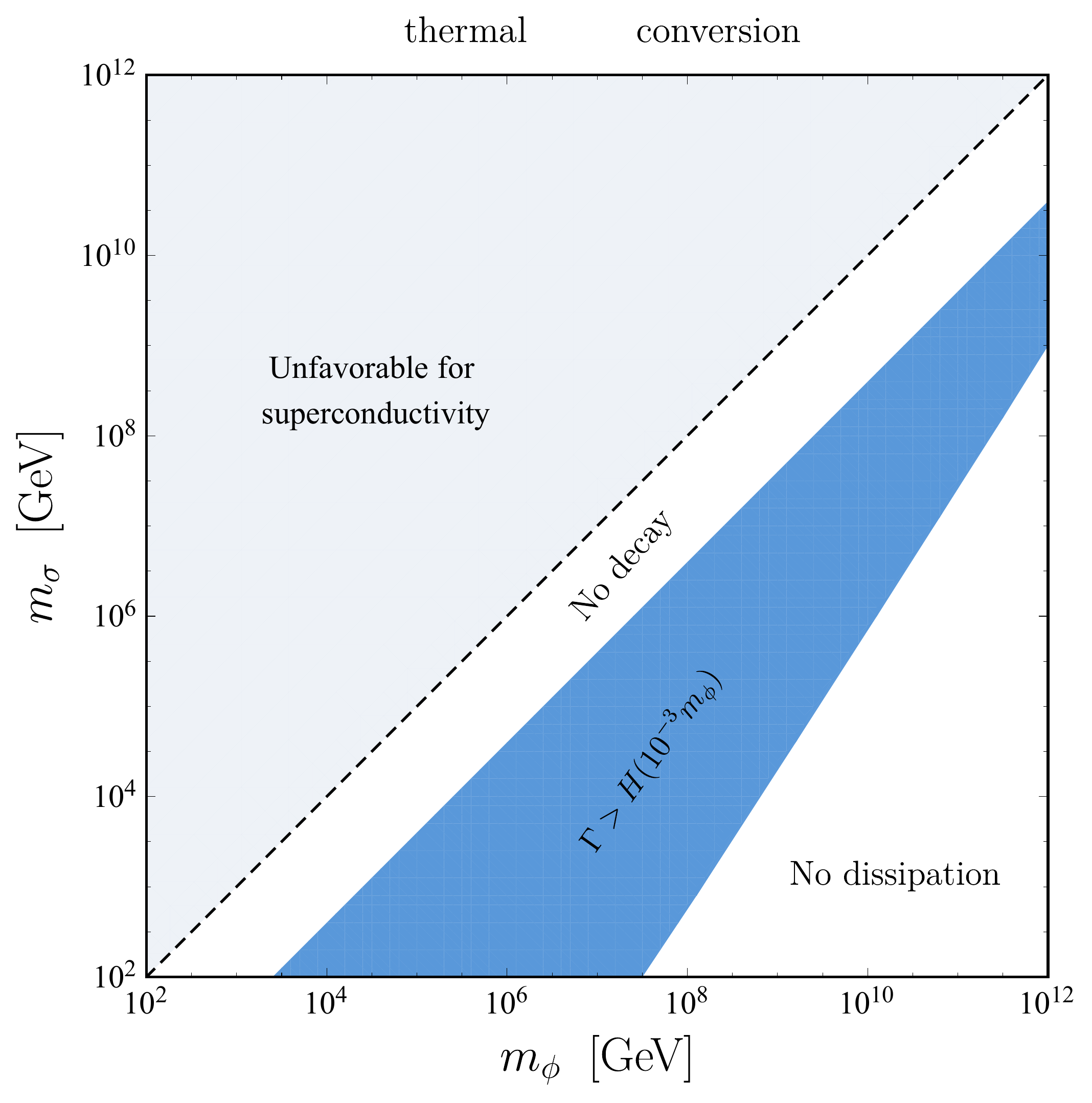}
  \\[4mm]
  \includegraphics[width=0.48\textwidth]{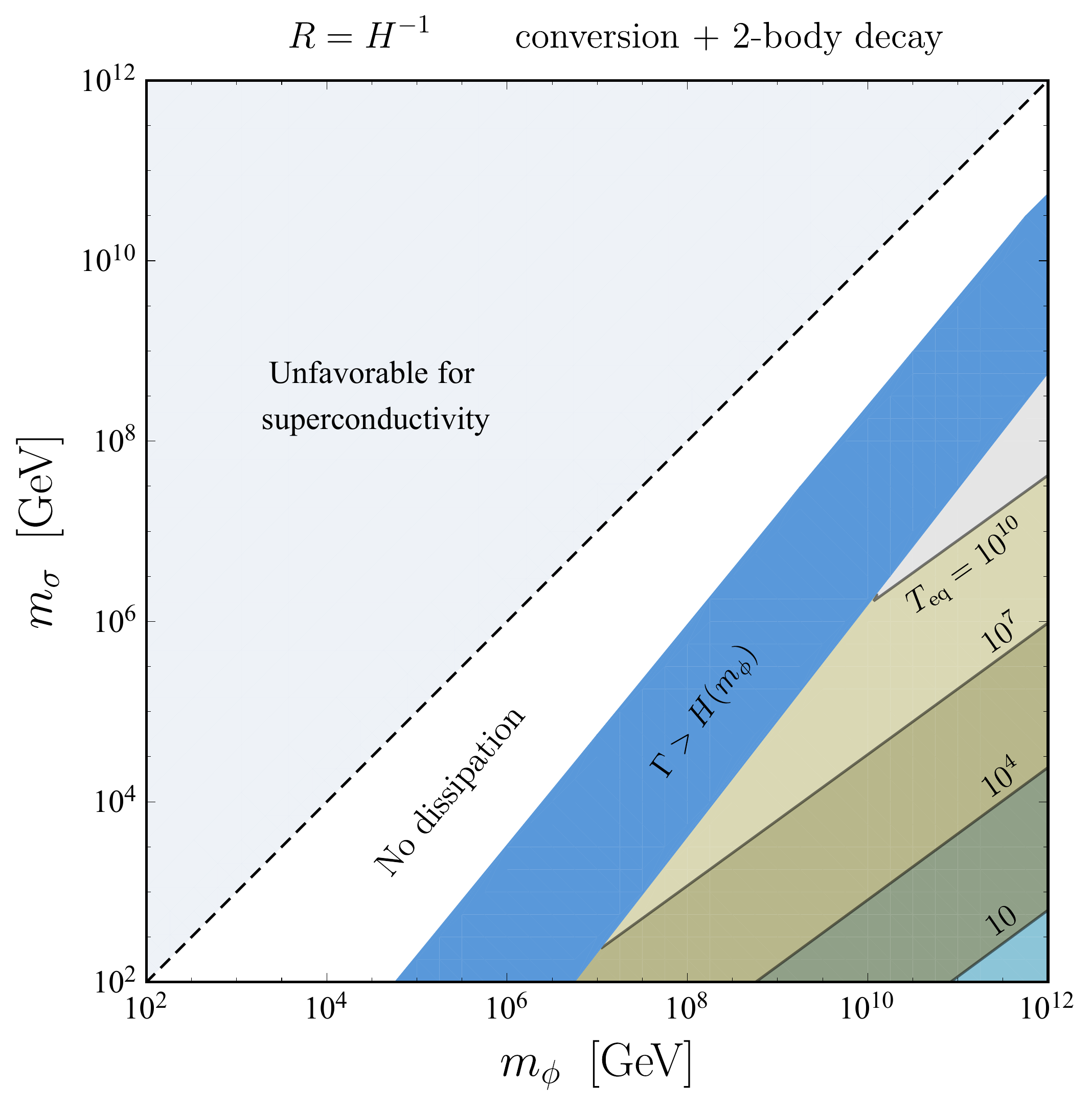}
  \quad
  \includegraphics[width=0.48\textwidth]{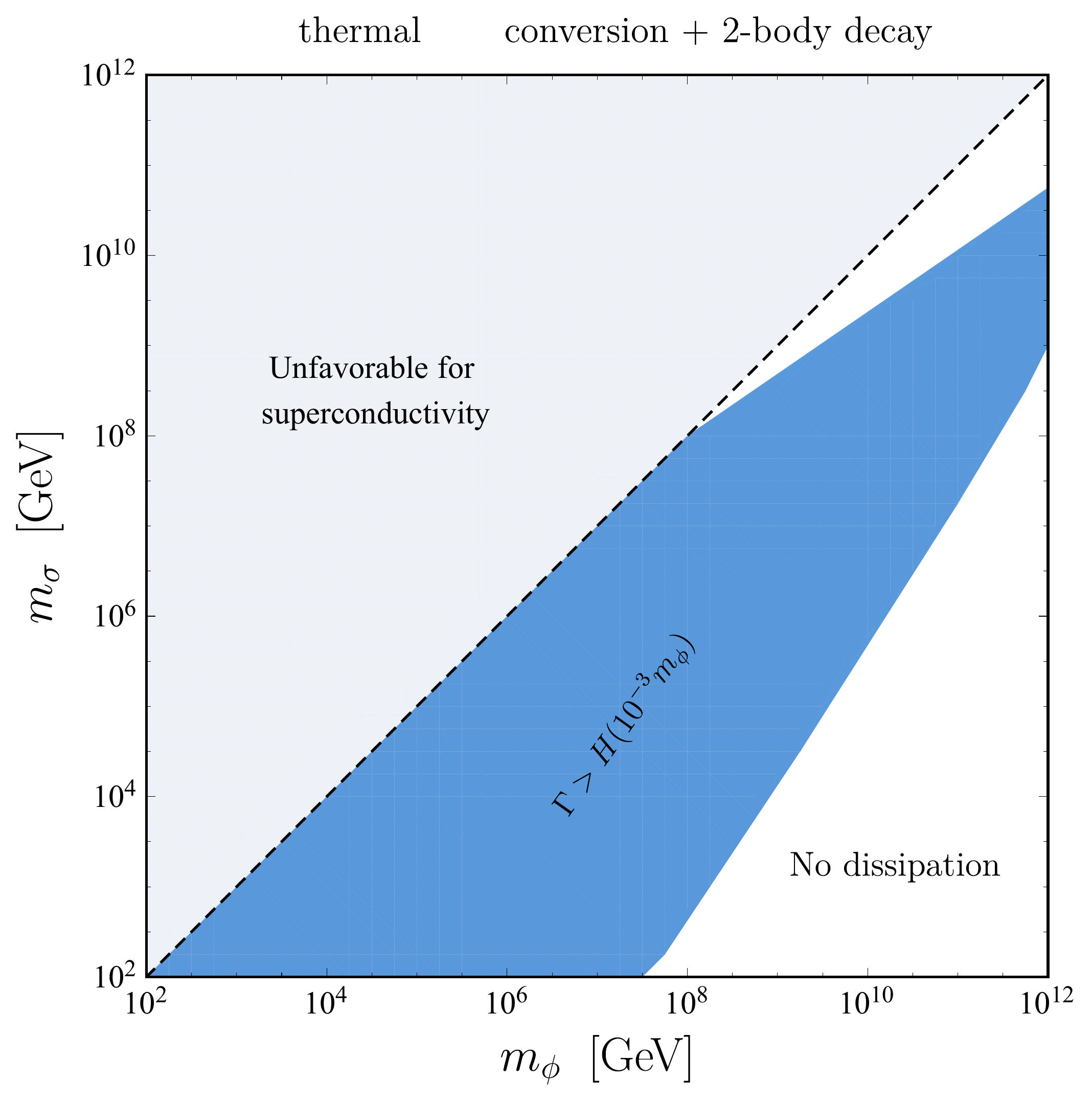}
  \caption{The parameter space where the decay of the zero mode with 
  $E=m_\phi$ is significant, i.e., $\Gamma(T) > H(T)$ holds at some 
  stage. These figures differ with respect to the type of 
  string curves and decay processes as indicated. 
  The contours correspond to the GeV-unit values of the 
  temperature $T_\mr{eq}$ at which the transition 
  from $\Gamma<H$ to $\Gamma>H$ occurs. 
  The blue regions means $\Gamma>H$ is already met at the 
  stage of the production of string network. The white 
  blank regions represent the parameter space in which the 
  zero mode survives at all stages, i.e., $\Gamma<H$ for 
  arbitrary temperature.
  }
  \label{fig:regionMphiMsigma}
  \bigskip
\end{figure}

In Fig.~\ref{fig:regionMphiMsigma}, we show such a parameter space
on the $(m_\phi, m_\sigma)$-plane with the $\Gamma(T)$ behavior.
Only the conversion process is taken into account in the top 
two panels while both processes are in the bottom two panels.
The left two panels correspond to the curve 
with $R=H^{-1}$ while the right two panels to the thermal 
one. The colored region in the left panels show the 
parameter space for $m_\sigma$ and $m_\phi$ in which the zero 
mode decay becomes significant, i.e., $\Gamma(T) > H(T)$ 
holds at some stage in the Universe. The contours correspond 
to when a transition from $\Gamma(T) < H(T)$ 
to $\Gamma(T) > H(T)$ occurs, i.e., the temperature 
$T_\mr{eq}$ such that $\Gamma(T_\mr{eq}) = H(T_\mr{eq})$.
The blue regions represent that $\Gamma(T) > H(T)$ is already 
met at $T = m_\phi$, i.e., at the stage of the production 
of strings. The white blank represents the parameter space 
in which the zero mode survives at all stages, 
i.e., $\Gamma(T) < H(T)$ for arbitrary $T$.
``No decay'' in the white blank indicates that the decay 
cannot occur kinematically since the threshold 
condition \eqref{eq:threshold} cannot be satisfied there, 
while ``No dissipation'' indicates that the decay can 
occur kinematically but $\Gamma(T) < H(T)$ holds at any 
stage, resulting in no significant dissipation. Since the 
ratio $\Gamma(T)/H(T)$ for the thermal curve 
is UV-dominated, whether or not the decay is significant 
is determined at the highest temperature $T=10^{-3}\, m_\phi$.
That is why other color than blue regions cannot be seen 
in the right panels.

From the above results, we obtain the cosmological 
consequences of the zero mode decay. If the Yukawa 
coupling $\lambda$ is tiny or absent, the two-body decay 
does not occur and the bottom panels 
in Fig.~\ref{fig:regionMphiMsigma} are irrelevant.
In this case, by taking into account both types of curves, 
superconducting strings cannot carry significant amount of 
current due to the rapid decay for 
$m_{\phi}\gg m_\sigma$ (typically, 
$m_{\phi}\gtrsim 10^2\, m_\sigma$). This can be 
intuitively understood since the zero mode 
energy $E=m_\phi$ suffices to produce the on-shell bulk modes 
with the mass $m_\sigma$ and to escape from the string.
On the other hand, if a sizable Yukawa 
coupling $\lambda$ is present, one should take into 
account also the bottom panels 
in Fig.~\ref{fig:regionMphiMsigma}. In this case, 
superconducting strings cannot carry significant current 
in almost all parameter region
because, even with $m_{\phi}\approx m_\sigma$, 
the two-body decay is kinematically allowed and the 
thermal curve with $T= 10^{-3}\, m_\sigma$ is sufficient 
to cause a rapid decay. Note that there is an exceptional region 
$m_\sigma \approx m_\phi \gtrsim 10^8$ GeV 
(upper-right blank region in the bottom-right panel).
The reason why the decay is suppressed for this region is
that $\Gamma/H$ at $T= 10^{-3}\, m_\phi$ is proportional 
to $M_P/m_\phi$ (see Eq.~\eqref{eq:Gamma-2body-thermal}).

\subsection{Vorton stability}

Throughout this paper, we only consider vortons that 
are sequentially produced from the string 
network~\cite{Peter:2013jj,Fukuda:2020kym,Auclair:2020wse}.
Note that there are also vortons produced at the phase 
transition of the string formation. The latter ones typically 
have shorter lifetime due to less numbers of charges than 
the former, and thus the discussion on the former type 
is sufficient to study the stability of vortons.

In the VOS model, closed loops of strings are generated with 
the length $\sim 1/H(T_\mr{gen})$ at $T=T_\mr{gen}$.
On these loops, the charge and current are randomly induced 
with the coherent length $1/T_\mr{gen}$ thermally. It is 
expected~\cite{Davis:1988jq} that the zero mode on closed 
loops becomes chiral (see Eq.~\eqref{eq:chiral-zeromode}) as 
the loops contract.
In the rest of the paper, we focus on such chiral vortons. 
Using the central limit theorem, we estimate the net 
charge value on the loop at the production, given as
\begin{align}
  Q(T_\mr{gen}) \,\sim\, 
  \sqrt{\frac{T_\mr{gen}}{H(T_\mr{gen})}} \,=\, 
  \biggl(
    \frac{90}{\pi^2 g_*}
  \biggr)^{1/4}
  \sqrt{\frac{M_P}{T_\mr{gen}}}.
  \label{eq:vorton-initial-Q}
\end{align}
At the classical level, after the production, the loop 
shrinks until the radius is stabilized, i.e., its size 
becomes $R_0\sim Q/m_\phi$ (see Eq.~\eqref{eq:vorton_radius}).
Since the energy of the zero mode is given by 
$E \sim Q/R_0 \sim m_\phi$, we consider the zero mode decay 
width for $E=m_\phi$.

Similarly to the case of the string network, the zero mode 
decay in the vorton is caused by two types of string 
curves. One comes from the fact that a vorton is 
curved, i.e., its radius of curvature is nothing but the 
radius of the stabilized loop $R_0$. For this case, we again 
take $L$ such that $\varepsilon =1$ to avoid breakdown of 
the perturbativity. 
The other is the thermal curve from the interaction 
with thermal plasma. We consider the decay width for each type.

For the former type curve with a vorton produced 
at $T=T_\mr{gen}$, the vorton radius depends on $T_\mr{gen}$
and so does the decay width $\Gamma$ as
\begin{align}
  \Gamma_{\conv}^\mr{vort} \,\approx\, 
  \Gamma_{2\body}^\mr{vort} \,\approx\, 
  \frac{m_\sigma^2 T_\mr{gen}^{1/4}}{m_\phi M_P^{1/4}}
  \Big[\!\log\Big( \frac{m_\sigma}{m_{\phi}} \Big)\Big]^2
  \qquad \text{for } T_\mr{gen} \gtrsim T_\text{trs}^\mr{vort} ,
  \label{eq:Gamma-2body-conv-vort}
\end{align}
and
\begin{align}
  \Gamma_{\conv}^\mr{vort} \,=\,0, \qquad 
  \Gamma_{2\body}^\mr{vort} \,\approx\, 
  \lambda^2 \frac{m_\phi^3 T_\mr{gen}^{3/4}}{m_\sigma^2 M_P^{3/4}}
  \Big[\!\log\Big( \frac{m_\sigma}{m_{\phi}} \Big)\Big]^2
  \qquad \text{for } T_\mr{gen} \lesssim T_\text{trs}^\mr{vort},
  \label{eq:Gamma-2body-vort}
\end{align}
where $T_\text{trs}^\mr{vort}=M_P m_\sigma^2/m_\phi^2$. 
We have used the series expansions for the Bessel 
functions $J_1(x)\sim x$ and $K_0(x)\sim (\log x)^2$, and 
dropped $\mathcal{O}(1)$ numerical coefficients.
(The above expressions are numerically confirmed 
in Appendix~\ref{app:Gamma-temperature}.)
As mentioned in the last subsection, the two partial decay 
widths $\Gamma_{\conv}$ and $\Gamma_{2\body}$ agree in the 
high temperature region 
$T_\mr{gen} \gtrsim T_\text{trs}^\mr{vort}$, while the 
conversion process is kinematically forbidden in the 
low temperature region. 

For the latter type curve with thermal plasma, the decay width 
does not depend on $T_\mr{gen}$ but on $T$. ($T_\mr{gen}$ labels 
each vorton as its production time while $T$ is used as the 
temporal variable.) The partial widths are the same as those in 
the last subsection (Eqs.~\eqref{eq:Gamma-2body-conv-thermal} and 
\eqref{eq:Gamma-2body-thermal} for their approximate expressions). 

After the zero mode decay occurs with the time 
scale $1/\Gamma$, its charge decreases by a unit charge,
which results in that the vorton shrinks and becomes 
balanced at a radius $R\sim (Q-1)/m_\phi$. Then the zero mode can 
decay again with the time scale $1/\Gamma$ in which the 
classically stable radius $R_0$ is replaced by $\sim (Q-1)/m_\phi$.
Thus the time evolution of the radius $R$ of a vorton 
produced at $T=T_\mr{gen}$ is described by the following equation:
\begin{align}
  \frac{\dot{R}}{R} \,=\, 
  \frac{\dot{Q}}{Q} \,=\, -\Gamma(R,T), 
  \label{eq:vorton-R-time-evol}
\end{align}
with the initial condition 
\begin{equation}
   R|_{T=T_\mr{gen}} =\, R_0 (T_\mr{gen}) \,\sim\, 
   \frac{Q(T_\mr{gen})}{m_\phi} \,.
\end{equation}
(The dotted quantities indicate their derivatives with 
respect to the cosmic time $t$.) After shrinking 
sufficiently with Eq.~\eqref{eq:vorton-R-time-evol}, the 
vorton eventually becomes to have a radius comparable to 
the string width $1/m_\phi$, which is regarded as 
the death of the vorton.

\begin{figure}
  \centering
  \includegraphics[width=0.48\textwidth]{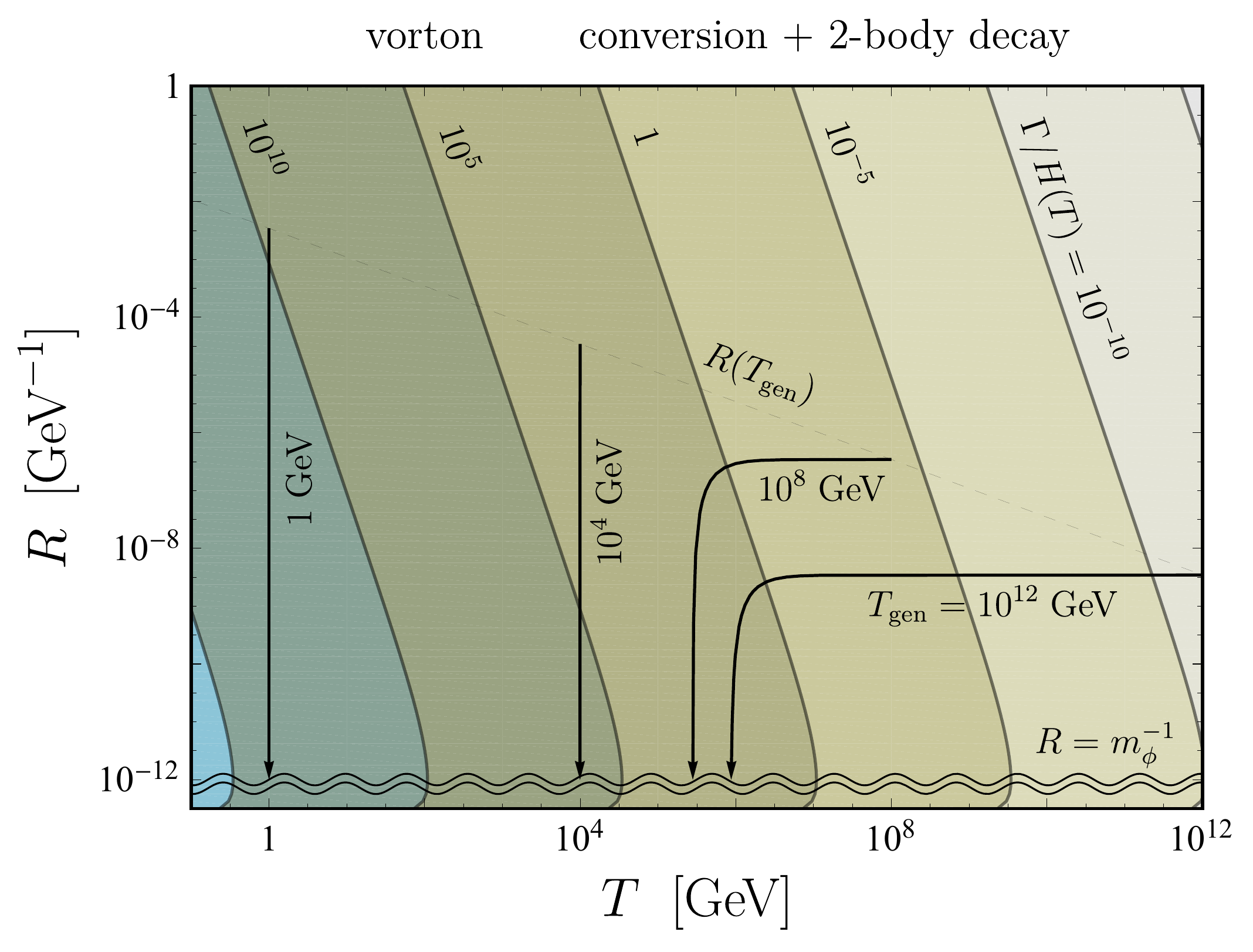}
  \quad
  \includegraphics[width=0.48\textwidth]{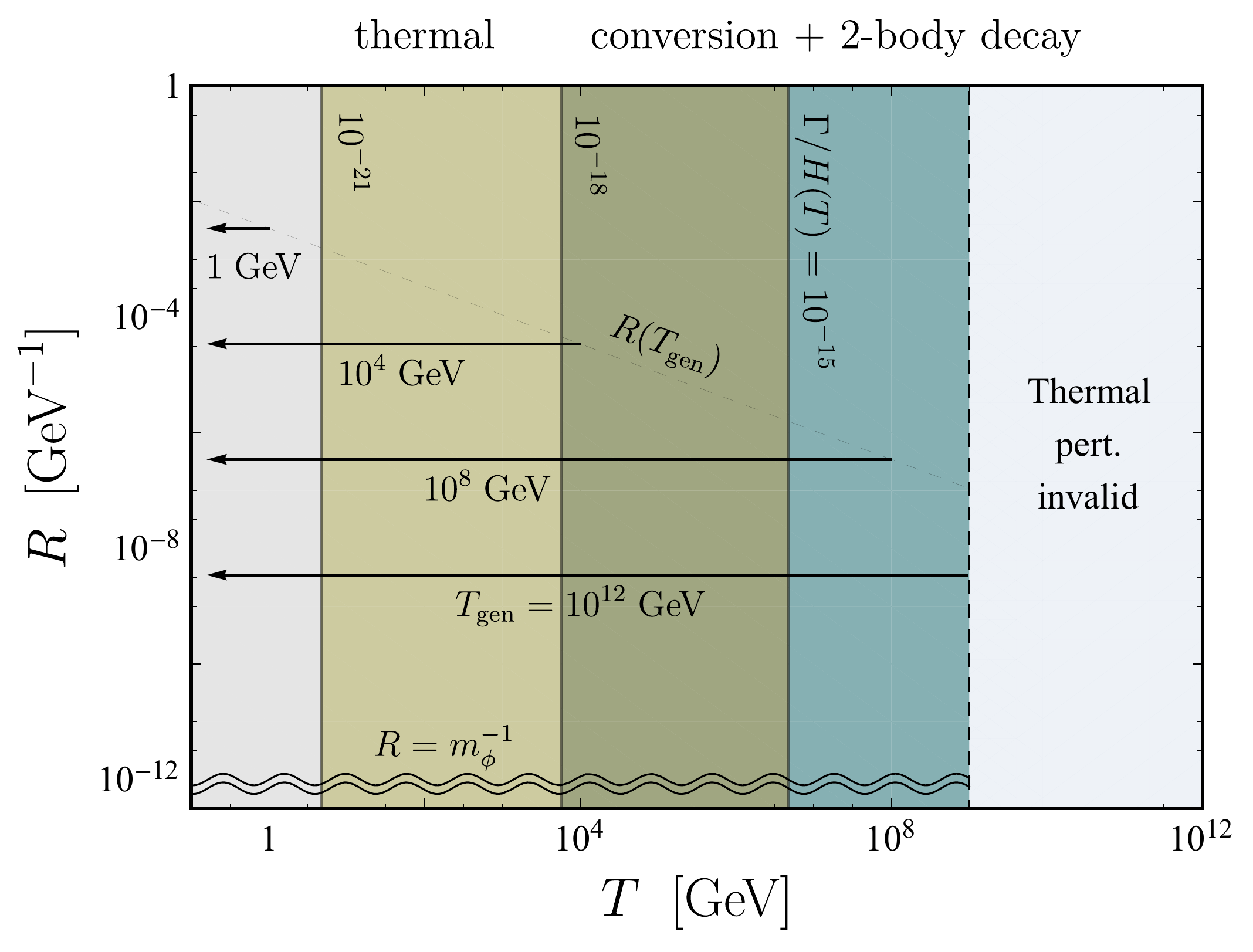}
  \caption{The time evolution of the vorton radius $R$ with 
  various production temperatures in the cases of the vorton 
  curve (left) and the thermal curve (right). The model 
  parameters are set to be $E=m_\phi=10^{12}~\mr{GeV}$ and 
  $m_\sigma=10^2~\mr{GeV}$. The black lines with arrows 
  indicate the trajectories of vorton radii produced 
  at $T=T_\mr{gen}$. 
  Each contour means the specified values of decay widths over 
  the Hubble parameter.
  }
  \label{fig:vortonRT}
  \bigskip
\end{figure}
\begin{figure}
  \centering
  \includegraphics[width=0.48\textwidth]{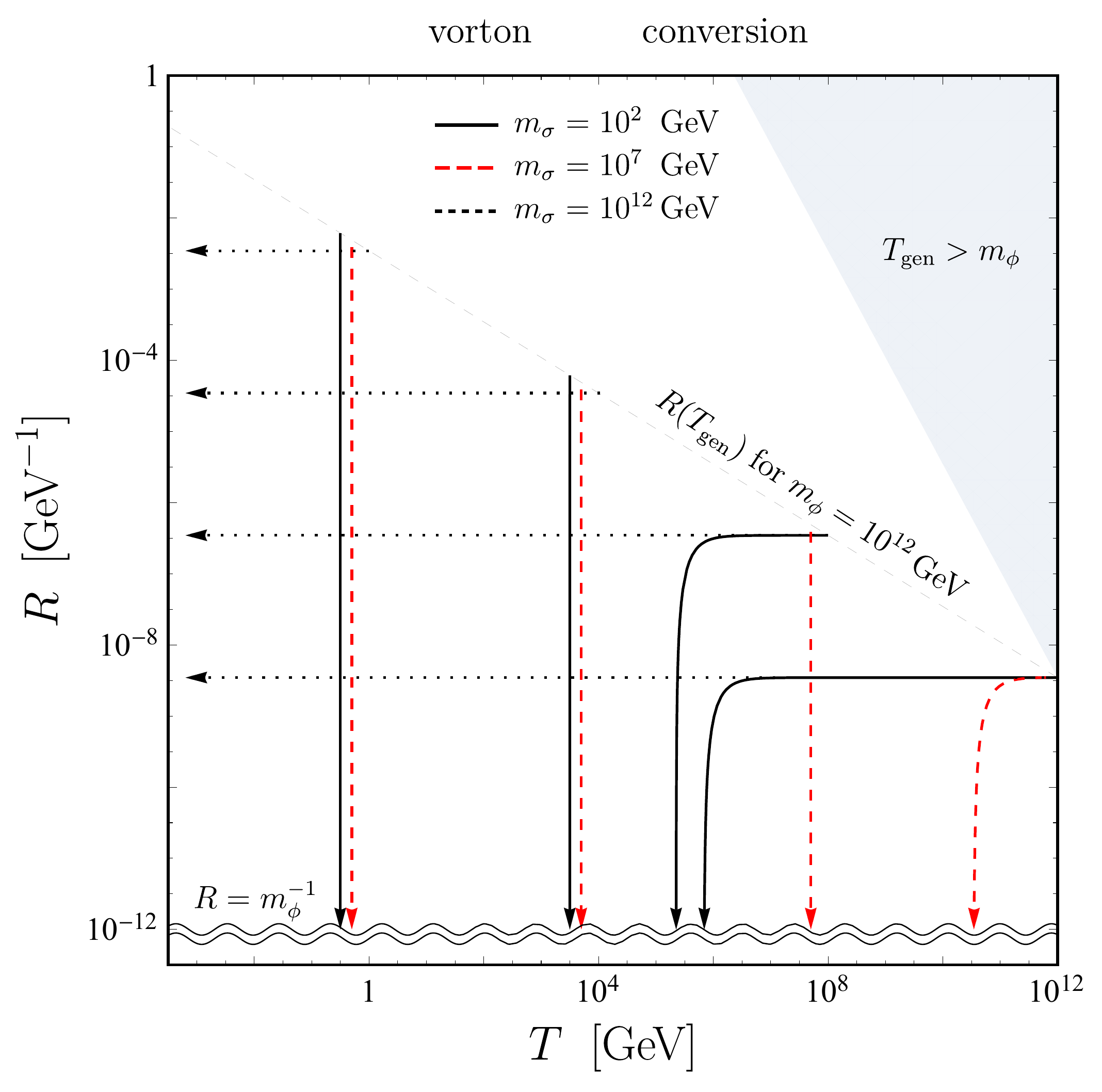}
  \quad
  \includegraphics[width=0.48\textwidth]{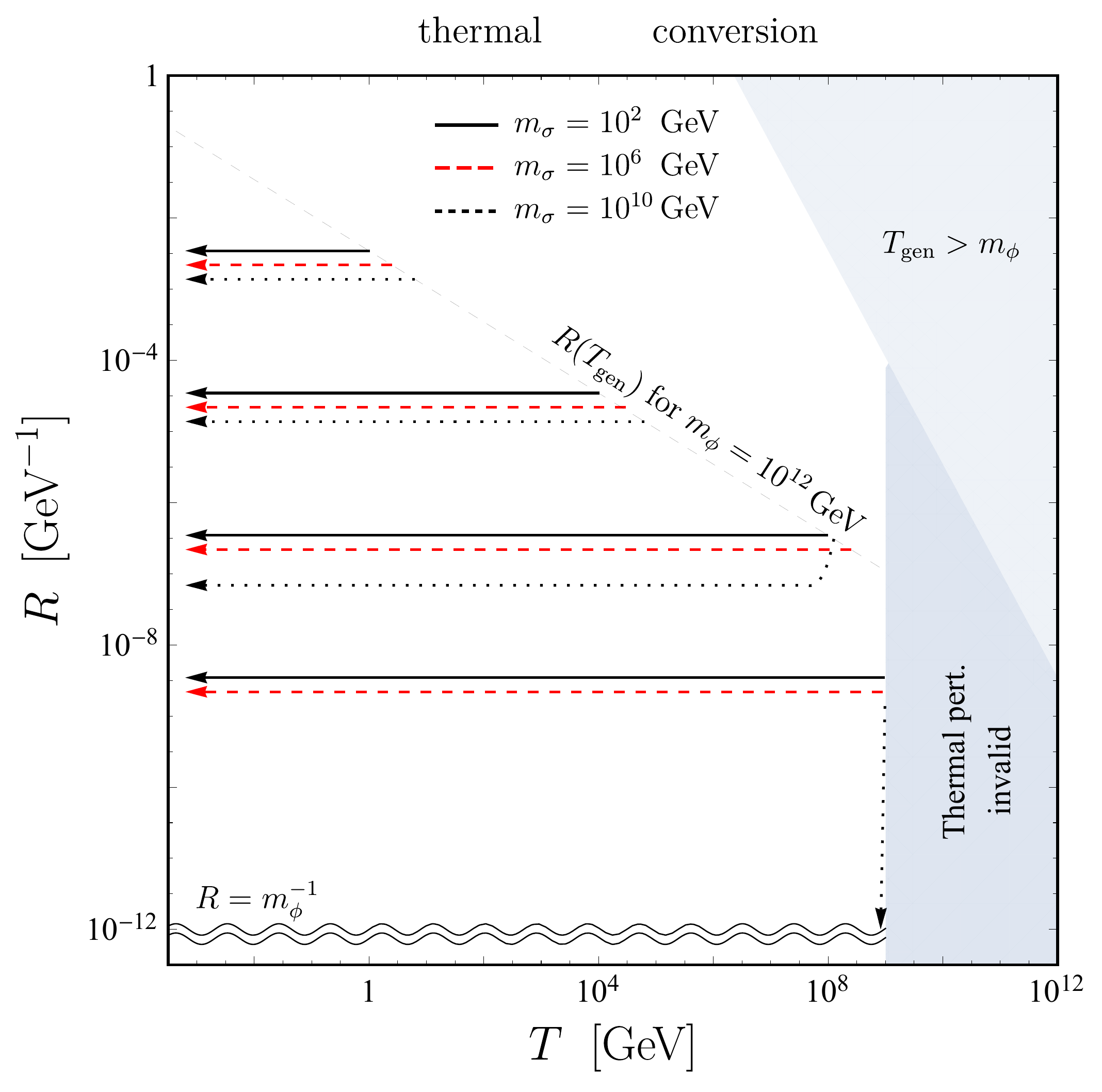}
  \\[4mm]
  \includegraphics[width=0.48\textwidth]{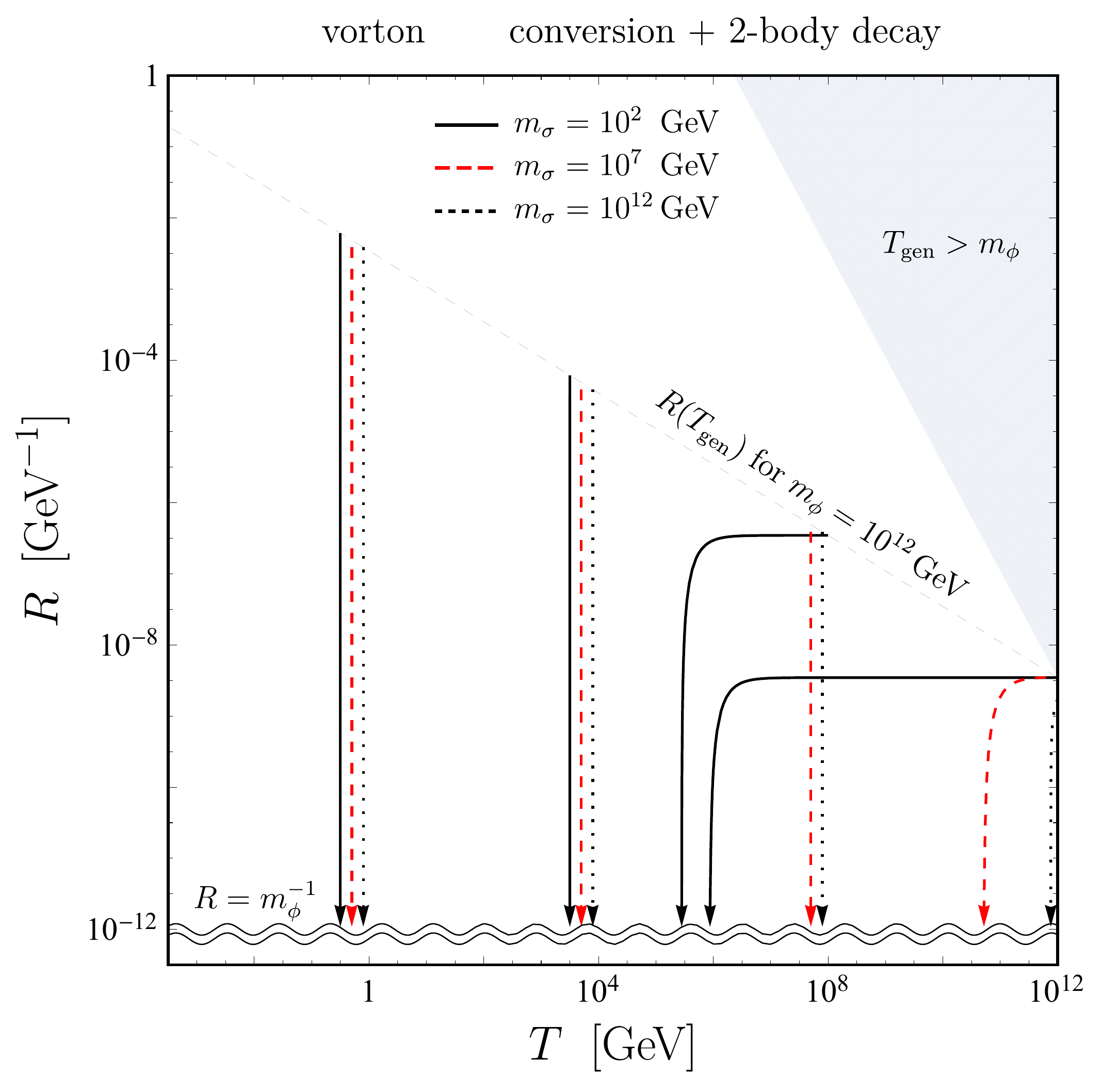}
  \quad
  \includegraphics[width=0.48\textwidth]{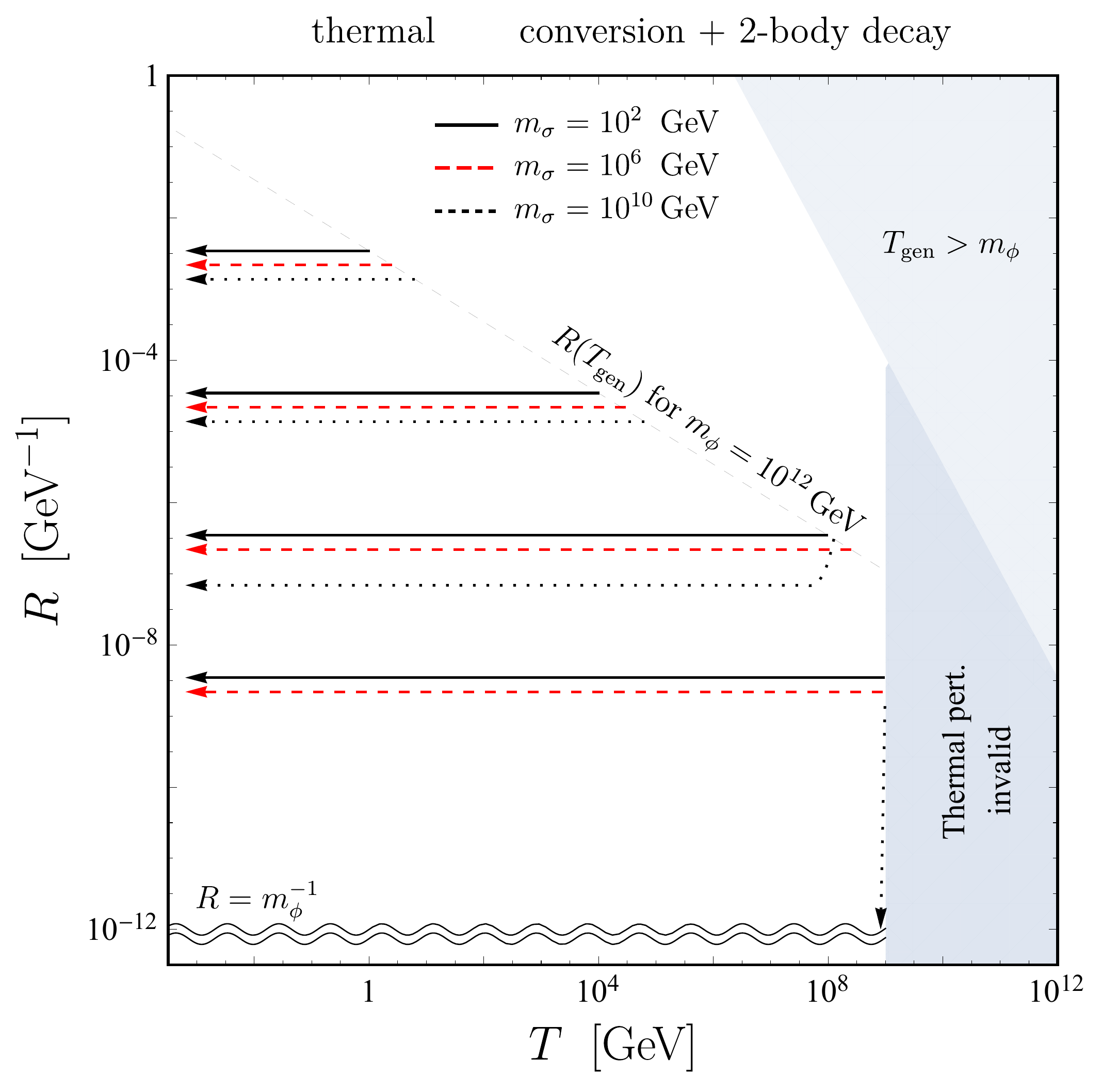}
  \caption{%
  The time evolution of the vorton radius $R$ with various 
  production temperatures and $E=m_\phi=10^{12}~\mr{GeV}$. The 
  lines with arrows indicates the 
  trajectories of the vorton radii for different $m_\sigma$ 
  and $T_\mr{gen}$. These figures differ with respect to the 
  type of string curves and decay processes as indicated.
  In the gray regions, the temperature $T$ does not satisfy 
  the condition Eq.~\eqref{eq:thermal_cond} and we assume 
  no decay process occurs in the regions.
  }
  \label{fig:vortonRT2}
  \bigskip
\end{figure}

Fig~\ref{fig:vortonRT} shows the time-evolution of the vorton 
radius with $m_\phi = 10^{12}~\mr{GeV}$ and 
$m_\sigma = 10^2~\mr{GeV}$. The black lines with arrows 
indicate the trajectories of the vorton radii with 
various $T_\mr{gen}$. The decay widths include both the 
processes of the conversion and the two-body decay 
(with $\lambda=1$). In the left panel, we consider the decay 
due to the vorton curve,  while the thermal one 
is considered in the right panel. Each contour means the 
value of $\Gamma/H$. 
We can find that once $\Gamma$ exceeds $H$, the vorton 
radius immediately shrinks into $R=m_\phi^{-1}$, the death 
of the vorton, which means that vortons die approximately 
when $\Gamma = H$. The contribution from the thermal 
curves is not dominant to the decay for these mass 
parameters because the string is too heavy to be deformed 
by thermal plasma.

Fig.~\ref{fig:vortonRT2} shows the time evolution of the 
vorton radius for various values of $m_\sigma$
with $m_\phi$ fixed to be $10^{12}$ GeV\@. The solid, dashed 
and dotted lines with arrows indicate the trajectories of 
the vorton radii for various $m_\sigma$ and $T_\mr{gen}$.
The top (bottom) two panels correspond to the conversion 
process (the conversion and the two-body decay 
with $\lambda=1$). The left panels correspond to the 
vorton curve while the right panels to 
the thermal one. In the gray region 
with ``Thermal pert.\ invalid'', the temperature $T$ does 
not satisfy the condition Eq.~\eqref{eq:thermal_cond} and 
hence the calculation is not reliable. Thus the vortons in 
such regions are assumed to be stable against the 
thermal curves. We find that particularly 
the two-body decay due to the vorton curves (bottom-left 
panel) is significant than the others. The thermal ones 
do not cause the strong decay since the thermal curves 
become effective at higher temperature.

\begin{figure}
  \centering
  \includegraphics[width=0.48\textwidth]{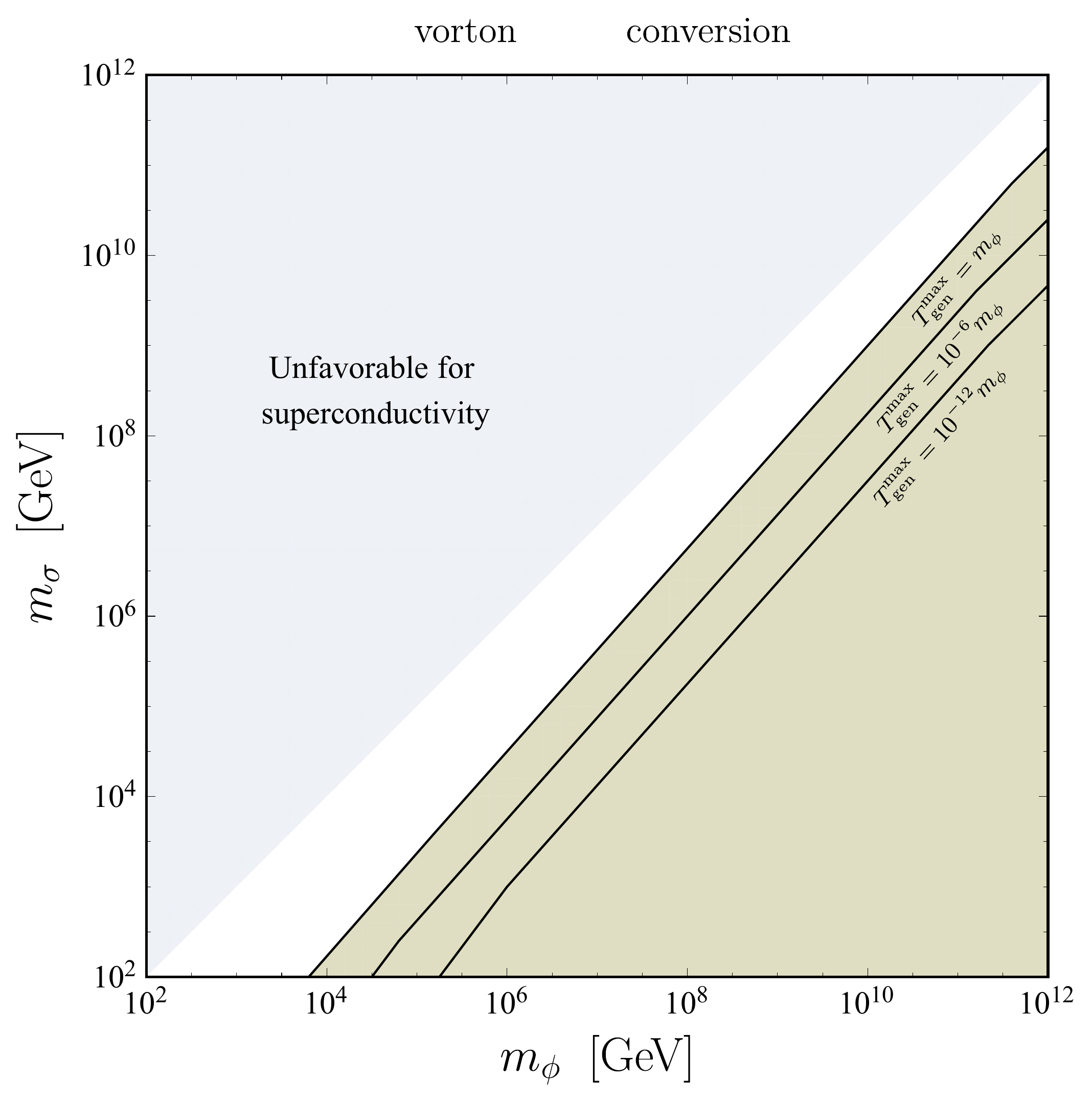}
  \quad
  \includegraphics[width=0.48\textwidth]{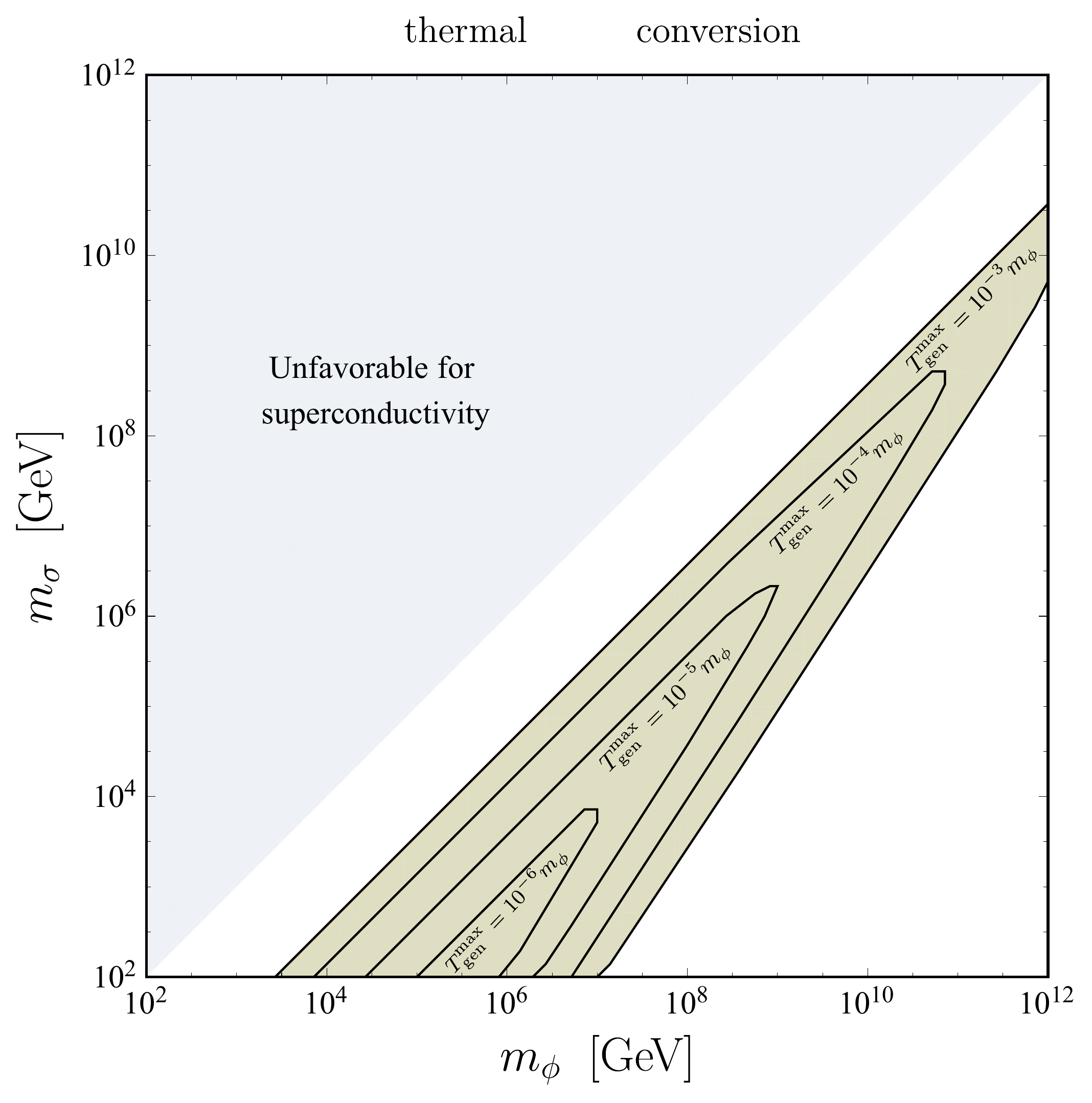}
  \\[4mm]
  \includegraphics[width=0.48\textwidth]{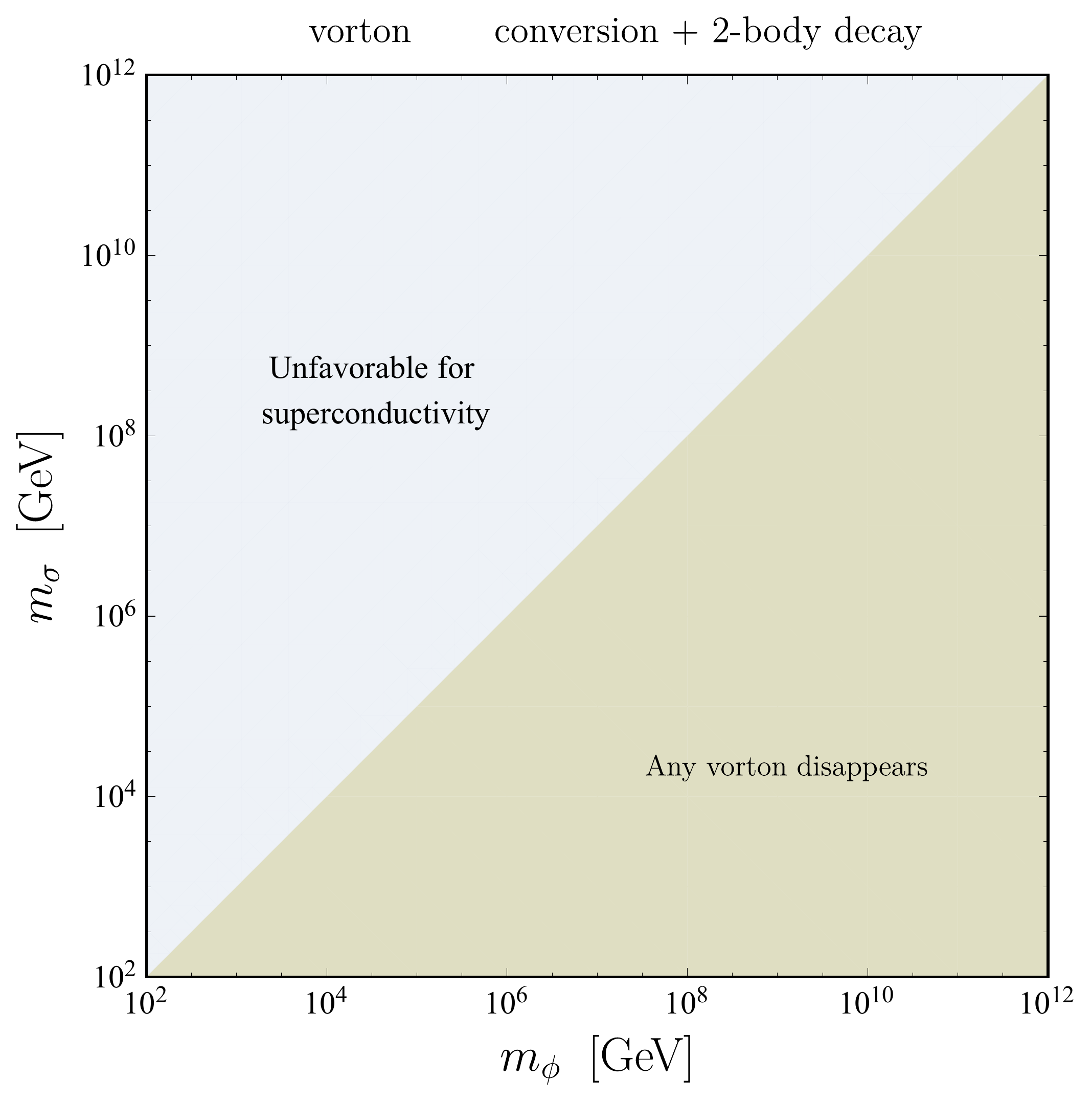}
  \quad
  \includegraphics[width=0.48\textwidth]{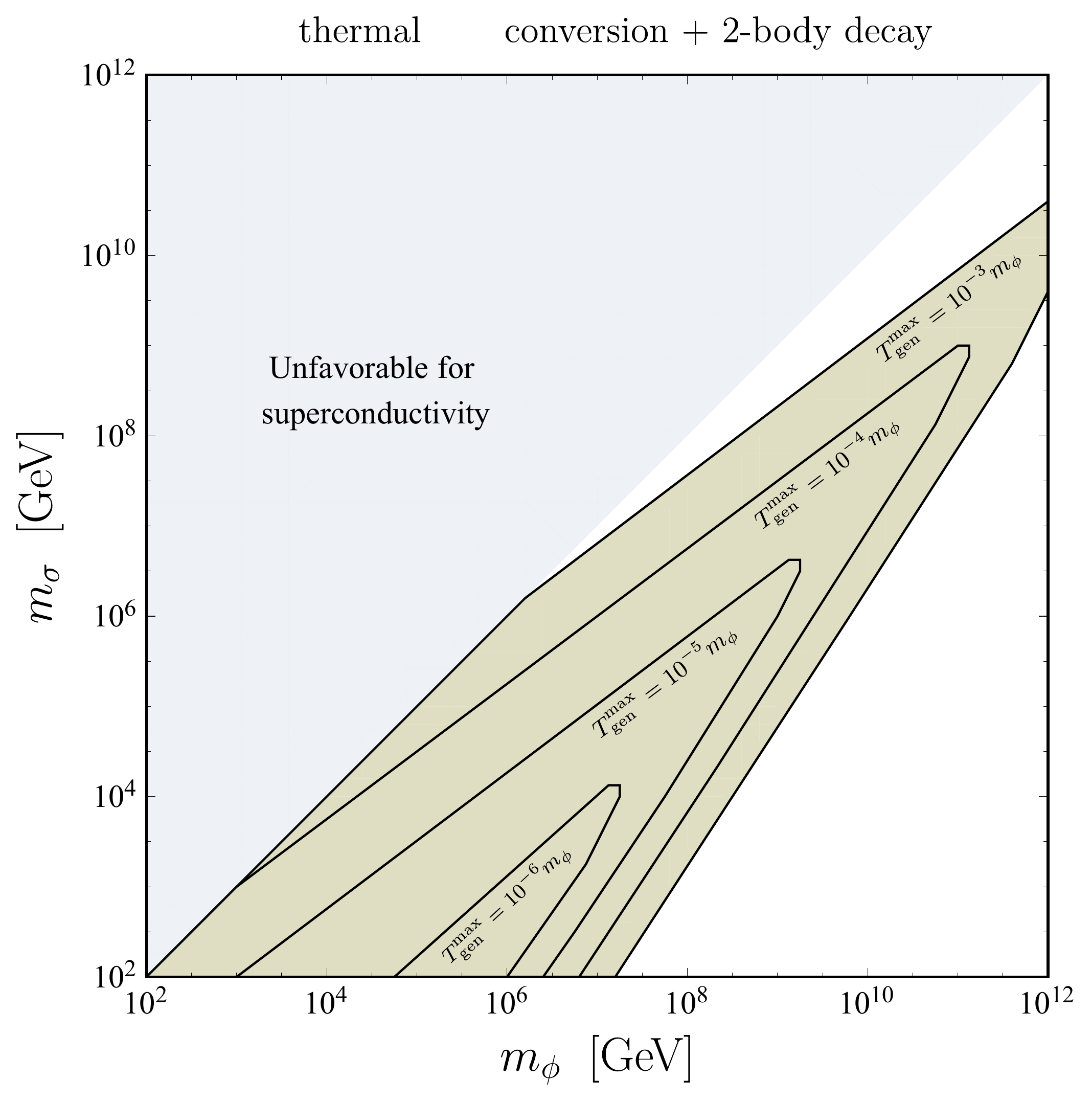}
  \caption{
  The parameter spaces for $(m_\sigma,m_\phi)$ where the 
  vortons are dead by now. We set $E=m_\phi$. The black contours 
  in the colored regions indicate the values of 
  $T_\mr{gen}^\mr{max}$, defined as the maximum $T_\mr{gen}$ 
  among the living vortons in the present Universe. In the white 
  blank regions, all vortons survive in the present Universe. 
  These figures differ with respect to the type of string curves 
  and decay processes as indicated.
  The upper-left gray regions in the plots are unpreferable to 
  the superconductivity.
  }
  \label{fig:regionMphiMsigma_vorton}
  \bigskip
\end{figure}

Fig.~\ref{fig:regionMphiMsigma_vorton} shows the parameter 
space $(m_\sigma,m_\phi)$ by colored regions in which some of 
the vortons die by now due to the decay of the current and 
charge. The black contours in the colored regions indicate 
the values of $T_\mr{gen}^\mr{max}$, defined as the 
maximum $T_\mr{gen}$ among the living vortons in the 
present Universe, i.e., the temperature at which the oldest 
living vortons were produced. In the white blank regions, 
all vortons are stable and survive in the present Universe.
The top (bottom) two panels correspond to the conversion 
process (the conversion and two-body decay processes). 
The left panels correspond to the vorton 
curves while the right panels to the thermal ones.
The upper-left gray regions in the plots are unpreferable to
the superconductivity. We can see that the vorton curve is 
more crucial for the vorton lifetime than the thermal one.

From these figures, we obtain the effects of zero mode decay 
on the vorton stability. Once one takes into account the 
two-body decay process with the (sizable) Yukawa coupling $\lambda$, 
the vortons cannot be stable due to the zero mode decay caused 
by the vorton curves (bottom-left panel).
On the other hand, the vortons can be stable if the model 
does not contain such a decay process and the mass 
parameters $m_\sigma$ and $m_\phi$ are not so hierarchical, 
$10^{-2}\lesssim m_\sigma/m_\phi \lesssim 1$, which 
forbids the decay kinematically, as seen in the white 
regions in the top two panels.

\bigskip

\section{Conclusion}
\label{sec:conclusions}

In this work, we studied the stability of zero modes traveling along the 
bosonic superconducting string. In particular, we focused on the zero 
mode quantum decay via the deformation of the straight string solution.
This deformation is treated as string curves of several types and the 
effective interaction vertices including the string deformation 
are evaluated. Then we calculated the decay widths for the conversion 
to bulk massive modes and the two-body decay to bulk light fermions.
The approximate forms of the decay widths were derived by regarding 
the scalar mass as a square well potential. We numerically evaluated 
the behavior of these decay widths and found the parameter space where 
stable superconducting current and vortons are able to exist.
These results are summarized in Figs.~\ref{fig:regionMphiMsigma} 
and \ref{fig:regionMphiMsigma_vorton}.

In the present $U(1) \times \tilde{U}(1)$ model, the hierarchy among 
mass parameters is not favored if superconducting strings survive to 
the present Universe. 
The vortons could have somewhat long lifetime but the current on the 
vortons disappear due to the quantum decay processes at some stage. 
Even in this case, the vortons and their decays may have impacts 
on cosmology. We considered the behavior of superconducting current 
in the model where two complex scalars are introduced. This minimal setup 
has a few number of couplings and the existence of the superconducting 
solution requires constraints on the parameters. 
That leads to a similar magnitude between the widths of string and of 
the potential trapping the zero mode.
In more general models admitting the bosonic superconductivity, 
these widths can be different and the parameter space for the existence 
of long-lived currents may be different from our results. 
For example, in the DFSZ model, two Higgs doublets and one Peccei-Quinn 
scalar are introduced, whose interactions are suppressed by a large 
Peccei-Quinn scale. This model has a possibility that the conditions 
for the existence of stable superconducting strings and vortons are 
relaxed, and there can be more long-lived objects in wider parameter 
space. It is worth considering cosmological impacts of these charged 
objects in the DFSZ model. We leave it as future work.

\section*{Acknowledgments}
\noindent
The authors would like to thank Keisuke Harigaya, 
Koji Hashimoto, and Kai Schmitz for useful discussions and comments.
This work is supported in part by JSPS KAKENHI Grant Numbers
JP21J01117 (YH) and JP20K03949 (KY). 

%%%%%%%%%%%%%%%%%%%%%%%%%%%%%%%%%%%%%%%%%%%%%%%%%%%%%%%%%%%%%%%%%%
\bigskip

\appendix

\section{Scalar Lagrangian}
\label{app:details}

The scalar Lagrangian of the $U(1) \times \tilde{U}(1)$ model is 
originally given by \eqref{eq:Lagrangian-U1xtU1} with the real 
couplings $\lambda_\phi$, $\lambda_\sigma$, and $\kappa$. 
Around the classical string background $\phi_\cl$ 
and $\sigma_\cl$, the scalar fields are described by 
$\Phi = \phi_\cl + \phi$ and $\Sigma = \sigma_\cl + \sigma$.
Substituting them into the Lagrangian and using the EOMs for 
$\phi_\cl$ and $\sigma_\cl$, we find the scalar Lagrangian 
for the quantum fields $\phi$ and $\sigma$:
\begin{align}
  \mc{L} &\,=\, 
  |\del_\mu \phi|^2 + |\del_\mu \sigma|^2 
  - \frac{\lambda_\phi}{4} |\phi|^4
  - \frac{\lambda_\sigma}{4} |\sigma|^4
  - \kappa |\phi|^2 |\sigma|^2   \nonumber  \\
 & \qquad
  - \frac{\lambda_\phi}{4} \Bigl[
   (\phi_\cl^* \phi + \phi_\cl \phi^*)^2 
    + 2 (|\phi_\cl|^2 - v_\phi^2) |\phi|^2 \Bigr] 
  \nn \\
  & \qquad
  - \frac{\lambda_\sigma}{4} \Bigl[ 
  ( \sigma_\cl^* \sigma + \sigma_\cl \sigma^*)^2
    + 2 (|\sigma_\cl|^2 - v_\sigma^2) |\sigma|^2 \Bigr]
  \nn \\[1mm]
  & \qquad
  - \kappa \Bigl[
    (\phi_\cl^* \phi + \phi_\cl \phi^*) 
    (\sigma_\cl^* \sigma + \sigma_\cl \sigma^*) 
    +|\sigma_\cl|^2 |\phi|^2 
    +|\phi_\cl|^2 |\sigma|^2 
  \Bigr]    \nonumber \\
  & \qquad\quad
  - \frac{\lambda_\phi}{2} (\phi_\cl^* \phi 
  + \phi_\cl \phi^*) |\phi|^2 
  - \frac{\lambda_\sigma}{2} (\sigma_\cl^* \sigma 
  + \sigma_\cl \sigma^*) |\sigma|^2   \nonumber \\
  & \qquad\quad
    -\kappa \Bigl[ 
    ( \sigma_\cl^* \sigma+ \sigma_\cl \sigma^*) |\phi|^2 + 
    (\phi_\cl^* \phi + \phi_\cl \phi^*) |\sigma|^2 
   \Bigr]  \nonumber \\
  & \qquad\qquad 
  +\frac{\lambda_\phi}{4} (|\phi_\cl|^4 - v_\phi^4)
  + \frac{\lambda_\sigma}{4} (|\sigma_\cl|^4 - v_\sigma^4)
  - \kappa |\phi_\cl|^2 |\sigma_\cl|^2  .
  \label{eq:full-Lagrangian}
\end{align}
The first line includes the kinetic terms and the 
genuine interactions of the quantum fields. The following 
three lines are quadratic in the quantum fields, that is, 
the mass terms in the string background. The next two line 
contains the scalar cubic interactions, and 
the last one just the constant.

When we decompose $\phi$ and $\sigma$ as 
\begin{align}
  \phi \,=\, \frac{\phi_R + i \phi_I}{\sqrt{2}},  \qquad 
  \sigma \,=\, \frac{\sigma_R + i \sigma_I}{\sqrt{2}},
\end{align}
and assume $\sigma_\cl$ is real-valued, the scalar mass matrix 
in the string background reads 
{
\setlength{\dashlinedash}{2pt}
\setlength{\dashlinegap}{2pt}
\begin{align}
  \cal{M} \,=\, \left(
 \begin{array}{c:c:c:c}
   \begin{minipage}{34mm}{\small
     ~~~$\tfrac{\lambda_\phi}{2}(|\phi_\cl|^2-v_\phi^2)$ \\
     $+\kappa\sigma_\cl^2$
     $+\lambda_\phi(\text{Re}\,\phi_\cl)^2$ } \end{minipage} & 
   \lambda_\phi(\text{Re}\,\phi_\cl)(\text{Im}\,\phi_\cl) & 
   2\kappa(\text{Re}\,\phi_\cl)\sigma_\cl &   \\ \hdashline
   \lambda_\phi(\text{Re}\,\phi_\cl)(\text{Im}\,\phi_\cl) & 
   \begin{minipage}{34mm}{\small
     ~~~$\tfrac{\lambda_\phi}{2}(|\phi_\cl|^2-v_\phi^2)$ \\
     $+\kappa \sigma_\cl^2$ 
     $+\lambda_\phi(\text{Im}\,\phi_\cl)^2$ } \end{minipage} &
     2\kappa(\text{Im}\,\phi_\cl)\sigma_\cl &  \\ \hdashline
   2\kappa(\text{Re}\,\phi_\cl)\sigma_\cl & 
   2\kappa(\text{Im}\,\phi_\cl)\sigma_\cl & 
   \begin{minipage}{23mm}{\small 
     ~$\tfrac{\lambda_\sigma}{2}(3\sigma_\cl^2-v_\sigma^2)$  \\
     $~~~+\kappa|\phi_\cl|^2$ } \end{minipage} & \\ \hdashline
   & & & 
   \begin{minipage}{19mm}{\small
    $\tfrac{\lambda_\sigma}{2}(\sigma_\cl^2-v_\sigma^2)$  \\
    $~~+\kappa|\phi_\cl|^2$ } \end{minipage}
 \end{array} \right)
\end{align}}%
in the $(\phi_R,\phi_I,\sigma_R,\sigma_I)$ basis.
One can see that $\sigma_I$, which contains the zero mode 
(the current carrier of superconductivity), is decoupled from 
the other scalars in the quadratic order. The other three 
modes mix with each other via the position-dependent 
quadratic couplings. Its diagonalization is not necessarily 
useful since it generally turns out to induce complex kinetic 
terms. We also notice that the Lagrangian has a $Z_2$ parity 
under which $\sigma_I$ is odd. It is originated from the 
CP symmetry in the $\Sigma$ sector, 
$\Sigma\leftrightarrow\Sigma^*$. That restricts the 
interaction form, in particular, there is no decay vertex of 
$\sigma_I$ (no term involving the first power of $\sigma_I$).

In the case of straight string, the position-dependent 
mass of $\sigma_I$ is defined by
\begin{align}
  M_I^2(r) \,=\, \frac{\lambda_\sigma}{2} 
  (\hat{\sigma}^2 - v_\sigma^2) + \kappa |\hat{\phi}|^2.
\end{align}
The string profile functions have the asymptotic behavior 
$f(r)\to 1$ and $h(r)\to 0$ far from the string, and then we 
find $M_I^2(r)\to m_\sigma^2$. In the same limit, $\sigma_R$ 
also has a mass $m_\sigma$. On the other hand, the $\phi$ 
sector contains two mass eigenvalues $0$ 
and $m_\phi^2=\lambda_\phi v_\phi^2$. 
The massless mode is the direction 
$(\text{Re}\,\hat{\phi})\phi_I-(\text{Im}\,\hat{\phi})\phi_R$,  
the Nambu-Goldstone boson associated with 
the $\tilde{U}(1)$ breaking.

\bigskip

\section{Propagator in the string background}
\label{app:propagator}

We present the formal expression of the four-dimensional 
propagator for the $\sigma_I(x)$ field that is defined by
\begin{align}
  \braket{0| \mr{T} \sigma_I(x) \sigma_I(x')|0} \,=\, G(x, x').
\end{align}
The propagator $G(x,x')$ is the inverse of the Klein-Gordon 
operator in the string background
\begin{align}
  [\, \square + M_I^2(r)\,] G(x,x') \,=\, \delta^4(x-x'),
\end{align}
where the position-dependent mass operator $M_I^2$ is 
defined by \eqref{eq:M_I^2}. Performing the Fourier 
transformation with respect to the $t$, $z$, $\theta$ 
variables, we have 
\begin{align}
  \Bigl[ \frac{-1}{r}\del_r(r\del_r) +\frac{l^2}{r^2} 
  +M_I^2(r) - \ms{k}^2 \Bigr] G_\ms{k}^l(r,r') \,=\,
  \frac{1}{r}\delta(r-r') ,
  \label{eq:Green_kl}
\end{align}
\begin{align}
  G(x, x') \,=\, \int\!\frac{d^2 \ms{k}}{(2\pi)^3}\,
  \sum_l G^l_{\ms{k}}(r, r') e^{- i \ms{k} 
  \cdot (\ms{x} - \ms{x}')} e^{i l (\theta- \theta')} .
\end{align}

First, dropping the right-handed side of \eqref{eq:Green_kl}, 
we solve the equation in the region $r \in [r_0,\,r_\infty]$ 
(e.g., $r_0$ and $r_\infty$ mean the UV and infrared (IR) 
cutoff, respectively). Before specifying the boundary 
conditions at $r=r_0$ and $r=r_\infty$, we have two 
independent solutions $\varphi_1(r)$ and $\varphi_2(r)$. 
These solutions depend on the momenta $\ms{k}$ and $l$. Here 
and hereafter we do not explicitly show these momentum 
indices for notation simplicity. With these solution at hand, 
the propagator is written down as
\begin{align}
  G_<(r,r') &=
  A_<(r')\varphi_1(r) + B_<(r')\varphi_2(r) 
  \qquad \text{for}\;\; r<r' \,,\\
  G_>(r,r') &=
  A_>(r')\varphi_1(r) + B_>(r')\varphi_2(r) 
  \qquad \text{for}\;\; r>r'\,.
\end{align}
The constants of integration $A_<$ $A_>$, $B_<$, $B_>$ are
determined by the boundary conditions and matching in the 
following way.

Let us consider the Neumann conditions at both boundaries. 
That implies
\begin{align}
  A_<(r')\varphi_1'(r_0) +B_<(r')\varphi_2'(r_0) &=\, 0 \,, \\
  A_>(r')\varphi_1'(r_\infty) +B_>(r')\varphi_2'(r_\infty) 
  &=\, 0 \,,
\end{align}
where $\varphi_i'$ denotes the derivative of $\varphi_i$. 
The propagators in the two regions ($r<r'$ and $r>r'$) are 
matched at $r=r'$ taking into account the right-handed side 
of \eqref{eq:Green_kl}. The continuity of wavefunctions 
and the discontinuity of the slopes which follows from
the integration of \eqref{eq:Green_kl} around $r=r'$ lead 
to the matching conditions
\begin{align}
  A_<(r')\varphi_1(r') +B_<(r')\varphi_2(r') &=
  A_>(r')\varphi_1(r') +B_>(r')\varphi_2(r') , \\
  A_<(r')\varphi_1'(r') +B_<(r')\varphi_2'(r') &=
  A_>(r')\varphi_1'(r') +B_>(r')\varphi_2'(r') +\frac{1}{r'} \,.
\end{align}
With all these conditions, we can fix the constants of 
integration and find the propagator 
\begin{align}
  G_\ms{k}^{l\,(\text{NN})}(r,r') =
  \frac{g_p(r_>,r_0)g_p(r_>,r_\infty)}{r'g_p(r',r')g_{pp}(r_\infty,r_0)},
  \label{eq:propNN}
\end{align}
where $r_<$ ($r_>$) stands for the lesser (greater) of $r$ and $r'$. 
The index $\text{NN}$ is attached to indicate that the propagator
satisfies the Neumann boundary conditions at $r=r_0$ and
$r=r_\infty$. The functions $g_p$ and $g_{pp}$ are defined as
\begin{align}
  g(r_1,r_2) &= \varphi_1(r_1)\varphi_2(r_2) 
   - \varphi_1(r_2)\varphi_2(r_1) \,, \\
  g_p(r_1,r_2) &= \del_{r_2} g(r_1,r_2)  \,, \\
  g_{pp}(r_1,r_2) &= \del_{r_1}\del_{r_2} g(r_1,r_2) \,.
\end{align}
It is easy to verify that $G_\ms{k}^{l\,(\text{NN})}$ satisfies 
$[\tfrac{-1}{r}\del_r(r\del_r) +l^2/r^2
+M_I^2(r) -\ms{k}^2 ] G_\ms{k}^{l\,(\text{NN})}(r,r') =$
$[\tfrac{-1}{r'}\del_{r'}(r'\del_{r'}) +l^2/{r'}^2
+M_I^2(r') -\ms{k}^2 ] G_\ms{k}^{l\,(\text{NN})}(r,r') =0$ 
at $r\neq r'$, noting that 
$[\tfrac{-1}{r}\del_r(r\del_r) +l^2/r^2
+M_I^2(r) -\ms{k}^2 ]g_p(r,r')=$ $[\tfrac{-1}{r}\del_r(r\del_r) 
+l^2/r^2 +M_I^2(r) -\ms{k}^2 ](rg_p(r,r))=0$.

The propagators for the other boundary conditions can be 
found in parallel ways as
\begin{align}
  G_\ms{k}^{l\,(\text{ND})}(r,r') &=
  \frac{g_p(r_<,r_0)g(r_>,r_\infty)}{r'g_p(r',r')g_p(r_\infty,r_0)}
  \, , \\
  G_\ms{k}^{l\,(\text{DN})}(r,r') &=
  \frac{g(r_0,r_<)g_p(r_>,r_\infty)}{r'g_p(r',r')g_p(r_0,r_\infty)} 
  \, , \\
  G_\ms{k}^{l\,(\text{DD})}(r,r') &=
  \frac{g(r_0,r_<)g(r_>,r_\infty)}{r'g_p(r',r')g(r_0,r_\infty)} 
  \, ,
\end{align}
where the superscript D denotes the Dirichlet boundary condition, 
for example, $G^{(\text{DD})}$ vanishes at both boundaries.

The mass spectrum in the effective theory is extracted 
from the poles of propagators. From \eqref{eq:propNN}, 
we find the pole condition
\begin{align}
  0 = g_{pp}(r_\infty,r_0) = \varphi_1'(r_\infty)\varphi_2'(r_0) 
   -\varphi_1'(r_0)\varphi_2'(r_\infty) \,.
  \quad (\text{for NN})
\end{align}
Remembering $\varphi_{1,2}$ depend on the momenta 
$\ms{k}$ and $l$, this equation determines $\ms{k}^2$ 
in terms of $l$, $r_{0,\infty}$, and other quantities. 
For the other-type of propagators, we have
\begin{align}
  0 &= g_p(r_\infty,r_0) = \varphi_1(r_\infty)\varphi_2'(r_0) 
   -\varphi_1'(r_0)\varphi_2(r_\infty) \,,
  \quad (\text{for ND}) 
  \label{eq:bcND}   \\
  0 &= g_p(r_0,r_\infty) = \varphi_1(r_0)\varphi_2'(r_\infty) 
   -\varphi_1'(r_\infty)\varphi_2(r_0) \,, 
  \quad (\text{for DN}) \\
  0 &= g(r_0,r_\infty) = \varphi_1(r_0)\varphi_2(r_\infty) 
   -\varphi_1(r_\infty)\varphi_2(r_0) \,.
  \quad\;\> (\text{for DD})
\end{align}
The zero mode ($\ms{k}^2=l=0$) is expected to be found 
in the spectrum \eqref{eq:bcND} (see the boundary condition 
for the string profile functions \eqref{eq:bc}).

As an application, we consider the propagator for an 
approximate mass operator \eqref{eq:approMI}. We first 
solve \eqref{eq:Green_kl} setting the right-handed side to 
zero. Two independent solutions are obtained by taking 
account of the continuities of wavefunctions and their 
slopes at $r=r_b$. The explicit forms of the solutions are
\begin{align}
  \varphi_1(r) &= 
  \begin{cases}
    J_l(\bar{M}r) + u(\bar{m}_\sigma,\bar{M}) H_l(\bar{M}r)
    &\quad \text{for}\;\; r<r_b \,,\\
    v(\bar{m}_\sigma,\bar{M}) H_l(\bar{m}_\sigma r) 
    &\quad \text{for}\;\; r>r_b \,,
  \end{cases}   \\[1mm]
  \varphi_2(r) &= 
  \begin{cases}
    v(\bar{M},\bar{m}_\sigma) H_l(\bar{M}r) 
    &\quad \text{for}\;\; r<r_b \,, \\
    J_l(\bar{m}_\sigma r) + u(\bar{M},\bar{m}_\sigma)
    H_l(\bar{m}_\sigma r)
    &\quad \text{for}\;\; r>r_b \,.
  \end{cases}
\end{align}
where $\bar{M}^2=M^2+\ms{k}^2$, 
$\bar{m}_\sigma^2=-m_\sigma^2+\ms{k}^2$, and 
$J_l$ ($H_l$) are the Bessel (Hankel) functions of the first 
kind. The functions $u$ and $v$ are defined by
\begin{align}
  u(\alpha,\beta) &= \frac{J_l(\beta r_b)\alpha H_l'(\alpha r_b)
  -\beta J_l'(\beta r_b)H_l(\alpha r_b)}{
  H_l(\alpha r_b)\beta H_l'(\beta r_b)
  -\alpha H_l'(\alpha r_b)H_l(\beta r_b)}   \,, \\
  v(\alpha,\beta) &= \frac{2i/\pi r_b}{
  H_l(\alpha r_b)\beta H_l'(\beta r_b)
  -\alpha H_l'(\alpha r_b)H_l(\beta r_b)} \,.
\end{align}
The propagator, e.g., $G_k^{l\,\text{(NN)}}$ for the 
Neumann boundary conditions at both sides, is obtained 
from \eqref{eq:propNN} with these solutions $\varphi_{1,2}$. 
The dominant contribution to the zero mode decay may be 
given by the mediator with the Neumann conditions at 
both boundaries, which means that it interacts with the 
zero mode in the string and can also escape to the outside 
of the string. Such process is described by the propagator 
between the UV and IR cutoff scales, which is found to take 
an approximate value
$G_k^{l\,(\text{NN})}(r_0,r_\infty)\sim 
(\bar{m}\bar{M}r_b r_0)^l/(\bar{m}r_\infty)^{1/2}$.

\bigskip

\section{Temperature dependence of $\Gamma$}
\label{app:Gamma-temperature}

\begin{figure}[p]
  \centering
  \includegraphics[width=0.48\textwidth]{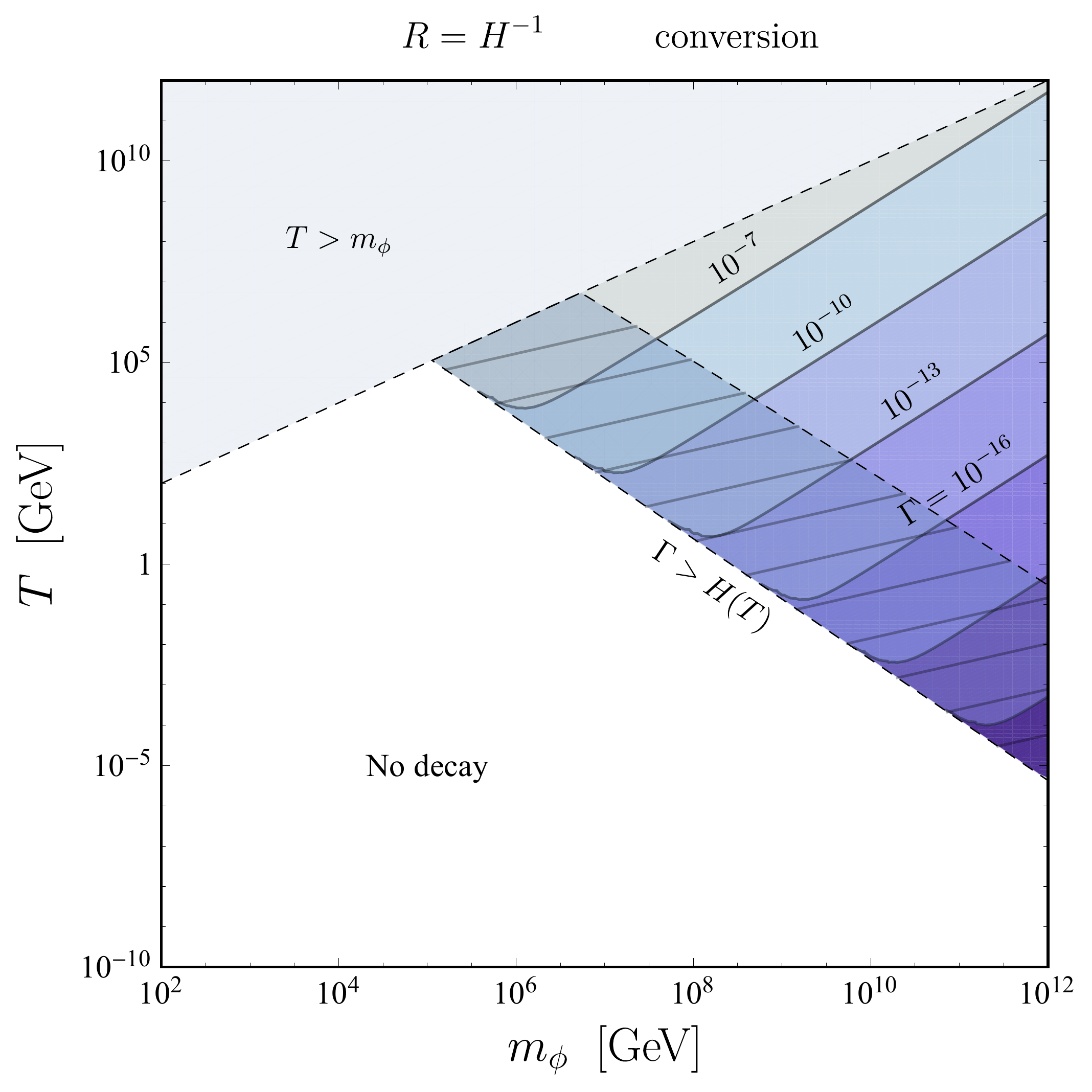}
  \quad
  \includegraphics[width=0.48\textwidth]{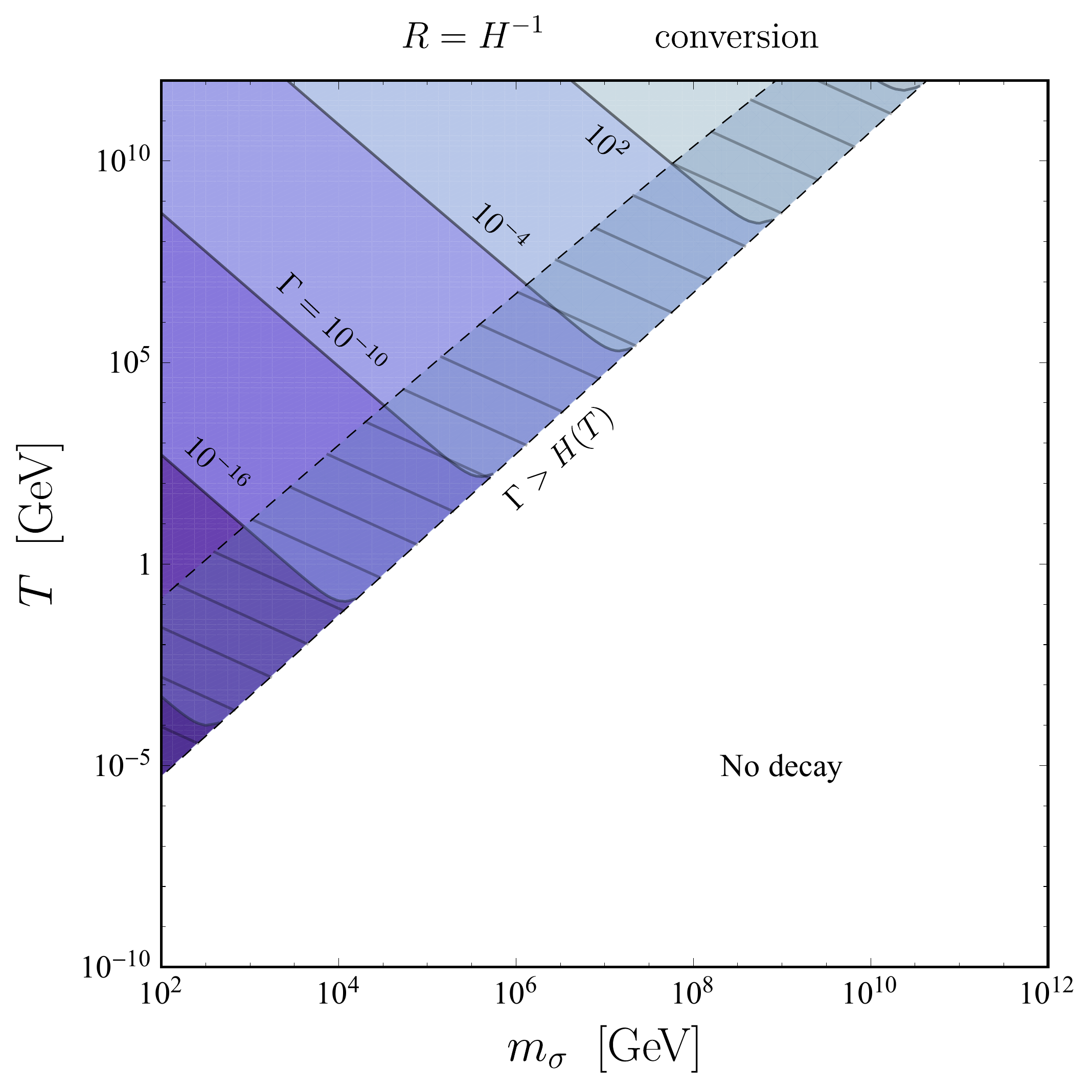}
  \\[4mm]
  \includegraphics[width=0.48\textwidth]{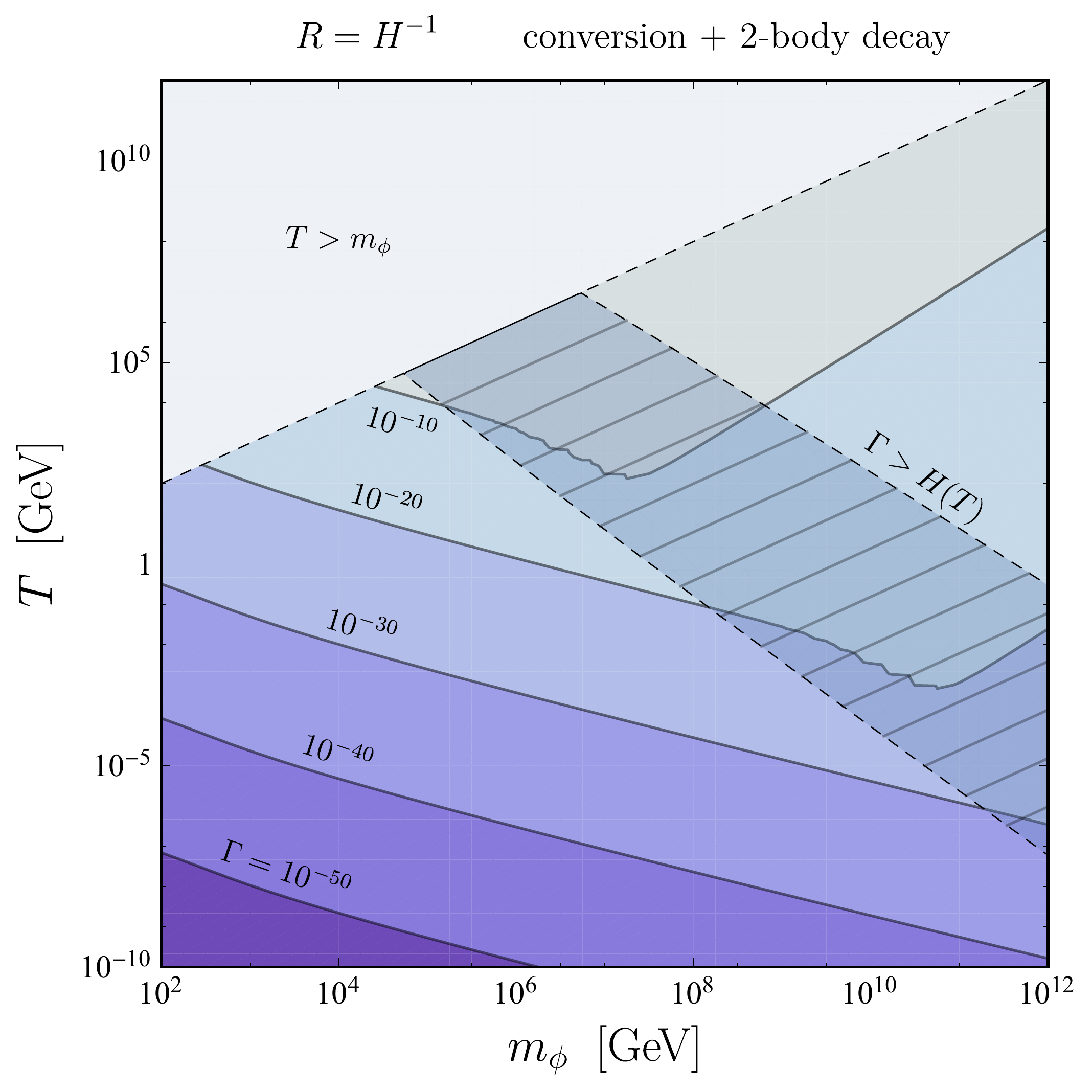}
  \quad
  \includegraphics[width=0.48\textwidth]{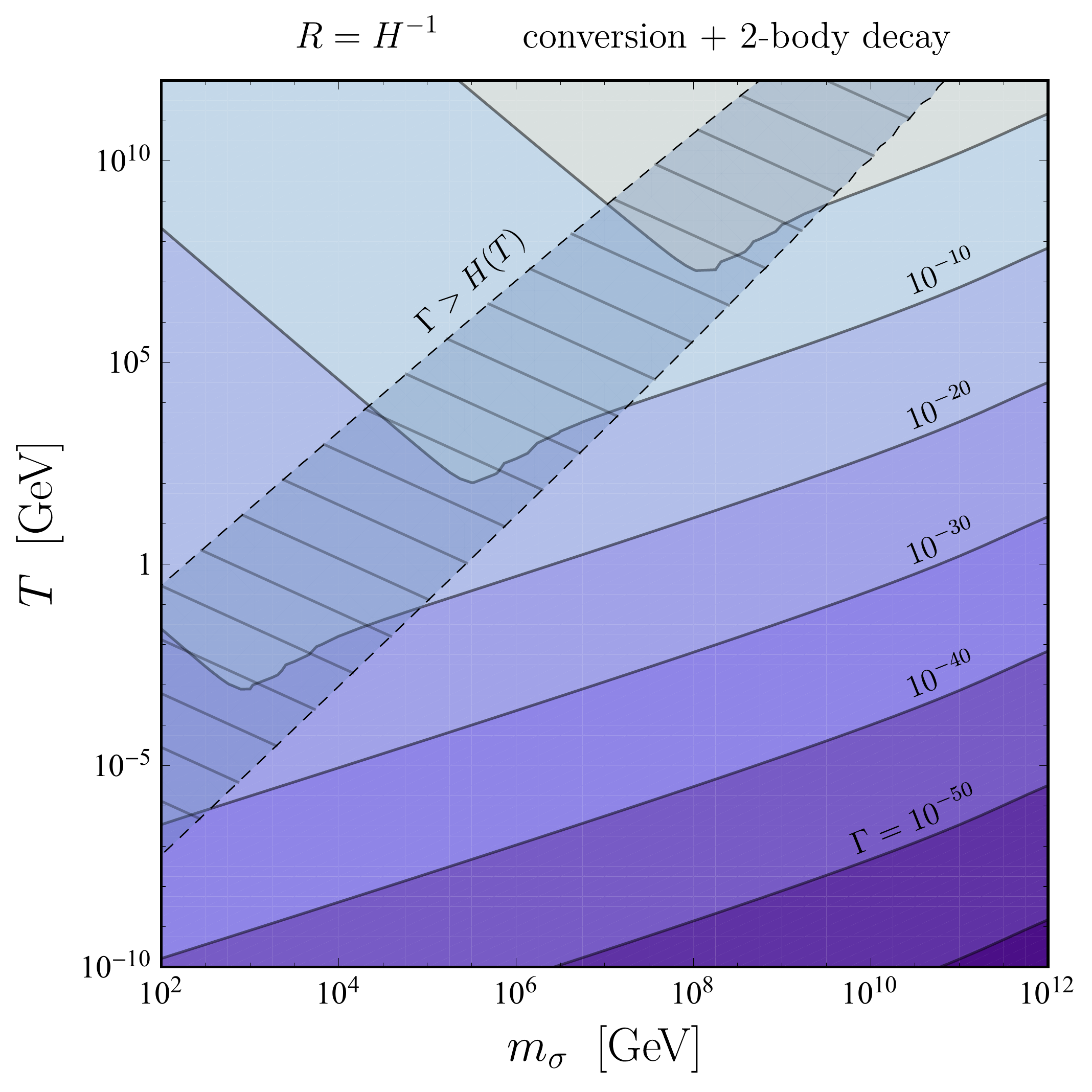}
  \caption{The decay width caused by the string curve with 
  $R=H^{-1}$ in the $(m_\phi,T)$-plane 
  ($m_\sigma=10^2~\mr{GeV}$) and in the $(m_\sigma, T)$-plane 
  ($m_\phi=10^{12}~\mr{GeV}$). 
  The top (bottom) two panels correspond to the conversion 
  process (the conversion and two-body decay processes). 
  The values around the contours are given in GeV unit. 
  In the white blank regions, the conversion process 
  is kinematically forbidden. The gray regions 
  with ``$T> m_\phi$'' is excluded since the string network 
  does not exist.
  }
  \label{fig:regionMT_Hubble}
  \bigskip
\end{figure}

\begin{figure}[p]
  \centering
  \includegraphics[width=0.48\textwidth]{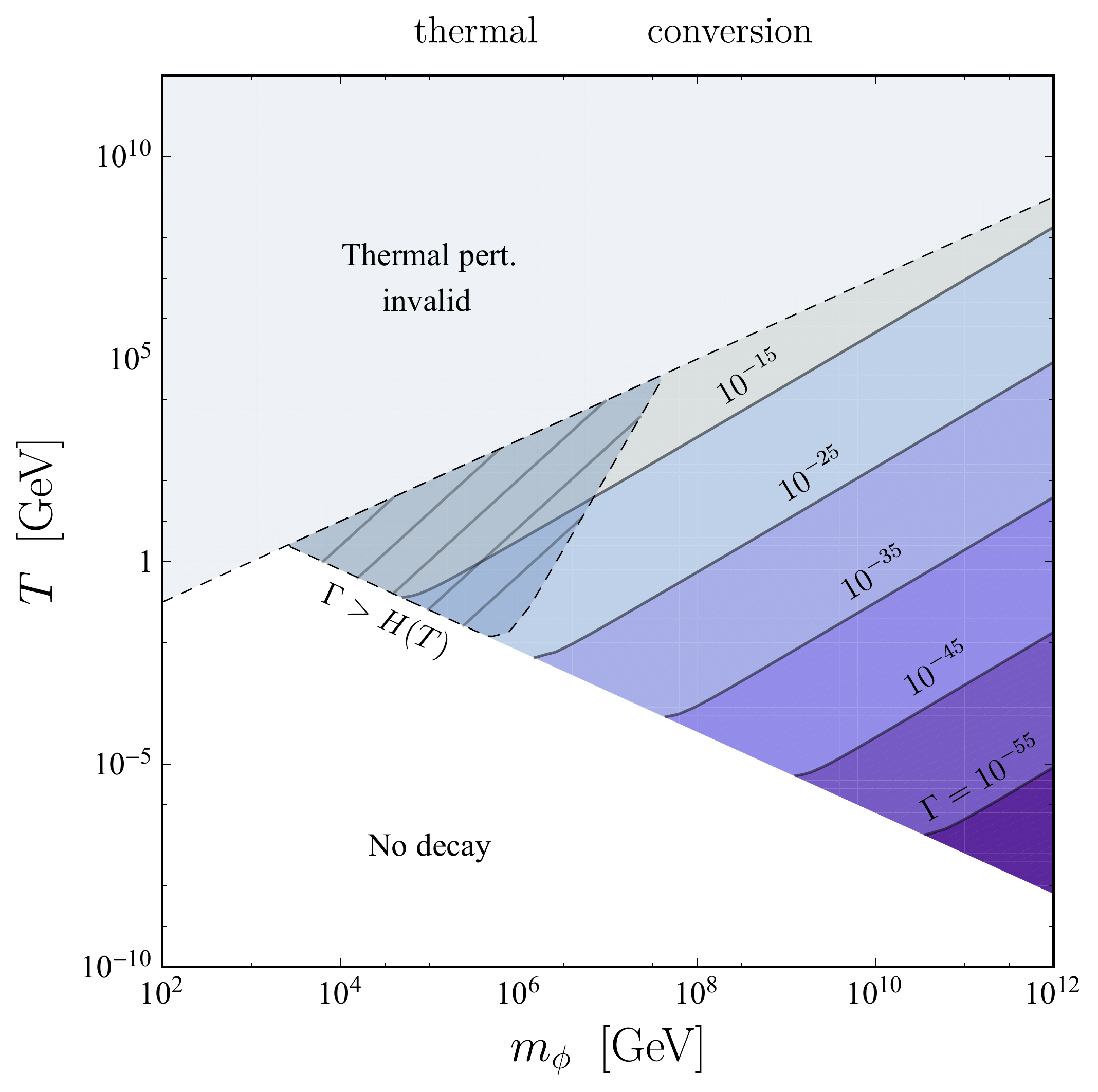}
  \quad
  \includegraphics[width=0.48\textwidth]{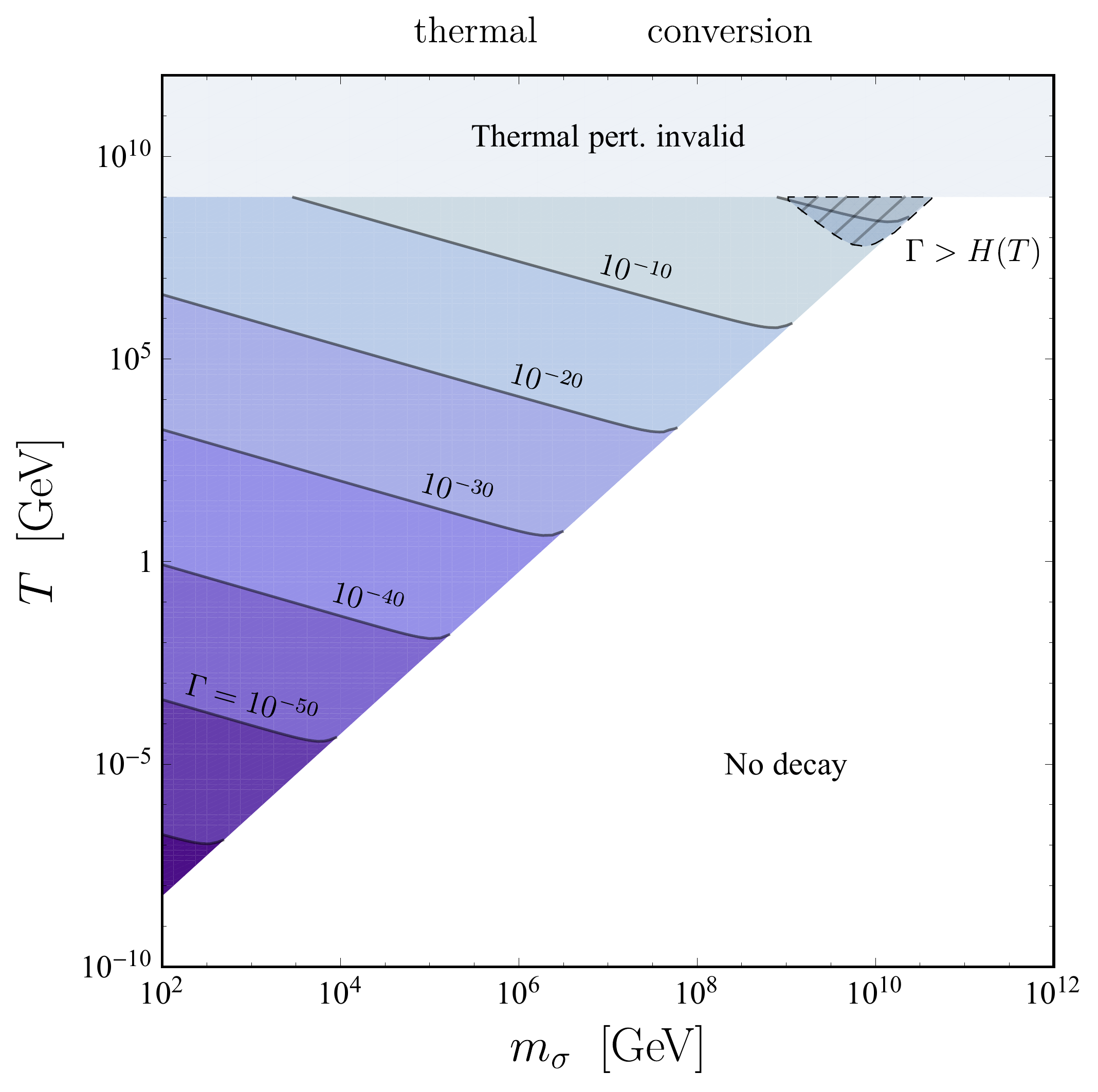}
  \\[4mm]
  \includegraphics[width=0.48\textwidth]{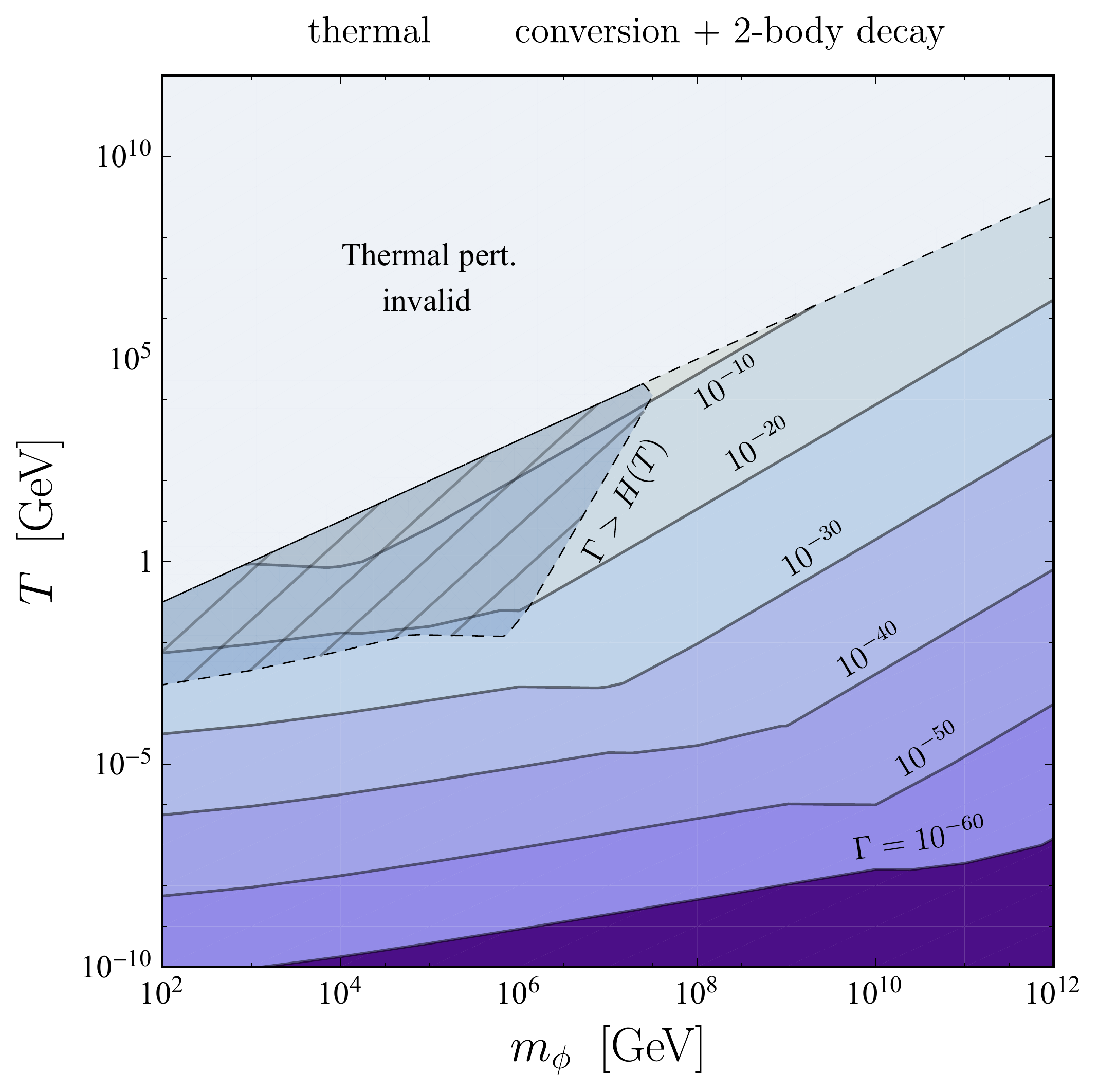}
  \quad
  \includegraphics[width=0.48\textwidth]{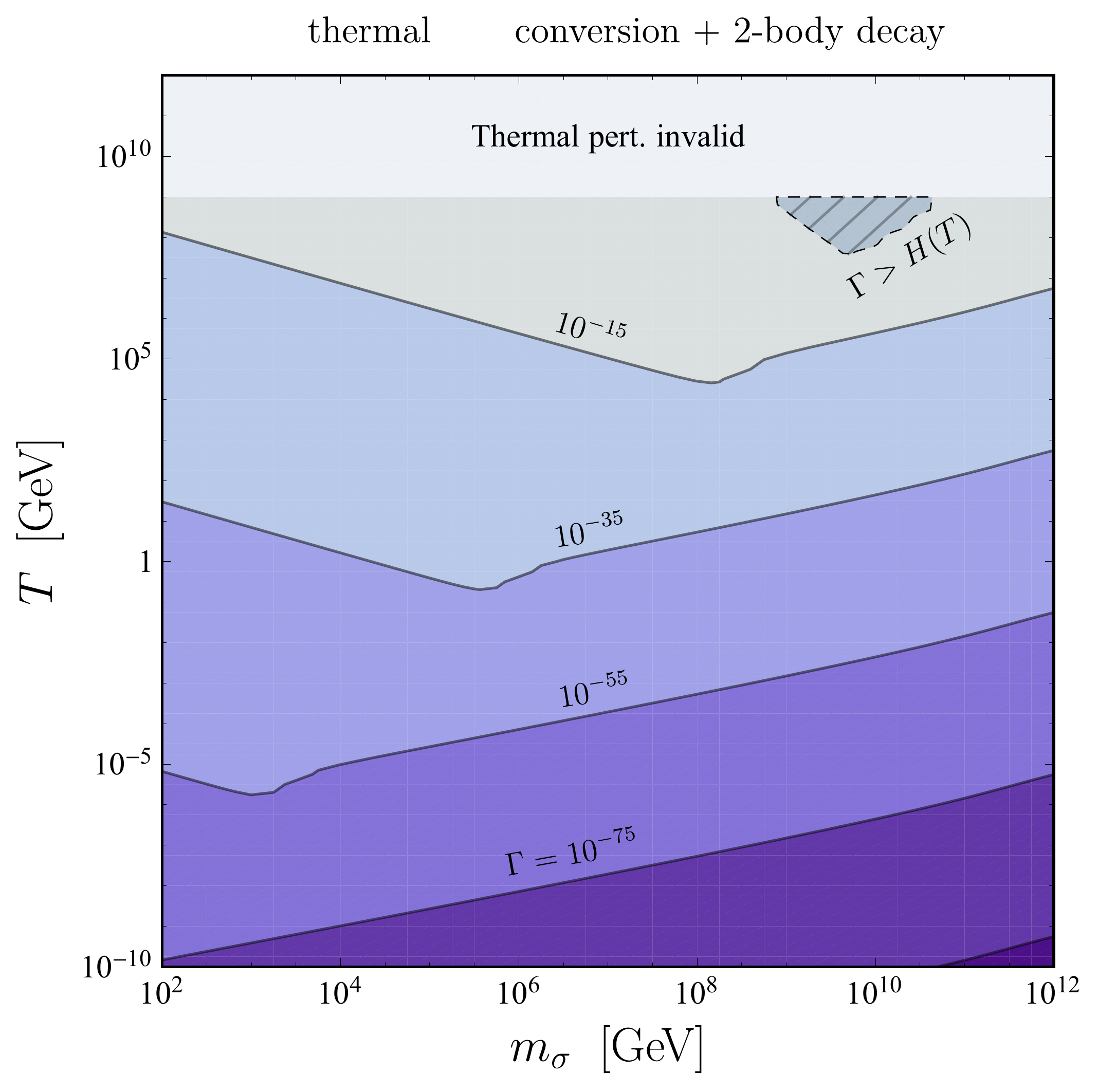}
  \caption{The decay width caused by the thermal curve 
  in the $(m_\phi,T)$-plane ($m_\sigma=10^2~\mr{GeV}$) and 
  in the $(m_\sigma, T)$-plane ($m_\phi=10^{12}~\mr{GeV}$). 
  The top (bottom) two panels correspond to the conversion 
  process (the conversion and two-body decay processes). 
  The values around the contours are given in GeV unit. 
  In the white blank regions, the conversion process 
  is kinematically forbidden. The gray regions 
  with ``Thermal pert.\ invalid'' is excluded since the 
  condition $T < 10^{-3} m_\phi$ is not satisfied.
  }
  \label{fig:regionMT_thermal}
  \bigskip
\end{figure}

We here present the numerical plots for the temperature 
$T$ dependence of the zero mode decay width $\Gamma$ 
for two types of string curves: the curve with $R=H^{-1}$ 
in the string network (Fig.~\ref{fig:regionMT_Hubble}) and 
the thermal one (Fig.~\ref{fig:regionMT_thermal}).
Both the figures show the values of $\Gamma$ 
in the $(m_\phi,T)$-plane with fixed $m_\sigma$ 
and the $(m_\sigma, T)$-plane with fixed $m_\phi$.
The top (bottom) two panels correspond to the conversion 
process (the conversion and two-body decay processes).
The shaded regions represent that the zero mode current 
cannot be significant, $\Gamma(T) > H(T)$, at the 
temperature $T$. The values around the contours are given 
in GeV unit. The gray regions with ``$T> m_\phi$'' 
and ``Thermal pert.\ invalid'' are excluded for the 
reasons that the string network does not exist and that 
the condition $T < 10^{-3} m_\phi$ is not satisfied, 
respectively. These plots agree well with 
the asymptotic expressions given in 
Eqs.~\eqref{eq:Gamma-2body-conv-Hub}--\eqref{eq:Gamma-2body-thermal}.


\begin{figure}
  \centering
  \includegraphics[width=0.48\textwidth]{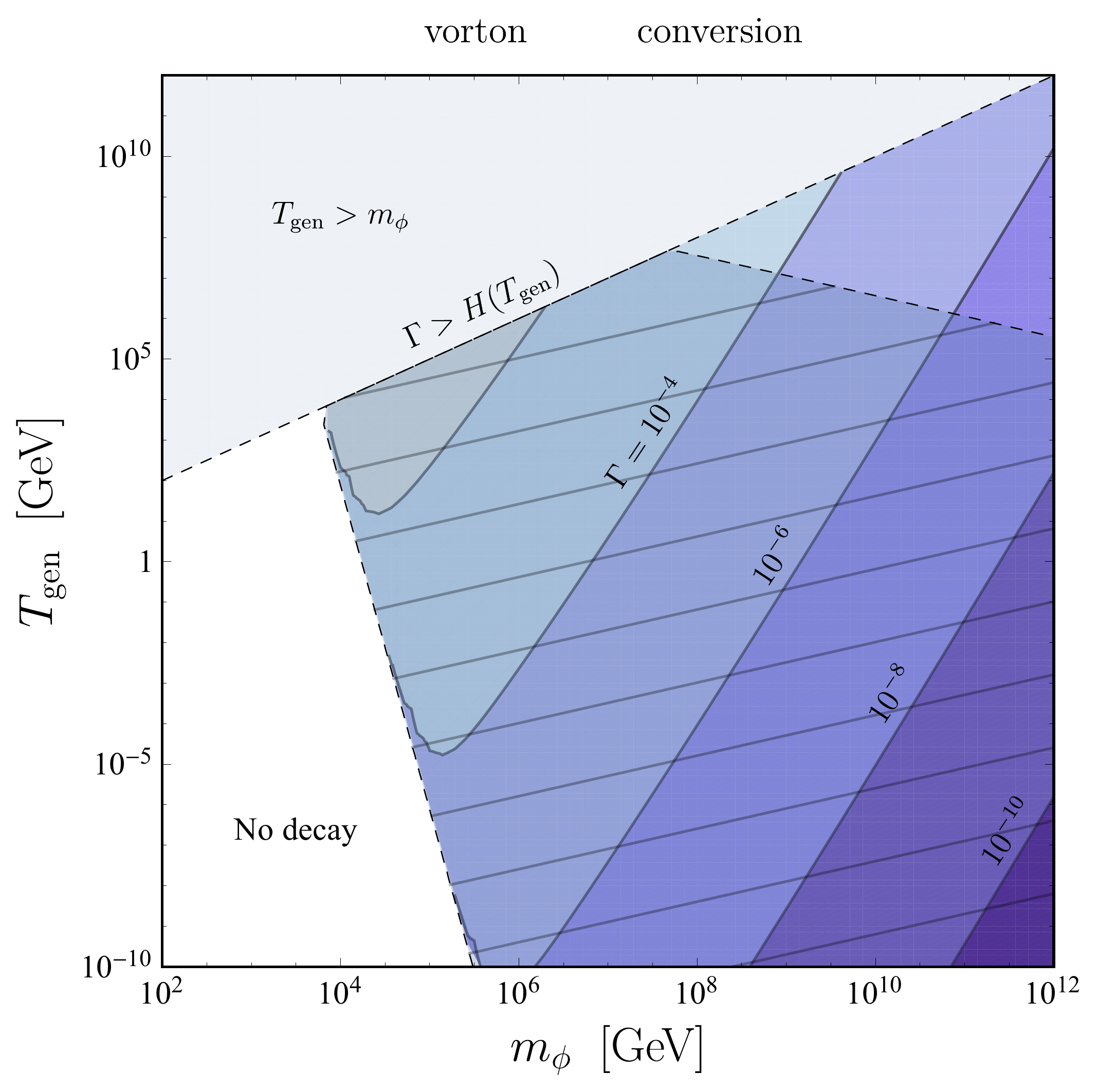}
  \quad
  \includegraphics[width=0.48\textwidth]{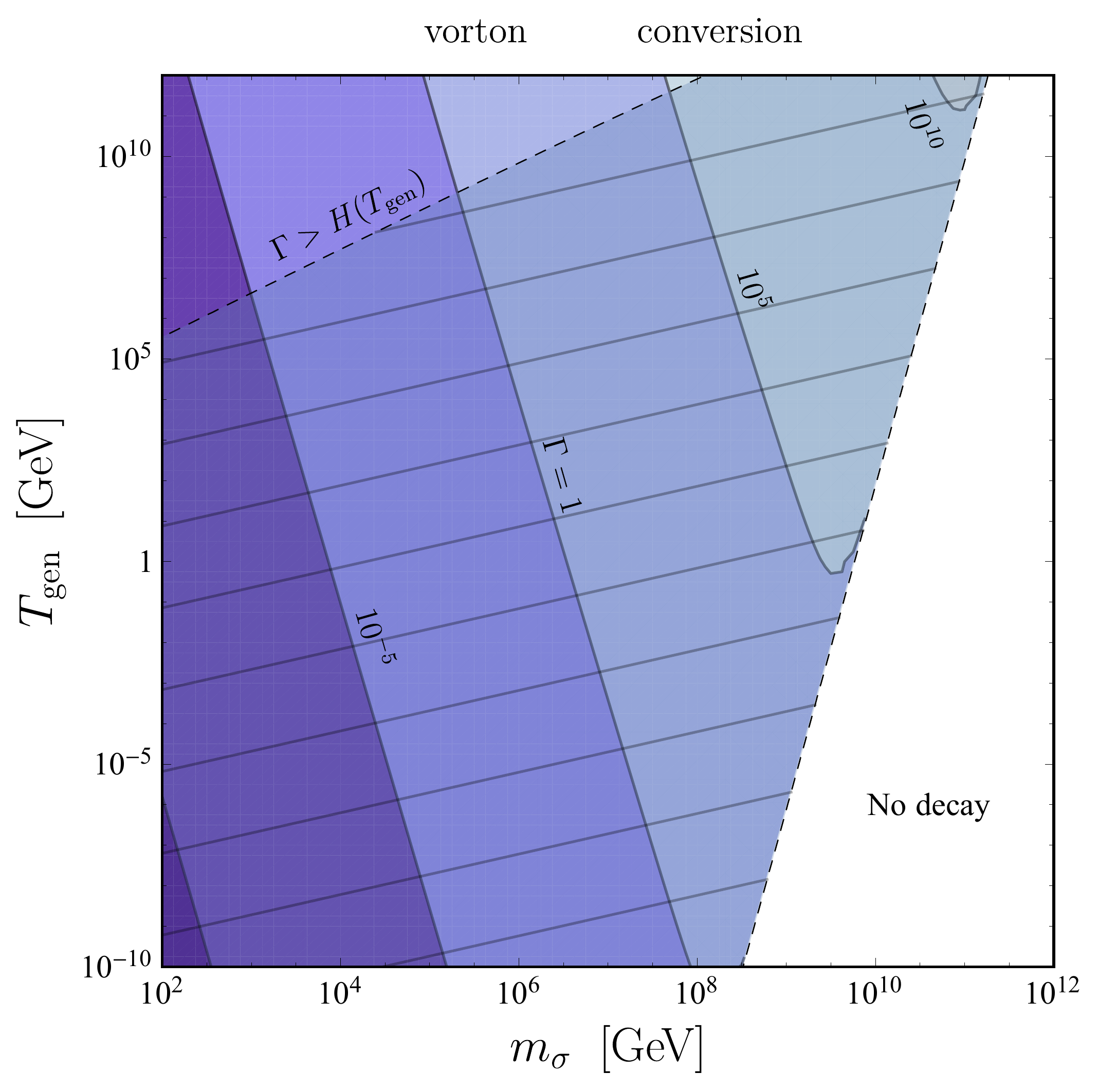}
  \\[4mm]
  \includegraphics[width=0.48\textwidth]{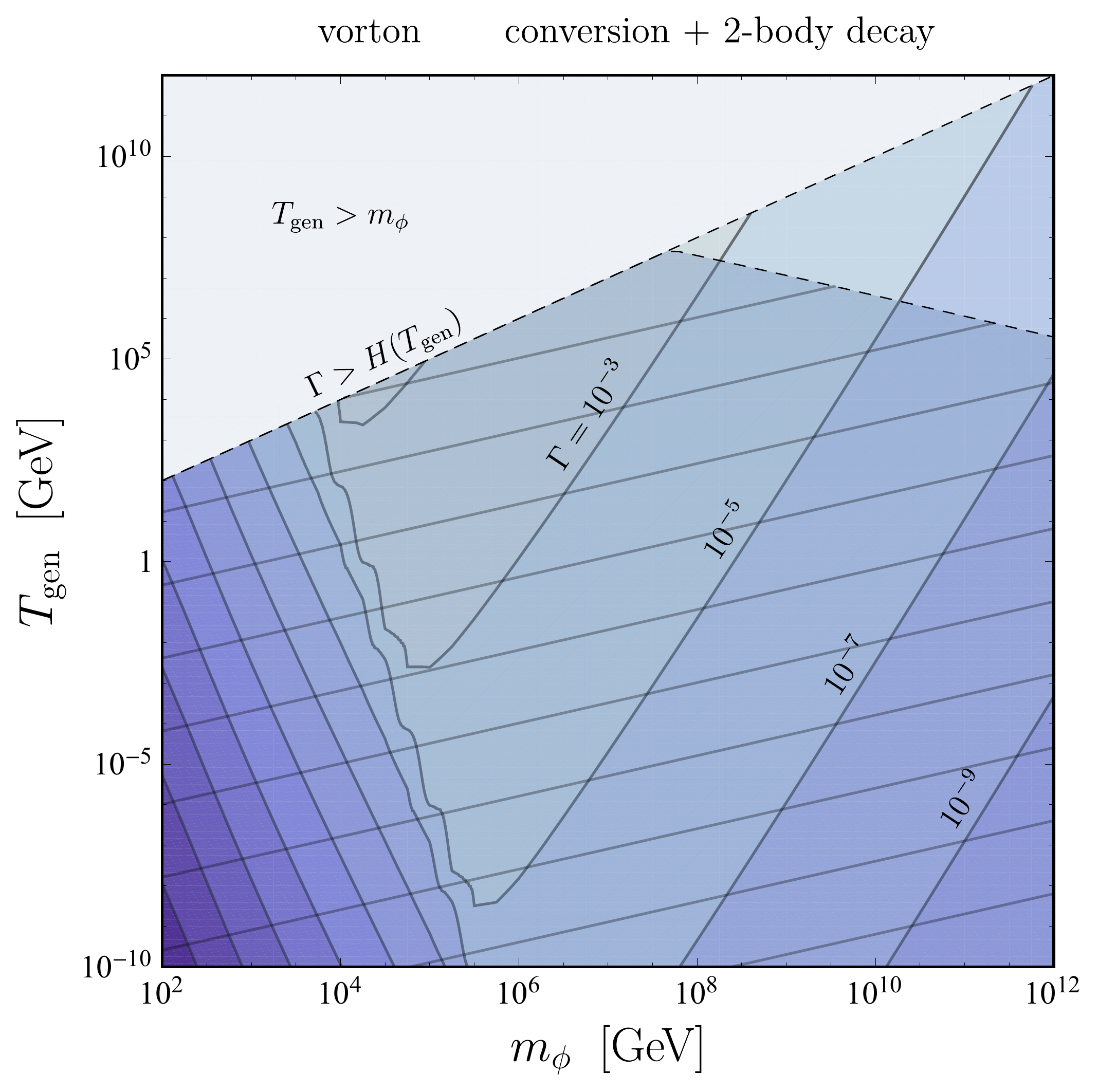}
  \quad
  \includegraphics[width=0.48\textwidth]{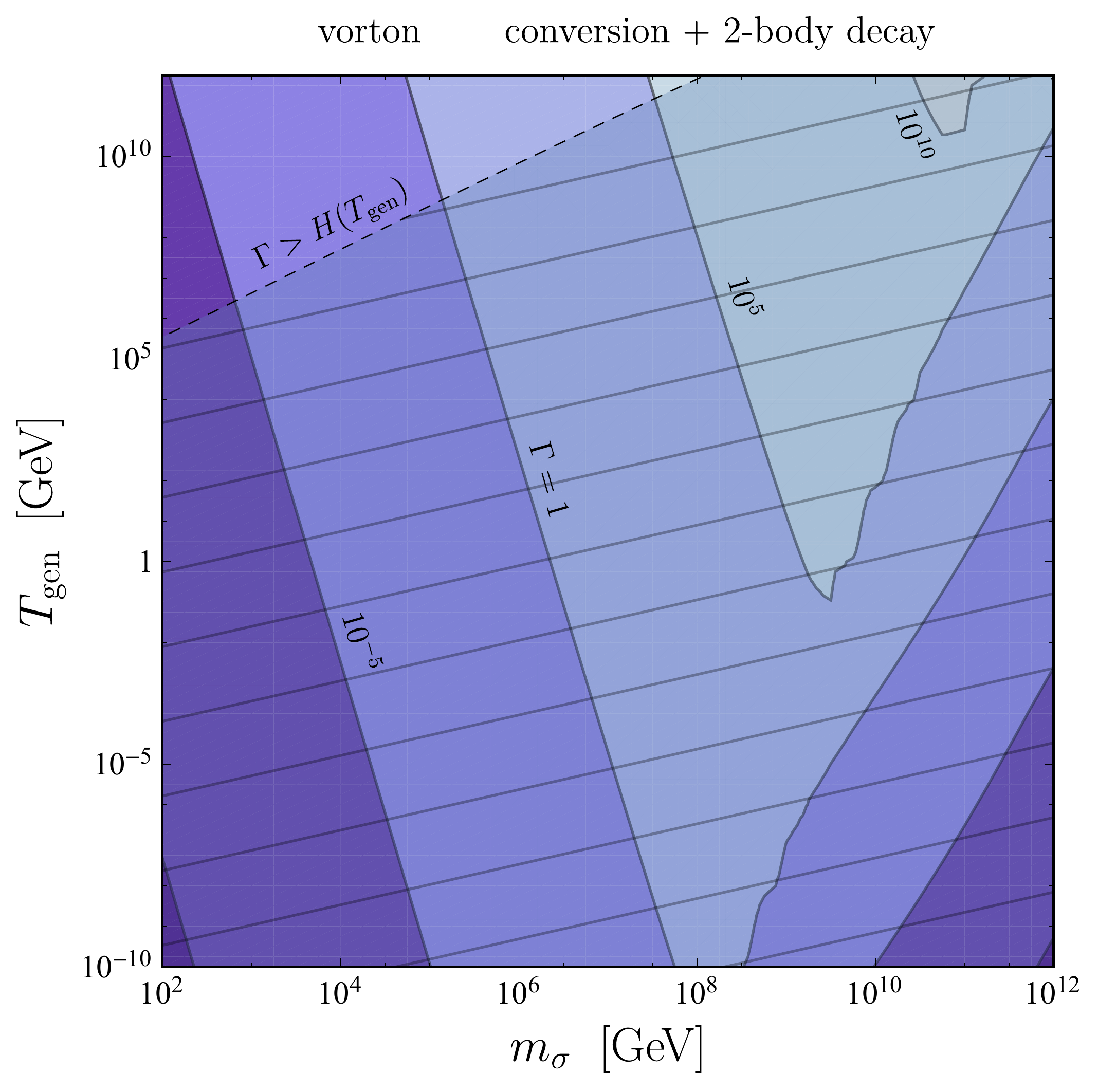}
  \caption{The values of $\Gamma$ in GeV unit for classically 
  stable vortons produced at the temperature $T=T_\mr{gen}$ are 
  shown in the $(m_\phi,T_\mr{gen})$-plane ($m_\sigma=10^2~\mr{GeV}$)
  and in the $(m_\sigma,T_\mr{gen})$-plane ($m_\phi=10^{12}~\mr{GeV}$).
  The decay is assumed to be caused by the vorton curvature 
  via the conversion (top panels) and the conversion and 
  two-body decay (bottom panels). 
  The white regions with ``No decay'' corresponds to the 
  region in which the conversion process is kinematically 
  forbidden.
  }
  \label{fig:regionMT_vorton}
  \bigskip
\end{figure}

On the other hand, the decay width $\Gamma$ due to the 
vorton curve does not depends on $T$ but on $T_\mr{gen}$, 
as mentioned in Sec.~\ref{sec:application}.
Fig.~\ref{fig:regionMT_vorton} shows $T_\mr{gen}$-dependence 
of $\Gamma$ for each vorton generated at $T=T_\mr{gen}$ 
in the $(m_\phi, T_\mr{gen})$ 
and $(m_\sigma, T_\mr{gen})$ planes 
in the top and bottom panels, respectively.
The left (right) panels are calculated by considering the 
conversion (the conversion and the two-body decay).
The contours correspond to the values of $\Gamma$ in GeV 
unit. The shaded regions indicate that the decay is 
significant compared to the Hubble parameter, 
$\Gamma(T_\mr{gen})>H(T_\mr{gen})$. In the 
white ``No decay'' region, the condition 
\eqref{eq:threshold} is not satisfied, forbidding the 
conversion kinematically. The upper-left gray regions 
in the top panels are excluded since the strings network 
does not exist when $T>m_\phi$. These plots agree with 
the asymptotic expressions in 
Eqs.~\eqref{eq:Gamma-2body-conv-vort} 
and \eqref{eq:Gamma-2body-vort}.

\clearpage

%% References
\newcommand{\arxivfont}{\rmfamily}
\bibliographystyle{yautphys}
\bibliography{references}

\end{document}